\documentclass[USenglish]{article}

\usepackage[utf8]{inputenc}%(only for the pdftex engine)
\usepackage[T1]{fontenc}
\usepackage{lmodern}
\usepackage[small]{dgruyter}
\usepackage{microtype}
\usepackage[bibstyle=biblatex-sp-unified, citestyle=sp-authoryear-comp, maxbibnames=99]{biblatex}
\usepackage{pdfpages}

\begin{document}

% authorship and affiliation details
\author*[1]{Jayden L. Macklin-Cordes}
\author[2]{Erich R. Round}
\runningauthor{Macklin-Cordes}
\runningtitle{Sampling and phylogenetic methods}
\affil[1]{Laboratoire Dynamique Du Langage (UMR 5596), CNRS / Université Lyon 2; Ancient Language Lab, The University of Queensland. E-mail: jayden.macklin-cordes@cnrs.fr}
\affil[2]{Surrey Morphology Group, University of Surrey; Ancient Language Lab, The University of Queensland; Max Planck Institute for the Science of Human History.}

% title details
  \title{Challenges of sampling and how phylogenetic comparative methods help}
  \runningtitle{Sampling and phylogenetic methods}
  \subtitle{With a case study of the Pama-Nyungan laminal contrast}

% abstract, keywords and PACS classification
  \abstract{Phylogenetic comparative methods are new in our field and are shrouded, for most linguists, in at least a little mystery. Yet the path that led to their discovery in comparative biology is so similar to the methodological history of balanced sampling, that it is only an accident of history that they were not discovered by a typologist. Here we clarify the essential logic behind phylogenetic comparative methods and their fundamental relatedness to a deep intellectual tradition focussed on sampling. Then we introduce concepts, methods and tools which will enable typologists to use these methods in everyday typological research. The key commonality of phylogenetic comparative methods and balanced sampling is that they attempt to deal with statistical non-independence due to genealogy. Whereas sampling can never achieve independence and requires most comparative data to be discarded, phylogenetic comparative methods achieve independence while retaining and using all data. We discuss the essential notions of phylogenetic signal; uncertainty about trees; typological averages and proportions that are sensitive to genealogy; comparison across language families; and the effects of areality. Extensive supplementary materials illustrate computational tools for practical analysis and we illustrate the methods discussed with a typological case study of the laminal contrast in Pama-Nyungan.}

% submission details
  \received{27 May 2021}
  \accepted{29 December 2021}

% journal details
  \journalname{Accepted for publication in \emph{Linguistic Typology}}
  \startpage{1}

\maketitle

\hypertarget{intro}{%
\section{Introduction}\label{intro}}

Linguistic typology examines the known diversity of languages with the aim of uncovering insights into the nature of human language itself. The task of cross-linguistic comparison is complicated, however, by the interwoven patterns of historical descent and contact between languages. These patterns of historical relatedness can manifest in shared forms and features in languages today. Consequently, there is widespread recognition that shared histories must be taken into account in typological analysis (see Section \ref{phylo-auto-ling}), and there is an abiding concern that the methods used in typology be attuned to the complications of genealogy to the best extent possible.

The non-independence of synchronic observations due to histories of shared descent is a fundamental concept not only in linguistics, but also in other fields where entities share common paths of descent, such as biology and anthropology. Nevertheless, there are a variety of lines of thought and responses that have developed in different fields over the course of a century of scholarship. Consequently, we begin our paper by considering this well-worn discussion within a cross-disciplinary scope. We find that all fields share, in origin, similar lines of development in the elaboration of sampling methodologies for producing phylogenetically independent samples. During this common phase, many independent developments in linguistics and biology have been uncannily parallel. However, biology is now pursuing a different set of solutions to challenges that we have long faced in common. It is instructive, therefore, to understand why a discipline that mirrored linguistic typology for so long has now shifted its approach, and to see how the factors that motivated the change in biology also exist in linguistics.

The paper proceeds as follows: Section \ref{phylo-autocorrelation} reviews literature on \emph{phylogenetic autocorrelation}---the tendency of languages to show similarities due to phylogenetic relatedness---and the methodological responses to it in linguistic typology and cognate fields (comparative biology, in particular). Section \ref{phylo-sig} then introduces the concept of phylogenetic signal, the degree of phylogenetic autocorrelation that is present in a comparative dataset, and describes statistical tools for quantifying it. Section \ref{phylo-uncertainty} addresses the topic of uncertainty in linguistic genealogies, and discusses ways in which phylogenetic comparative methods enable a nuanced, explicit examination of how inferences that are drawn from cross-linguistic data are affected by hypotheses about genealogy. In Section \ref{weighting}, because two of the most common types of scientific finding in typology are cross-linguistic averages of typological variables and proportions of languages that have particular properties, we describe phylogenetic methods for the calculation of averages and proportions that take genealogy into account. In Section \ref{laminals} we present a typological case study of the laminal places of articulation in the Pama-Nyungan languages of Australia. Here we illustrate both the principles and methods introduced earlier, and produce some new insights about this facet of Australian phonological typology that are obtainable only with phylogenetic comparative tools. To discuss and conclude, Section \ref{discussion} returns to the topics of mass comparison and deep-time language relateness, and language contact and areality, in the light of the foregoing discussions, and in Section \ref{conclusions} we offer a concluding outlook.

\hypertarget{phylo-autocorrelation}{%
\section{Phylogenetic autocorrelation: The consequences of relatedness}\label{phylo-autocorrelation}}

Phylogenetic autocorrelation is common to many comparative fields of science. It is a potential problem for comparative study, because shared phylogenetic histories limit the independence of observations in a comparative dataset. Observations from more closely related entities will tend to show less variation than more distantly related entities, because they share a longer period of common history and have had less time to diverge since the splitting up of their most recent common ancestor. If this tendency towards similarity due to shared phylogenetic history is not taken into account, it will introduce bias into the dataset and consequently affect statistical analysis. This section discusses phylogenetic autocorrelation and the history of responses to it in different fields. We emphasise some remarkable parallels across disciplines in their independent lines of thinking, especially around the issue of data sampling. However, we also highlight a significant distinction that has emerged since the uptake of quantitative phylogenetic comparative methods in comparative biology. We begin with some cross-disciplinary background (Section \ref{phylo-auto-omni}) then focus in particular on linguistics (Section \ref{phylo-auto-ling}) and biology (Section \ref{phylo-auto-bio}). We unpack the key methodological breakthrough that lies behind phylogenetic comparative methods (Section \ref{PICs}) and then discuss its uptake in disciplines beyond biology (Section \ref{beyond-biology}).

\hypertarget{phylo-auto-omni}{%
\subsection{Phylogenetic autocorrelation across the sciences}\label{phylo-auto-omni}}

Different fields have their own lines of literature grappling with phylogenetic autocorrelation extending back many decades. In comparative anthropology, this issue was noted as early as 1889 by Sir Francis Galton in the context of cross-cultural datasets, which lack independence due to shared histories of cultural innovation and exchange between societies \autocite[15]{naroll_two_1961}. This phenomenon, known as \emph{Galton's Problem}, is now more precisely understood as a form of statistical \emph{autocorrelation}, i.e., similarity between observations that correlates with their proximity, in this case, their proximity in evolutionary time. The same phenomenon has been recognised in comparative biology too. A seminal study concerning comparative studies of phenotypes, \textcite{felsenstein_phylogenies_1985} demonstrates that data from species cannot be assumed to be independently drawn from the same distribution, because species are related to one another via a branching, hierarchical phylogeny, thus, statistical methods that assume independent, identically-distributed observations will inflate the significance of the test (discussed further in Section \ref{phylo-auto-bio} below). Linguists, it was argued, had been somewhat slower than those in other fields to acknowledge exposure to Galton's problem, or phylogenetic autocorrelation \autocite[293]{perkins_statistical_1989}. However, this is a central concern of \textcite[259]{dryer_large_1989} and has been addressed in a considerable body of linguistic typological literature since then.

Statistical non-independence due to shared history is thus no new revelation, not in comparative anthropology, not in comparative biology, nor in linguistic typology. However, there are many possible approaches to dealing with its challenges and a sizeable body of literature on the topic. As we will see, although precise strategies are varied, a notable commonality to all fields is a history of first attempting to address phylogenetic autocorrelation through the development of sampling methods for the creation of phylogenetically independent---or phylogenetically balanced---samples. The most striking differences between disciplines emerges only later, following the uptake in comparative biology of phylogenetic comparative methods.

\hypertarget{phylo-auto-ling}{%
\subsection{Phylogenetic autocorrelation in linguistics}\label{phylo-auto-ling}}

In linguistic typology, the use of phylogenetically balanced language samples remains the predominant way of accounting for phylogenetic autocorrelation and literature on this topic extends back several decades. \textcite[145--149]{bell_language_1978} argues that common strategies which simply ensure equally-weighted representation of ``all major families'' or all continents is inadequate due to differing rates of divergence among families. He estimates the number of language groups separated by more than 3,500 years of divergence and uses it as a heuristic for estimating genealogical biases in a selection of proposed language samples. He concludes that European languages tended to be overrepresented and Indo-Pacific languages underrepresented in typological language samples at his time of writing. He attributes this to a corresponding over/under-representation among quality language resources, which is a persistent problem for comparative linguistics. Perkins \autocites*{perkins_evolution_1980}{perkins_covariation_1988} creates a sample of 50 languages, later adapted by \textcite{bybee_morphology_1985}, which attempts to account for both genealogical and areal biases by selecting no more than one language from each language phylum following \textcite{voegelin_index_1966} and no more than one language from each cultural and geographic area following work in comparative anthropology \autocites{kenny_numerical_1975}{murdock_ethnographic_1967}. This method attempts to account for non-independence due to areal spread, unlike Bell's heuristic measure which accounts only for genealogical bias, however it does not account for differing ages of divergence and size of language phyla in the way Bell does.

Balanced sampling methods seek to produce linguistic samples that are independent, by selectively excluding the vast majority of attested languages, as necessitated by their extensive, inherent non-independence. As typologists have developed these methods, they have confronted two main complications.

The first complication is that it may be difficult to find criteria for the inclusion/exclusion of languages which truly remove all dependencies, or which are uncontroversial. \textcite[261]{dryer_large_1989} refers to the example of the inclusion of three languages in Perkins' sample (Ingassana, Maasai and Songhai) which potentially are related as part of the Nilo-Saharan family, and thus non-independent, although these relationships are remote and subject to debate. One aspect of this problem is that the maximal extent of presently established language families is partially a product of the extent of adequate documentation and scholarly attention, rather than a reflection of the fullest extent to which the family may be reconstructed \autocite{levinson_universal_2011}. Two languages which are presently understood to be unrelated, and therefore statistically independent, may in fact belong to a shared larger grouping, which has not yet been identified due to poor documentation or lack of historical-comparative study. A second aspect is that language families undoubtedly share deep-time relationships that are currently beyond the reach of the comparative method, even if all extant languages were documented and compared completely. Both challenges can lead to languages being deemed as independent when in reality they are not. \textcite[263]{dryer_large_1989} raises a related concern, which is that languages selected on the basis of genealogical independence may nonetheless share characteristics due to non-genealogical processes---language contact and borrowing. This motivates the use of areal criteria in addition to genealogical ones when constructing an independent sample. As \textcite[284]{dryer_large_1989} acknowledges however, linguistic areas may be also subject to the same concerns about undetected historical non-independence and it is possible that the whole world may, in effect, function as a single linguistic area, such that the distribution of certain linguistic features may reflect extremely remote areal or genealogical patterns rather than some true tendency of human language.

The second complication is that once all genealogical and areal criteria are adhered to, the resulting sample may be too small for use in statistical analysis \autocites{cysouw_quantitative_2005}{jaeger_mixed_2011}{piantadosi_quantitative_2014}. In response, linguists have proposed various procedures for constructing samples which, if not fully independent, at least have a high degree of independence. Dryer's proposed solution is to build a sample of languages of approximately equal relative independence (at the level of major subfamilies within Indo-European, such as Romance, Germanic, and so on) for each of five large linguistic areas which are assumed to be independent, or at least sufficiently independent for statistical purposes. Any statistical test can then be applied to each of the five areas and only if the same result is replicated in all five areas is it considered statistically significant. If the same result is replicated in four of five areas, this falls short of statistical significance, although \textcite[272--273]{dryer_large_1989} considers such cases to be evidence of a ``trend''. \textcite[41]{nichols_linguistic_1992} uses Dryer's area-by-area testing method as part of a three-pronged approach. For any given question, Nichols first conducts a chi-square test of the world sample and then re-tests the significance of the finding using either Dryer's method or by running the same test on only the sample of ``New World'' languages (comprising North, Central and South America). \textcite{rijkhoff_method_1993} and \textcite{rijkhoff_language_1998} develop another approach to account for the possibility of non-independence across large linguistic areas and large, as-yet-undetected families. They permit multiple languages within a family to be included but develop a measure, based on the density of nodes in a known language phylogeny, to determine how many languages should be included. In this way, they also aim to account for the fact that some language families will have greater internal diversity than others \autocites[see also][]{bakker_language_2011}{miestamo_sampling_2016}.

Another approach is to include/exclude languages based on their typological profile. Following the logic that historical relatedness and interactions tend to result in elevated similarity, these methods bias their sample in favour of typological diversity, as a proxy for independence. \textcite{dryer_wals_2013} propose setting a minimum threshold of typological distance between languages, calculated from the \emph{World Atlas of Language Structures} (WALS), such that languages must be sufficiently typologically distinct from others in the sample to warrant inclusion. \textcite{bickel_refined_2009} develops an alternative algorithm based on \textcite{dryer_large_1989}, which allows all uniquely-valued data points within a family to be included in the sample, but then reduces the weighting of data points in the final analysis where a particular value is over-represented within a family. In other words, if all the languages in a particular family share the same value for a typological variable of interest, those observations may be reduced to a single data point.

In these ways, developments in typological methodology have treated historical non-independence between languages as a challenge to be addressed through sampling. Earlier researchers sought to maintain the independence of their sample by maximising the genealogical distance between the languages in their sample, such that no two languages were known to belong to the same family. Later, with subsequent acknowledgement of the possibility of non-independence from very large language families, as well as large-scale areal diffusion and effects from as-yet undetected or unconfirmed historical relations, it became apparent that it may be impossible to create a sample which is simultaneously independent and sufficiently large to generate statistical significance. As discussed above, typologists have primarily responded to this dilemma by developing a variety of robustness checks, even bootstrapping-like processes, whereby languages are sampled at an approximately equal relative level of independence and the sample is then subdivided in some way and a statistical test replicated over each subdivision. More recent years have seen the continued evolution of statistics and robustness checking methods \autocite[for an overview, see][]{roberts_robust_2018}, although balanced sampling remains a common element of modern, large-scale comparative linguistic studies \autocites[for example,][]{everett_climate_2015}{everett_languages_2017}{blasi_grammars_2017}.

Before turning to biology, it is worth underscoring how linguistic typology has arrived at its current mode of response to phylogenetic autocorrelation. The starting point is that many conventional statistical methods require observations that are independent, yet languages are non-independent. For four decades, the response has been to change the dataset, by means of balanced sampling, so that it better corresponds to the requirements of the statistics. Doing so requires excluding the vast majority of documented languages from the dataset and hence from the analysis, and even then, the result is still not truly independent. In the next section, we will see that biology initially followed the same path. The key breakthrough, though, was to invert the response to the original problem that phylogenetic autocorrelation posed: to change not the dataset to suit the statistics, but the statistics to suit the dataset. Those changed statistics are phylogenetic comparative methods.

\hypertarget{phylo-auto-bio}{%
\subsection{Phylogenetic autocorrelation in comparative biology}\label{phylo-auto-bio}}

Comparative biology faces the same issue of phylogenetic autocorrelation as comparative linguistics. Many conventional statistical methods assume that observations are independent, which is problematic since observations come from species, which are related to one another through shared evolutionary histories.

Earlier approaches to phylogenetic autocorrelation in biology are in a similar vein to the sampling methods in linguistic typology discussed in the previous section. \textcite[346--347]{harvey_comparisons_1982} seek to find a taxonomic level to sample from, which strikes the right balance in terms of being sufficiently statistically independent without being so conservative that sample sizes become prohibitively small, an aim similar to \textcite{dryer_large_1989}. Their proposed solution is to identify and sample from the lowest taxonomic level which can be ``justified on statistical grounds''. One method of doing this is suggested by \textcite[6--8]{clutton-brock_primate_1977}, who conduct a nested analysis of variance and then select the taxonomic level containing the greatest level of variation. Similar to the methods of \textcite{dryer_wals_2013} and \textcite{bickel_refined_2009}, this approach makes reference to diversity in the traits of the species (cf.~diversity in typological traits) to guide the sampling procedure.\footnote{Once \textcite{clutton-brock_primate_1977} identify their taxonomic level of interest, they average out data for all species within a given genus for which they have data. In other words, the unit of analysis has shifted from individual species to genera, and each data point represents a genus in the form of an averaged representation of all the species within the genus. This genus-level averaging process is in contrast to balanced sampling methods discussed in the previous section, where an unaltered observation from a single exemplar language is taken as representative of its given family, subfamily or other defined grouping, though has affinities with \textcite{bickel_refined_2009}, which also reduces with-family observations to a smaller number of data points (albeit of a different kind to an average).}

As in linguistics, areality is also an issue in biology. Geographical and ecological proximity can lead to similarities in taxa (i.e., species or languages) which is causally separate from the effects of genealogy. Two distinct, causal scenarios can be distinguished. In the first scenario, material is passed directly between taxa, such as lateral transfer of genetic material between species, especially but not exclusively in prokaryotic life forms such as bacteria \autocite{keeling2008horizontal}, or borrowing between languages. In the second scenario there is no direct transfer of material, rather a shared environment leads to similar developments in taxa, such as parallel dwarfism on islands or, in some cases more contentiously, parallel conditioning of language by its environment \autocites{everett_climate_2015}{everett_languages_2017}{blasi_grammars_2017}{everett_sound_2021}. In both kinds of scenario, there is a causal, areally-correlated contribution to similarity which is separate from the contribution due to shared genealogy. While it is true that modern, genomic studies can circumvent some of the difficulties due to the second scenario in biology, it should be noted that phylogenetic comparative methods in biology predated the emergence of widespread genomic sequencing, and for many species including those attested only as fossils, genetic data is still not available. Consequently, the problem of convergent evolution due to areality was and still is a genuine, hard problem that comparative biology has faced, and should not be misunderstood as a problem specific to linguistics. In an approach with strong conceptual similarities to the area-by-area robustness checking of \textcite{dryer_large_1989} and \textcite{nichols_linguistic_1992}, \textcite[85--86]{baker_evolution_1979} discussed how the causal effects of ecological areas might be addressed while constructing a sample which is genealogically balanced. To do so, \textcite{baker_evolution_1979} replicate their analysis within individual families as well as within different ecological areas, with the assumption that if the same associations are observed within different areas as across the dataset as a whole, then one can discount the possibility that the full analysis is simply picking up differences between different families or different ecological areas.

In essence, both linguistics and biology face the same phenomenon of phylogenetic autocorrelation including the complication of areality, and for several decades explored strikingly similar methodological responses based on sampling. However, in recent decades the primary methods in linguistic typology and biology have diverged as biology has undergone a fundamental shift. While typologists continue to focus on sampling procedures as the response to phylogenetic autocorrelation, comparative biologists have moved to a more direct, statistical solution. Since the solution addresses phylogenetic autocorrelation, not areality, our focus will narrow now to the genealogical aspects of taxon relatedness. We return to the separate and additional problem of areality in Section \ref{areality}.

\hypertarget{PICs}{%
\subsection{Phylogenetically independent contrasts}\label{PICs}}

\begin{figure}
\includegraphics[width=0.45\linewidth]{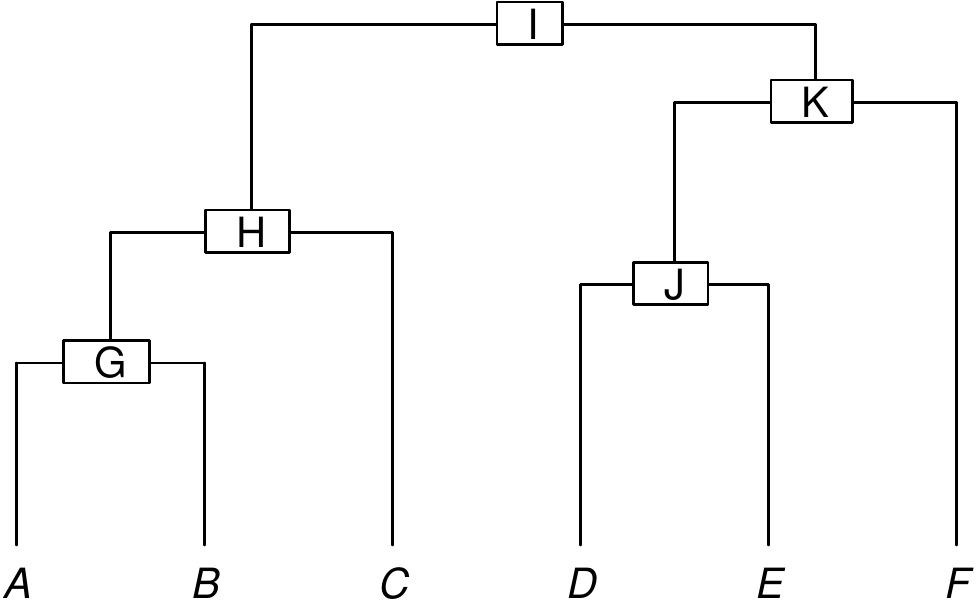} \caption{A phylogeny of six species or languages}\label{fig:phy0}
\end{figure}

\textcite{felsenstein_phylogenies_1985} demonstrates that it is possible to account for phylogenetic non-independence in a statistical model without the need to remove data or compromise the unit of analysis (for example, by collapsing or averaging observations within a subgroup). Felsenstein's breakthrough insight is that this can be achieved not by directly comparing non-independent observations but by comparing \emph{phylogenetically independent contrasts} (PICs) between observations. His method has become, by one estimate, the most widespread in comparative biology \autocite[p.~162]{nunn_comparative_2011}. The essential insight is relatively straightforward. Consider the tree in Figure \ref{fig:phy0}. Any traits of A and B will be non-independent observations, since much of their evolutionary history is shared: all of the evolutionary change between points I and H, and between H and G, has contributed equally to both A and B. However, any differences (or in biological parlance, \emph{contrasts}) between A and B have the particular status that they must have arisen after the split at point G. That period of development, after split G until the modern species (or languages) A and B is not shared with any other part of the tree. It is independent. Felsenstein's insight is that by examining phylogenetic contrasts such as this, one can obtain observations that truly are independent. It is then possible to apply standard statistical tests to the phylogenetically independent contrasts (rather than directly to observed values) without phylogenetic autocorrelation introducing bias into the results.

In the remainder of this subsection we discuss some finer technical points of Felsenstein's notion for readers who are interested. Others may wish to skip ahead directly to the next subsection.

In order to calculate PICs not only between sister tips of a tree such as A and B, but also between sister interior nodes such as H and K, or node-tip sisters such as G and C, one requires in addition to a phylogeny, a model according to which the variable evolves. As a starting point, Felsenstein assumes a \emph{Brownian motion} model of evolution, since Brownian motion is one of the simplest and most fundamental of all stochastic processes. In a Brownian motion model, an evolving quantitative trait can wander positively or negatively with equal probability, and each new time step is independent from the last, with the resulting effect that displacement of the variable over time will be drawn from a normal distribution with a mean of zero and variance proportional to the amount of elapsed time \autocite[p.~8]{felsenstein_phylogenies_1985}. An observed contrast can be scaled by dividing it by the standard deviation of its expected variance. This gives a statistically independent contrast of expectation zero and unit variance (i.e.~variance equal to 1). This process can be repeated for all adjacent tips in the tree. Contrasts can then be extracted from adjacent nodes in the tree, where the value of the node is an average of the observed values of the tips below it. In the end, there will be a collection of phylogenetically independent contrasts, all of expectation zero and unit variance, to which statistical analysis can be applied.

One drawback of Felsenstein's initial method is the reliance on the assumption of Brownian motion as a model of variable evolution. \textcite{grafen_phylogenetic_1989} subsequently devises a similar method, \emph{the phylogenetic regression}, which has the flexibility to incorporate models of evolution other than Brownian motion. Further, Grafen's method is able to be applied in situations where phylogenetic information is incomplete (for example, where the phylogeny is an incomplete work-in-progress rather than an accepted gold-standard). This method is a phylogenetic adaptation of \emph{generalised least squares} (GLS). In this model, the value of a dependent variable, \(y_{i}\), is predicted by the equation \(y_{i} = \alpha + \beta x_{i} + \epsilon\), where \(\alpha\) is the intercept, \(\beta\) is the regression slope, \(x\) is the independent variable and \(\epsilon\) is an error term \autocite[p.~164]{nunn_comparative_2011}. Phylogenetic information can be incorporated into the error term, in the form of a variance-covariance matrix of phylogenetic distances between tips in a tree. PICs and GLS are mathematically equivalent when a Brownian motion evolutionary model is assumed and the reference tree is fully bifurcated, so PICs are essentially a special case of GLS where these assumptions are met \autocite{nunn_comparative_2011}.

\hypertarget{beyond-biology}{%
\subsection{Phylogenetic comparative methods beyond biology}\label{beyond-biology}}

Linguistic typology and comparative anthropology have long faced the same essential problem of phylogenetic autocorrelation that comparative biology contends with. Initially, all three disciplines followed similar trajectories, responding to phylogenetic autocorrelation through the development of increasingly elaborate methods of balanced sampling. By historical accident it was in biology that the breakthrough of examining PICs occurred, but the breakthrough is a solution to an inherent problem that transcends disciplinary boundaries. Anthropologists, recognising the same problem in kind, followed this breakthrough in biology with their own uptake of phylogenetic comparative methods around 10--20 years later \autocites[e.g][]{mace_comparative_1994}{holden_spread_2003}{holden_phylogenetic_2009}{jordan_matrilocal_2009}{nunn_comparative_2011}, and recently there has been growing interest in the application of phylogenetic comparative methods in linguistics \autocites[e.g][]{maslova_dynamic_2000}{maslova_stochastic_2000}{dunn_evolved_2011}{maurits_tracing_2014}{verkerk_diachronic_2014}{birchall_comparison_2015}{zhou_quantifying_2015}{calude_typology_2016}{dunn_dative_2017}{verkerk_phylogenetic_2017}{bentz_evolution_2018}{cathcart2020numeral}{macklin-cordes_phylogenetic_2021}{jager2021phylogenetic}.

One of our motivations for this paper, however, is that despite the increasing uptake of phylogenetic comparative methods in linguistics, there has been little attempt until now to explain why phylogenetic comparative methods can best be understood as a continuation of a tradition of inquiry that typology is greatly invested in. Previously, that tradition of inquiry, whether in comparative biology, comparative anthropology or linguistics, had led to methods of balanced sampling. Like balanced sampling methods, phylogenetic comparative methods are a response to phylogenetic autocorrelation, one of the central and most persistent problems of linguistic typology. Methodologists working on balanced sampling have striven to generate samples that come as close as possible to phylogenetic independence, but the goal cannot be fully attained even with the most elaborate sampling procedures, and in the meantime procedures of balanced sampling require the exclusion of the vast majority of documented languages from the dataset and hence from the analysis. As it turns out, the solution is to be found not in phylogenetically independent samples, but in phylogenetically independent contrasts (PICs). By focussing on PICs, Felsenstein unlocked a method for obtaining truly independent observations, without excluding data. This is why typologists have every reason to be keenly interested in phylogenetic comparative methods: they solve a problem which has stood at the centre of our discipline for decades.

In the sections that remain, we shift our focus away from theory and onto practicality: how can typologists begin making use of phylogenetic comparative methods? In Sections \ref{phylo-sig}--\ref{weighting} we introduce key phylogenetic concepts and techniques that typologists can employ, followed by a phylogenetic typological case study in Section \ref{laminals}. In Section S1 of the Supplementary Materials, we provide an extended practical introduction to a suite of computational tools that have been designed with the typologist in mind \autocites{glottoTrees}{phyloWeights}, enabling phylogenetic comparative methods to be used in everyday typological research. In Section \ref{discussion} we return to the topic of areality.

\hypertarget{phylo-sig}{%
\section{Phylogenetic signal: The extent to which synchronic distributions mirror genealogy}\label{phylo-sig}}

As discussed in Section \ref{phylo-autocorrelation}, phylogenetic comparative methods are applicable in linguistic typology when phylogeny is a causal factor that has shaped the distribution of a linguistic variable. The previous section described the means by which phylogenetic comparative methods are able to take such a phylogeny into account in statistical analysis. However, some variables may not evolve through descent with modification and consequently may not pattern phylogenetically. Others may be subject not only to descent with modification, but to other causal factors in addition such as areality, and thus may pattern phylogenetically only weakly. How, then, does one determine for a variable of interest whether a phylogeny may have contributed to the cross-linguistic distribution of diversity? In the last twenty years, an advance in this area has been the advent of methods for explicitly quantifying the degree of \emph{phylogenetic signal} in comparative data \autocites{freckleton_phylogenetic_2002}{blomberg_testing_2003}. Phylogenetic signal refers to the tendency of phylogenetically-related entities to resemble one another \autocites{blomberg_tempo_2002}[p.~717]{blomberg_testing_2003}. This resemblance is more technically defined as statistical non-independence among observation values due to phylogenetic relatedness between taxa \autocite[p.~591]{revell_phylogenetic_2008}. This concept of phylogenetic signal has important applications in comparative linguistics. Here we argue that for many purposes, measuring phylogenetic signal should be considered as a first step in a phylogenetically aware comparative methodology, since it can determine empirically whether phylogenetic comparative methods are required or whether regular statistical methods may suffice \autocite[as in][]{irschick_comparison_1997}.\footnote{Note, however, that the absence of phylogenetic signal does not necessarily indicate that non-phylogenetic statistical methods are appropriate in all cases, in particular for phylogenetic generalised least squares (PGLS) \autocites{revell_phylogenetic_2010}{symonds_primer_2014}.} Further, the result of a phylogenetic signal test can contribute to evolutionary hypotheses in its own right, as we will see in the case study in Section \ref{laminals}.

This section describes fundamental methods for measuring phylogenetic signal in variables with continuous values (Section \ref{phylo-sig-quant}) and with discrete binary values (Section \ref{phylo-sig-bin}). The discussion below will get technical, but we have included it because we expect that some readers will be interested in the details and the underlying logic. For others, who may prefer to skim over the denser technical passages here or skip directly to Section \ref{phylo-uncertainty}, it will suffice to make note of the core message, that testing for phylogenetic signal provides insight into how strongly genealogy may be shaping the data. This is useful knowledge in itself and it enables a more nuanced, judicious use of other phylogenetic comparative methods. For these reasons, testing for phylogenetic signal as part of a research workflow is good practice and is widely employed in phylogenetic studies.

\hypertarget{phylo-sig-quant}{%
\subsection{Phylogenetic signal in continuous variables}\label{phylo-sig-quant}}

\textcite{blomberg_testing_2003} provide a suite of tools for quantifying phylogenetic signal, which have become somewhat of a standard in the field (cited 3780 times as of September 2021, according to Google Scholar).\footnote{In the R statistical programming language \autocite{Rcore} the tests described here are implemented in the \texttt{phylosig} function of the \emph{phytools} package \autocite{phytools}.} Recent comparative studies using these tools include \textcite{balisi_dietary_2018}, \textcite{hutchinson_contemporary_2018} and \textcite{macklin-cordes_phylogenetic_2021}. \textcite{blomberg_testing_2003} present a descriptive statistic, \(K\), which is generalizable across phylogenies of different sizes and shapes. In addition, they provide a randomisation test for checking whether the degree of phylogenetic signal for a given dataset is statistically significant. \(K\) can be calculated using either phylogenetically independent contrasts (PICs) \autocite{felsenstein_phylogenies_1985} or generalised least squares (GLS) \autocite{grafen_phylogenetic_1989} (see Section \ref{PICs}). In a Brownian motion model, where variable values can wander up and down with equal probability through time, PIC variances are expected to be proportional to elapsed time. Among more closely related languages, where there has been less divergence time for variable values to wander, the variance of PICs is expected to be low. The randomisation test works by comparing whether observed PICs are lower than the PIC values obtained by randomly permuting the data across the tips of the tree. The process of permuting data across tree tips at random is repeated many times over. If the real variances, with data in their correct positions on the tree, are lower than 95\% of the randomly permuted datasets, then the null hypothesis of no phylogenetic signal can be rejected at the conventional 95\% confidence level. In other words, closely related languages resemble one another to a statistically significantly greater degree than would be expected by chance.

The descriptive statistic, \(K\), quantifies the strength of phylogenetic signal. As with the randomisation procedure above, the input is a set of observed values, where each observation is associated with a tip of the reference tree. \textcite[722]{blomberg_testing_2003} give an explanation of the calculation of the \(K\) statistic. To summarise briefly, \(K\) is calculated by, firstly, taking the mean squared error (\(MSE_0\)), as measured from a phylogenetic mean,\footnote{We discuss the phylogenetic mean further in Section \ref{weighting} below. Simply taking a non-phylogenetic mean of a variable would be misleading in cases where members of a particularly large clade happen to share similar values at an extreme end of the range.} and dividing it by the mean squared error (\(MSE\)) calculated using a variance-covariance matrix of phylogenetic distances between tips in the reference tree (the same variance-covariance matrix of phylogenetic distances incorporated into the error term in GLS-based phylogenetic regression, as discussed in the previous section). This latter value, \(MSE\), will be small when the pattern of covariance in the data matches what would be expected given the phylogenetic distances in the reference tree, leading to a high \(MSE_0/MSE\) ratio and vice versa. Thus, a high \(MSE_0/MSE\) ratio indicates higher phylogenetic signal. Finally, the observed ratio can be scaled according to its expectation under the assumption of Brownian motion evolution along the tree. This gives a \(K\) score which can be compared directly between analyses using different tree sizes and shapes. Where \(K = 1\), this suggests a perfect match between the covariance observed in the data and what would be expected given the reference tree and the assumption of Brownian motion evolution. Where \(K < 1\), close relatives in the tree bear less resemblance in the data than would be expected under the Brownian motion assumption. \(K > 1\) is also possible---this occurs where there is less variance in the data than expected, given the Brownian motion assumption and divergence times suggested by the reference tree. In other words, close relatives bear greater resemblance than would be expected, given the overall phylogenetic diversity.

As discussed, the assumption of a Brownian motion model of evolution, where a variable is free to wander up or down, with equal probability, as time passes, is central to quantification of phylogenetic signal with the \(K\) statistic. \textcite[726--727]{blomberg_testing_2003} extend their approach to cover two different modes of evolution as well. This is achieved by incorporating extra parameters into the variance-covariance matrix to reflect different evolutionary processes. The first evolutionary model alternative is the Ornstein-Uhlenbeck (OU) model \autocites{felsenstein_phylogenies_1988}{garland_phylogenetic_1993}{hansen_translating_1996}{lavin_morphometrics_2008} whereby variables are still free to wander up or down at random, but there is a central pulling force towards some optimum value. The second alternative is an acceleration-deceleration (ACDC) model, developed by \textcite{blomberg_testing_2003} where a variable value moves up or down with equal probability (like Brownian motion) but the rate of evolution will either accelerate or decelerate over time.

Other statistics for quantifying phylogenetic signal have been proposed and warrant mention. \textcite{freckleton_phylogenetic_2002} propose using the \(\lambda\) (lambda) statistic, based on earlier work by \textcite{pagel_inferring_1999}. As for \textcite{blomberg_testing_2003}, this approach works with a variance-covariance matrix showing the amount of shared evolutionary history between any two tips in the tree (the diagonal of the matrix, the variances, will indicate the total height of the tree; the off-diagonals, the covariances, will indicate the amount of shared evolutionary history between two given entities, before they diverge in the tree). The statistic, \(\lambda\) is a scaling parameter which can be applied to this variance-covariance matrix. Scaling the values in the matrix by \(\lambda\) transforms the branch lengths of the tree, from \(\lambda = 1\), where branch lengths are left unscaled, to \(\lambda = 0\), where all covariances in the matrix will be zero, in other words, no covariance through shared evolutionary history is indicated between any tips, thus all tips will be joined at the root by branches of equal length (a star phylogeny). \textcite{freckleton_phylogenetic_2002} present a method for finding the \(\lambda\) parameter that maximises the likelihood of a set of observations arising, given a Brownian motion model of evolution. If \(\lambda\) is close to 1, this indicates high phylogenetic signal, where the data closely fit expectation given the shared evolutionary histories in the tree and a Brownian motion model of evolution. Further measures which have been proposed are \(I\) \autocite{moran_notes_1950}, a spatial autocorrelation measure which was adapted for phylogenetic analyses by \textcite{gittleman_adaptation:_1990}, and \(C_{mean}\) \autocite{abouheif_method_1999}, which is a test for serial independence \autocite[for an overview, see][]{munkemuller_how_2012}. In an evaluation of different methods \textcite{munkemuller_how_2012} find that, assuming a Brownian motion model of evolution, \(C_{mean}\) and \(\lambda\) generally outperform \(K\) and \(I\). However, \(C_{mean}\) considers only the topology of the reference tree (i.e., the order of the branches from top to bottom), but not branch length information, and the value of the \(C_{mean}\) statistic is partially dependent on tree size and shape, so it lacks comparability between different studies. In addition, \(\lambda\) shows some unreliability with small sample sizes (trees with \textless{}20 tips).

\hypertarget{phylo-sig-bin}{%
\subsection{Phylogenetic signal in binary variables}\label{phylo-sig-bin}}

The methods so far described concern continuously-valued data. Other methods have been proposed for quantifying phylogenetic signal in binary and categorical variables too. \textcite{abouheif_method_1999} presents a simulation-based approach for testing whether discrete values along the tips of a phylogeny are distributed in a phylogenetically non-random way. Although this method is useful for testing whether the phylogenetic signal in a set of discretely-valued data is statistically significant, it does not provide a quantification of the level of phylogenetic signal which is comparable between different datasets. Although specific to binary data only, \textcite{fritz_selectivity_2010} present a statistic, \(D\), which quantifies the strength of phylogenetic signal for binary variables.

The \(D\) statistic is based on the sum of differences between sister tips and sister clades, \(\Sigma d\). To summarise, following \textcite{fritz_selectivity_2010}, differences between values at the tips of the tree are summed first (all tips will either share the same value, 0 or 1, with 0 difference; or one will be 0 and the other will be 1, for a difference of 0.5). Nodes immediately above the tips are valued as an average of the two tips below (either 0, 0.5 or 1) and the differences between sister nodes is summed. This process is repeated for all nodes in the tree, until a total sum of differences, \(\Sigma d\), is reached. At two extremes, data may be maximally clumped, such that all 1s are grouped together in the same clade in the tree and likewise for all 0s, or data may be maximally dispersed, such that no two sister tips share the same value (every pair of sisters contains a 1 and a 0, leading to a maximal sum of differences). Lying somewhere in between will be both a phylogenetically random distribution and a distribution that is clumped to a degree expected under a Brownian motion model of evolution. A distribution of sums of differences following a phylogenetically random pattern, \(\Sigma d_r\), is obtained by shuffling variable values among tree tips many times over. A distribution of sums of differences following a Brownian motion pattern, \(\Sigma d_b\) is obtained by simulating the evolution of a continuous trait along the tree, following a Brownian motion process, many times over. Resulting values at the tips above a threshold are converted to 1, values below the threshold are converted to 0. The threshold is set to whatever level is required to obtain the same proportion of 1s and 0s as observed in the real data. Finally, \(D\) is determined by scaling the observed sum of differences to the means of the two reference distributions (the expected sums of differences under a phylogenetically random pattern and under a Brownian motion pattern).

\begin{equation}
D = \frac{\Sigma d_{obs} - mean\left( \Sigma d_{b} \right)}{mean\left( \Sigma d_{r} \right) - mean\left( \Sigma d_{b} \right)}
\end{equation}

Scaling \(D\) in this way provides a standardised statistic which can be compared between different sets of data, with trees of different sizes and shapes, as with \(K\) for continuous variables. One disadvantage of \(D\), however, is that it requires quite large sample sizes (\textgreater{}50), below which it loses statistical power, increasing the chance of a false positive result (type I error).

Although we have restricted our focus to continuous and binary data here, some recent developments in testing for phylogenetic signal in other kinds of data warrant brief mention also. For example, \textcite{borges_measuring_2019} have developed a statistic, \(\delta\), for quantifying phylogenetic signal in multivalued categorical variables. Other developments concern multivariate and multidimensional data. \textcite{zheng_new_2009} present a multivariate version of the \(K\) statistic discussed in Section \ref{phylo-sig-quant}, for measuring phylogenetic signal in groups of related variables. Their statistic also incorporates measurement error. Finally, \textcite{adams_generalized_2014} presents \(K_{mult}\), a statistic for detecting phylogenetic signal in multivariate traits, i.e.~conceptually unitary evolutionary traits that are defined by multiple values (e.g.~in biology, a set of measurements that together define skull shape).

In this section we have introduced the fundamental notion of phylogenetic signal---the degree to which the distribution of synchronic diversity reflects the shape of a phylogeny---and some key methods for estimating it. Of course, doing this requires a phylogeny to begin with, and typologists may have questions about the suitability of current linguistic trees for such purposes. It is to this important topic that we turn next.

\hypertarget{phylo-uncertainty}{%
\section{Approaches to uncertainty in linguistic trees}\label{phylo-uncertainty}}

A reasonable concern that typologists may have is whether currently available language trees are of sufficient quality to support the use of quantitative phylogenetic methods. Fortunately, there is a clear, technically sound response to this concern. However, the response is not necessarily intuitive, so here we examine it through both logical argumentation and an example.

Not by accident, a parallel concern about the quality of available phylogenies was raised directly by \textcite[14]{felsenstein_phylogenies_1985} in his seminal work on phylogenetic comparative biology.\footnote{It should be remembered that phylogenetic comparative methods arose in biology \emph{before} the widespread availability of high-quality phylogenies based on genome sequencing.} In response to this concern, Felsenstein stresses that logically, because genealogies are fundamental to comparative biology (as they are to comparative linguistics), they are also inescapable: ``there is no doing {[}comparison{]} without taking them into account''. No matter what methods we choose to use, if we make comparisons in biology or linguistics, we will inevitably implicate some genealogy, because genealogies are an inherent component of the real-world causal structure that underlies the data. The question, then, will always be not whether to use trees, but which trees to use. Methods of comparison which purport to operate independently of genealogies actually will implicate a phylogeny covertly.

\begin{figure}
\includegraphics[width=0.9\linewidth]{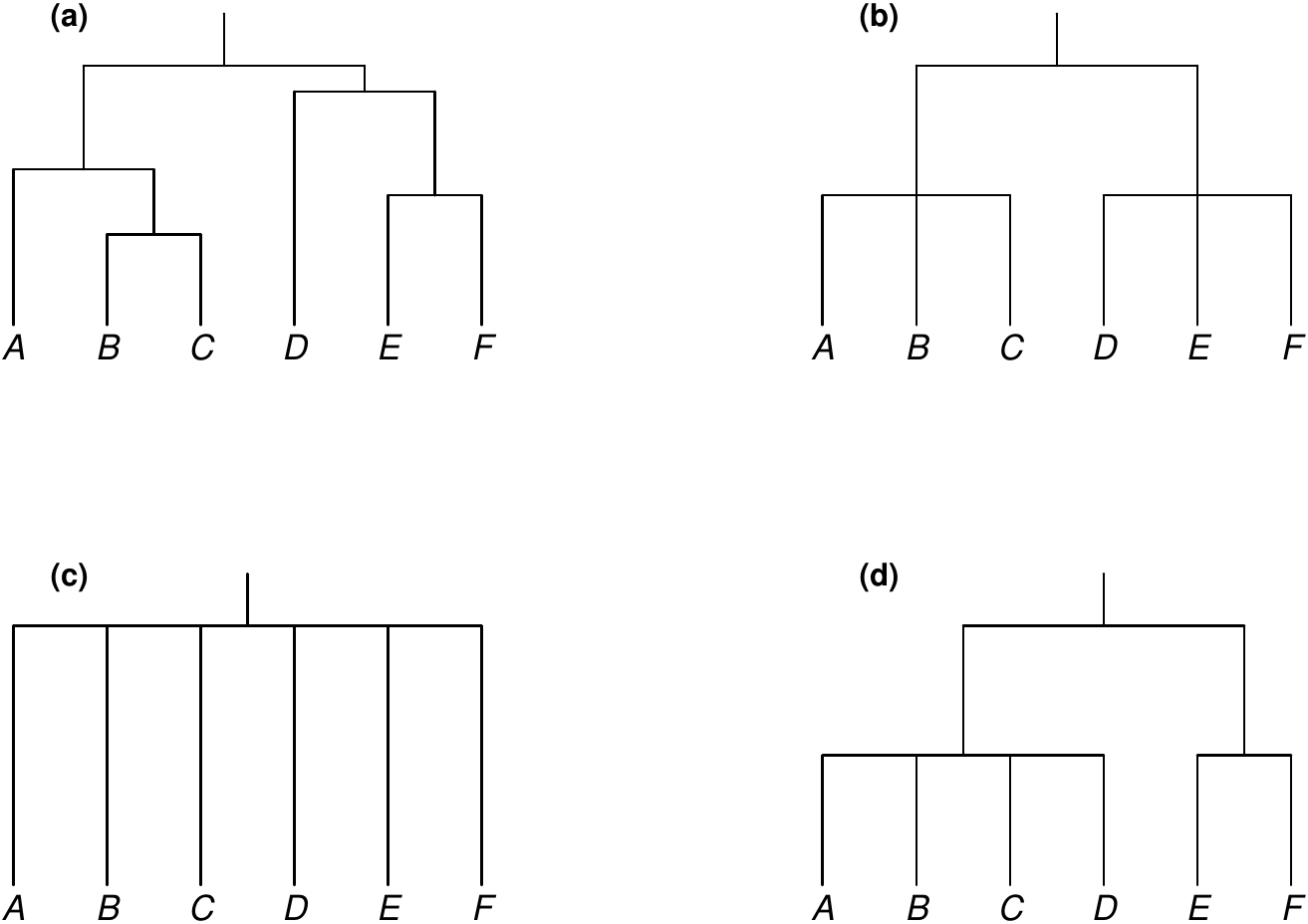} \caption{Four phylogenies of six languages: (a) with detailed branch lengths and topology (nesting structure), (b) with less detail, (c) a star phylogeny (rake phylogeny), (d) an alternate phylogeny with little detail}\label{fig:phy6}
\end{figure}

To take a concrete example, consider a situation where the true phylogenetic history of six languages is as shown in Figure \ref{fig:phy6}a, but that currently, this true history is only partially understood. Such is the case for almost any language family. Linguists may possess only a preliminary hypothesis of subgrouping, as in Figure \ref{fig:phy6}b, with little certainty about how deep in time the major splits are. Phylogeny \ref{fig:phy6}b is therefore a sub-optimal representation of \ref{fig:phy6}a and understandably, concern may arise over using it. However, using the tree in Figure \ref{fig:phy6}b would still be preferable to using no tree at all. Technically speaking, it is not possible to use `no tree'. When phylogeny is ignored entirely, then all languages are set on equal footing, which is equivalent to hypothesising a star tree, also called a rake tree, as in Figure \ref{fig:phy6}c \autocite{purvis_polytomies_1993}. Consequently, the choice between using the tree in Figure \ref{fig:phy6}b and `no tree' is in fact a choice between two trees: Figure \ref{fig:phy6}b or \ref{fig:phy6}c, and the former is almost certainly the better approximation of the true phylogeny, Figure \ref{fig:phy6}a. Evaluative studies have shown that even when phylogenies are incomplete, lacking branch length information, or subject to a degree of error, phylogenetic comparative methods still typically out-perform equivalent non-phylogenetic comparative methods, which effectively assume a star phylogeny in this way \autocites{grafen_phylogenetic_1989}{purvis_truth_1994}{symonds_effects_2002}{rohlf_comment_2006}. By using Figure \ref{fig:phy6}b with phylogenetic methods, it is possible to derive results that are `state-of-the-art' in the sense that they reflect the best of current knowledge; this is not true when using a star phylogeny.

Once it is recognised that using `no tree' is technically not possible, the question still remains of which tree to use. Linguistic trees are often subject to ongoing debate. For instance, different expert analyses may group six languages not only as Figure \ref{fig:phy6}b, but also as Figure \ref{fig:phy6}d. Expert debates such as this are reflective of the \emph{phylogenetic uncertainty} that currently exists about the details of the tree. In these cases, phylogenetic methods can be applied to multiple, alternative trees and the result interpreted critically. Applying phylogenetic methods to multiple trees enables us to move beyond merely disagreeing over phylogenetic hypotheses, towards clarifying what the implications are of adopting different genealogical hypotheses: some results may pivot crucially upon which phylogeny is assumed, while others are largely independent of the choice. Because modern phylogenetic methods are principally computational, there is little practical impediment to examining multiple, alternative tree hypotheses whenever the methods are used. Modern methods of tree inference \autocites[e.g.][]{bouckaert2012mapping}{chang2015ancestry}{kolipakam2018bayesian}{bouckaert_origin_2018} produce large sets termed \emph{tree samples}, of alternative, highly-likely trees, all of which can be used.\footnote{Even if only one phylogeny appears in a published diagram, studies of this kind will almost certainly have produced a full tree sample.} In our case study in Section \ref{laminals} below, we demonstrate this approach by using a tree sample of 100 highly-likely phylogenies to investigate the typology of laminal place of articulation contrasts in Pama-Nyungan languages.

In this section on phylogenetic uncertainty, we have framed our discussion primarily in terms of the kind of uncertainty that can surround the tree of a single language family. However, in linguistics we currently possess many separate trees, for many separate language families. The question arises, how can phylogenetic comparative methods be applied across multiple, distinct language families when there is no known, deep-time tree that links them together? We return to this issue in Section \ref{deep-time}, however the reader may already discern what the response will be, considering that our lack of a global linguistic tree is itself a matter of uncertainty: very likely, many if not all known language families in reality are genealogically linked. If this is true, then even though we are highly uncertain about what their deep-time genealogical links are, it will technically not be possible to use `no tree' when comparing across them, since in reality their genealogical relationships are an inherent component of the real-world causal structure behind the global typological diversity that we wish to analyse. We return to this matter in Section \ref{deep-time}.

\hypertarget{weighting}{%
\section{Genealogically-sensitive averages and proportions}\label{weighting}}

A perennial task in typology is the characterisation of frequencies of traits of interest among the world's languages. The scientific interest of such questions typically lies not merely in the contingent facts of today's particular languages and language families, rather the goal is to characterise the nature of human language in general, using today's contingent empirical data as evidence. Because of this, we are striving ideally for an answer that takes into account the unequal representation of different families and subgroups. Phylogenetic comparative methods can assist in achieving this recurrent and indispensable objective of typological research. In this section we describe methods for deriving genealogically-sensitive averages and proportions.

\begin{figure}
\includegraphics[width=0.9\linewidth]{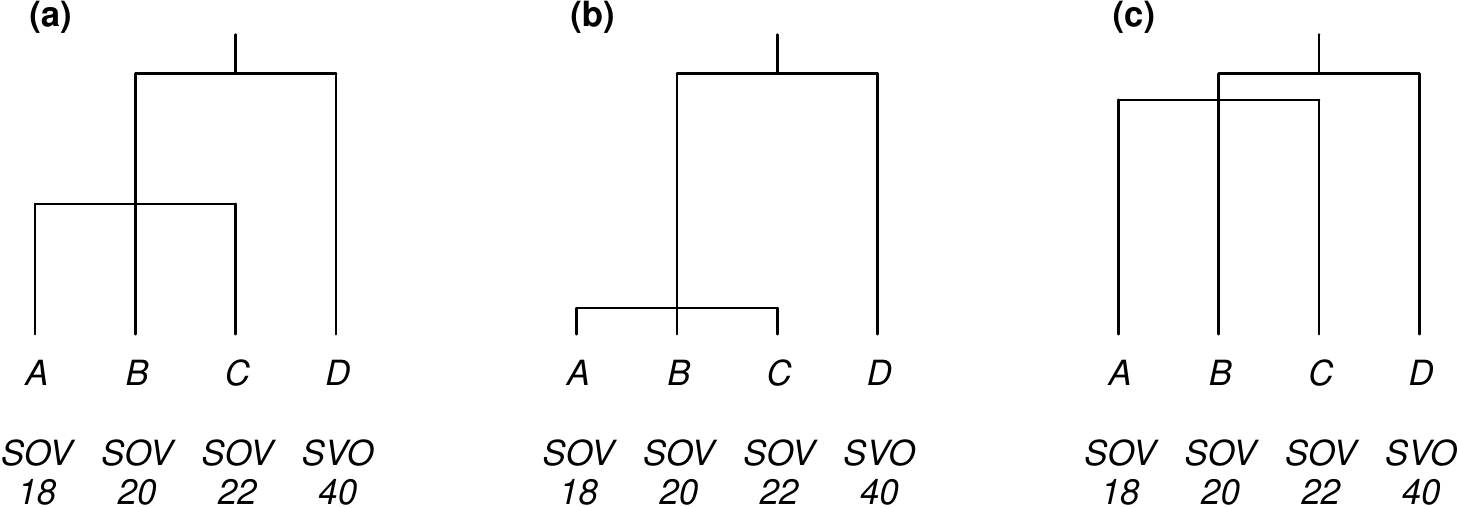} \caption{Three minimally different phylogenies of the same four languages, indicating their dominant word order and number of consonant phonemes}\label{fig:phy4}
\end{figure}

The essential challenge of formulating meaningful averages and proportions when languages are related will be well familiar to typologists. Figure \ref{fig:phy4} shows three, minimally different phylogenies for a set of four languages, together with the languages' dominant word order pattern and their number of consonant phonemes. If asked what proportion of these languages are SOV, a literal reply would be 75\%. However, that answer will strike us as less than satisfactory because languages A--C are more closely related to one another than to D. Merely tallying up the languages allows one of the two major branches in the tree to count three times more than the other. Moreover, the degree to which this answer seems unsatisfactory can vary between phylogenies \ref{fig:phy4}a,b,c. For instance, the answer `75\%', which is unsatisfactory for Figure \ref{fig:phy4}a, is arguably worse for Figure \ref{fig:phy4}b, since now A--C are very closely related indeed. Conversely, a reply of 75\% for Figure \ref{fig:phy4}c is still imperfect but arguably less unsatisfactory, since although A--C are more closely related to one another than to D, the difference is only slight. This example illustrates the fact that when quantifying the proportion of languages that have some property, any satisfactory method will need to take into account at least two facts about the phylogeny: its topology (i.e., the hierarchical embedding of subgroups) and its branch lengths (note that differing branch lengths are all that distinguish Figures \ref{fig:phy4}a,b,c). The same issues arise if we are seeking not a proportion but an average, such as the `average' size of the consonant inventories in these languages. The literal mean, \((18+20+22+40)/4 = 25\), is unsatisfactory for the same reason, that it accords much more weight to one major branch than the other. And similarly, it is even more unsatisfactory for Figure \ref{fig:phy4}b than for Figure \ref{fig:phy4}a, though less so for Figure \ref{fig:phy4}c.

There already exists a substantial literature on how to obtain principled values for proportions and averages that are sensitive to genealogy. Here we present two of the methods that have been developed. Before we do, it is useful to recall that even within non-phylogenetic statistics, there are multiple ways of formulating and defining an average, including means, medians, modes, harmonic means, geometric means, and so forth. Each of these operationalises a slightly different concept of the `representative middle value', or \emph{central tendency}, of some set of observations. Different averages have different properties which may prove advantageous or not, depending on the objectives and datasets at hand. For instance, means can be sensitive to outliers while medians are less so. It should be no surprise, then, that comparable issues arise in the formulation of phylogenetic averages, and the technical literature has discussed them at length \autocites{altschul_weights_1989}{vingron_weighting_1993}{stone_constructing_2007}{de_maio_phylogenetic_2020}. Here we will emphasise important properties of phylogenetic averages, in relation to the tasks that typologists face.

One way of construing different kinds of averages is in terms of the relative weight they accord to each observation. For instance, a simple mean accords every observation the same weight. Other kinds of averages can be expressed in terms of the slightly different weights they accord to each data point. This approach, of describing averages in terms of a list of weights for each observation, has also been used in the literature on phylogenetic averages, and we will adopt it here. We can also note that a proportion can be re-expressed as an average. Asking for the proportion of languages that are SOV is equivalent to asking for the mean of \(x\), where \(x=1\) if a language is SOV and \(x=0\) if it is not. Correspondingly, a method for constructing weighted averages will extend directly to the construction of weighted proportions. To take an example, suppose we assigned the four languages in Figure \ref{fig:phy4}a the weights \(\{0.2, 0.2, 0.2, 0.4\}\), which sum to \(1\). The weighted average of the consonant inventory sizes would then be \((0.2 \times 18 + 0.2 \times 20 + 0.2 \times 22 + 0.4 \times 40)/(0.2 + 0.2 + 0.2 + 0.4) = 28\). The correspondingly weighted proportion of SOV languages would be \((0.2 \times 1 + 0.2 \times 1 + 0.2 \times 1 + 0.4 \times 0)/(0.2 + 0.2 + 0.2 + 0.4) = 0.6\) or \(60\%\). Any method which can assign weights to a set of languages in a phylogenetically judicious manner will therefore enable us to calculate genealogically-sensitive averages and proportions.

\begin{figure}
\includegraphics[width=0.4\linewidth]{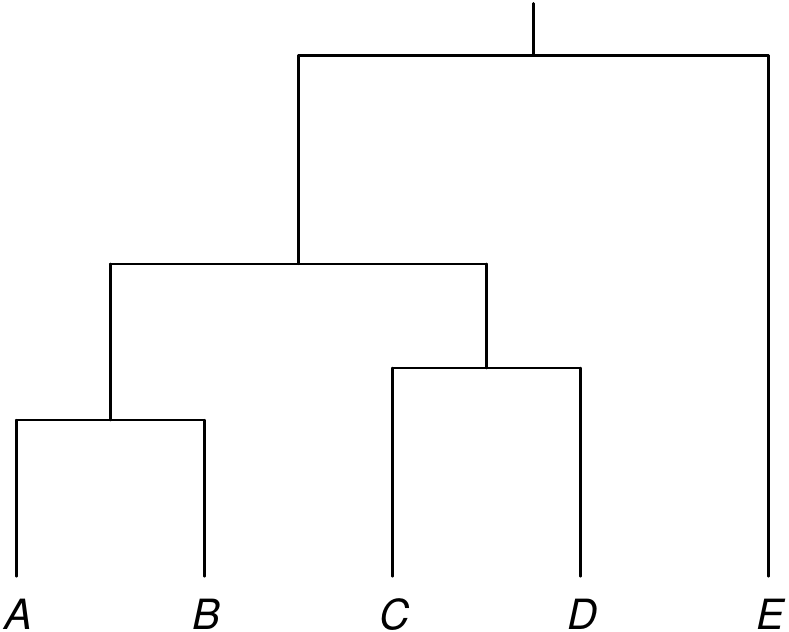} \caption{A phylogeny in which E is an outlier}\label{fig:phy5-outlier}
\end{figure}

The nearest phylogenetic equivalent to a simple mean is obtained by what is known as the `ACL' method presented by \textcite{altschul_weights_1989}. This kind of genealogically-sensitive average is often referred to as the \emph{phylogenetic mean}. It provides an unbiased estimate of the central tendency of a set of observations, taking into account tree topology and branch lengths. Nevertheless, the ACL method, like non-phylogenetic means, is known to be sensitive to outliers \autocite{stone_constructing_2007}. In a phylogeny, an outlier is a language (or subgroup) located on an early branch, only distantly related to the rest of the tree, such as language E in Figure \ref{fig:phy5-outlier}. Because the ACL method accords a high weight to outliers, its results can be particularly sensitive to the highest-level structure in a phylogeny. This can be of concern when confidence in the highest-order branching of the tree is low, as is often the case in linguistics, where the deepest splits in a family's history are also the murkiest or most contested by scholars. For that reason, it is prudent to consider another phylogenetic average, which was designed with this problem in mind.

The BranchManager (BM) method of \textcite{stone_constructing_2007} is also an unbiased estimate of the central tendency of a set of observations, taking into account tree topology and branch lengths. However, it is mathematically formulated to accord less extreme weight to high-order branching, in comparison to the ACL method. Arguably, this makes it a more conservative choice in cases where a phylogeny is especially uncertain at its greatest time depths. Moreover, it is possible to use both the ACL method and the BM method to estimate phylogenetically-sensitive proportions and averages, and then to compare them. The comparison will offer an indication of how the implied central tendency of the dataset changes, as we invest a greater or lesser degree of confidence in the correctness of the deepest levels of the tree structure. We make use of this approach in our case study, to which we now turn.

\hypertarget{laminals}{%
\section{A phylogenetic comparative case study: Laminal contrasts in Pama-Nyungan}\label{laminals}}

Phonemic systems are inherited with modification from ancestral languages into their descendants. Consequently, they are expected to contain considerable phylogenetic signal. In Australia, however, for one aspect of phonemic systems it has long been supposed that this is not the case. Australian languages contrast between four and six superlaryngeal places of articulation \autocites{evans_current_1995}{round_segment_2022}: bilabial, dorsal-velar and either one or two apical places (articulated with the tongue tip) and either one or two laminal places (articulated with the tongue blade). In this case study we focus on the laminals, and whether languages possess a contrast between two laminal places -- laminal dentals and laminal pre-palatals -- or just one. We introduce some long-standing claims about the distribution of this contrast across the continent, and then apply the kinds of analyses introduced in Sections \ref{phylo-sig}--\ref{weighting} above.

\begin{figure}
\includegraphics[width=0.49\linewidth]{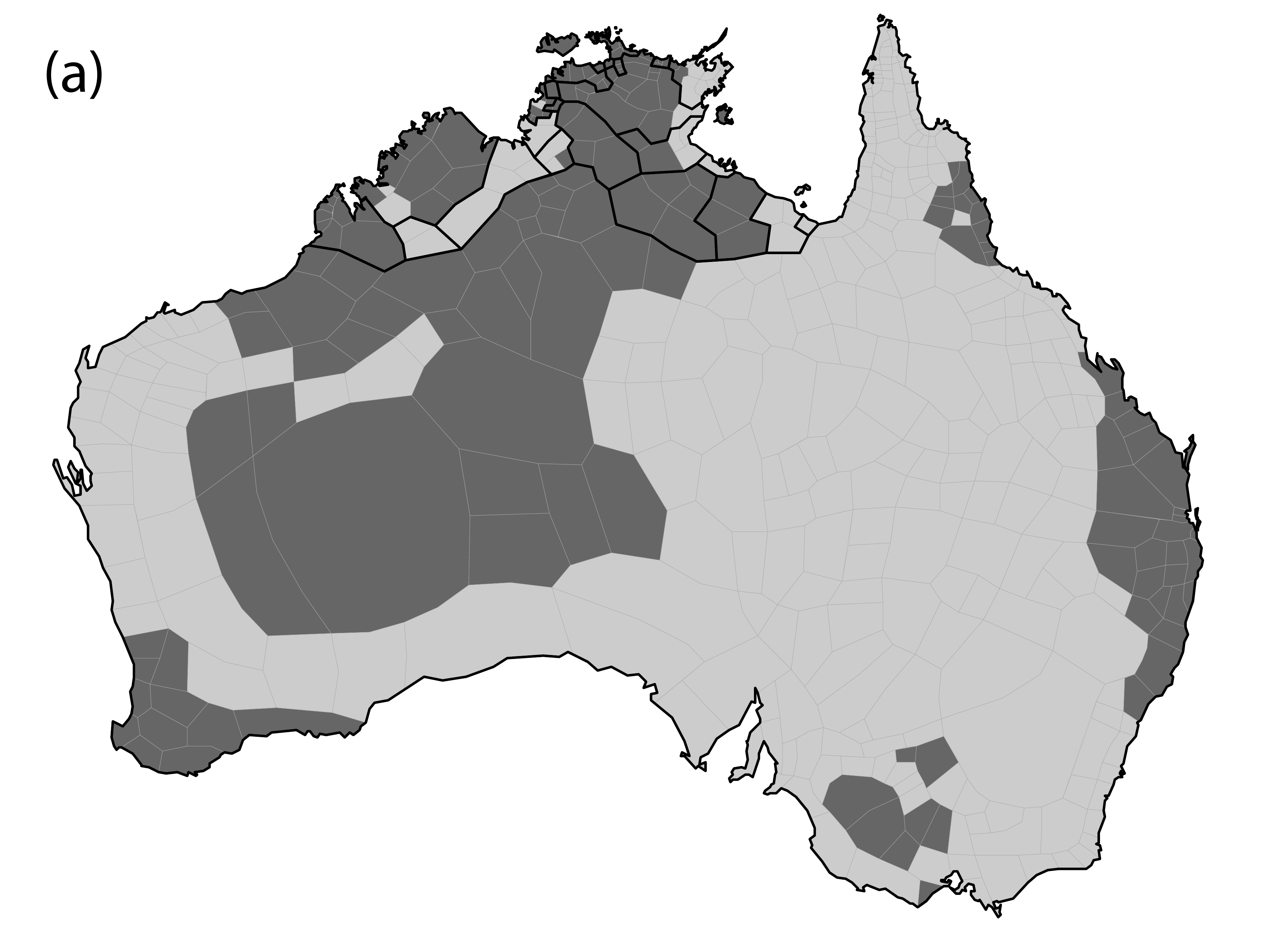} \includegraphics[width=0.49\linewidth]{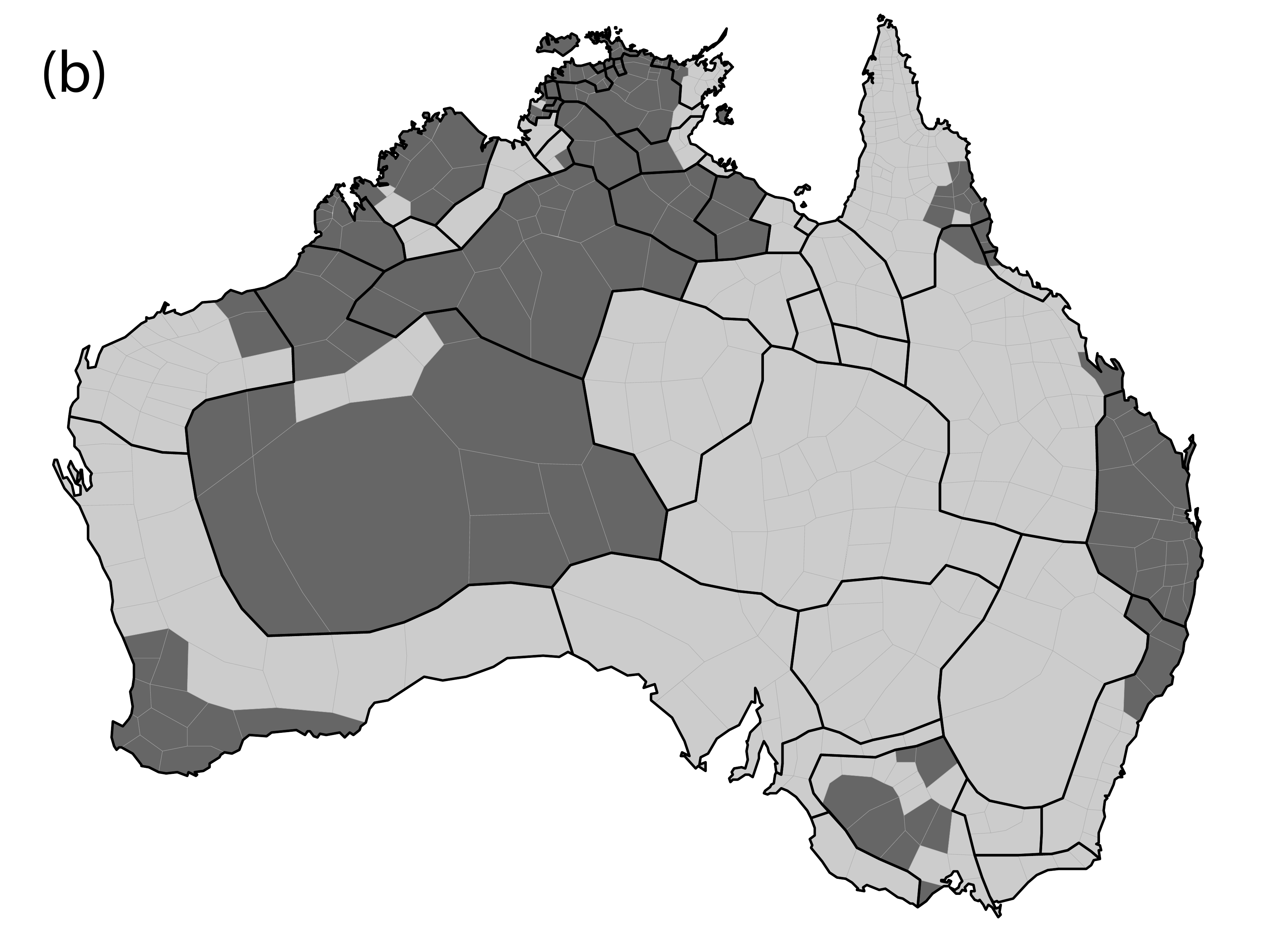} \caption{The distribution of the presence (light) and absence (dark) of a laminal place of articulation contrast in Australian languages. Dark lines indicate (a) language family boundaries, (b) major subgroups of Pama-Nyungan also.}\label{fig:maps}
\end{figure}

If we express the figure as a simple proportion, then around 62\% of Australian languages have a laminal contrast, according to data in \textcite{round_phonemic_2019}. The geographic distribution of the contrast is shown in Figure \ref{fig:maps}a, along with the boundaries of Australia's 25 mainland language families. The geographic distribution covers large contiguous swathes of the continent and can appear to exhibit little regard for the boundaries of language families. Understandably, this striking aspect of the distribution has been emphasised repeatedly in the literature on Australian phonological typology \autocites{dixon_proto-australian_1970}{dixon_languages_1980}{evans_current_1995}. However, here we ask, does this distribution also contain phylogenetic signal?

We begin by adding some additional information to our map. Figure \ref{fig:maps}b shows the same information as Figure \ref{fig:maps}a, but adds the boundaries of major subgroups of the Pama-Nyungan language family which dominates the continent. The reader may find that the effect of the map has changed: the distribution of the laminal contrast is largely organised neatly within the major phylogenetic units across the continent. Inspecting maps in this fashion can suggest potential conclusions about phylogenetic signal, but a more secure line of analysis is to use quantitative methods. Here we will focus on Pama-Nyungan. Within Pama-Nyungan, 73\% of languages have a laminal contrast, expressed as a simple proportion. In the remainder of the section, we first estimate the degree of phylogenetic signal in the distribution of the laminal contrast using the \(D\) statistic we introduced in Section \ref{phylo-sig-bin}, which measures phylogenetic signal in binary variables. We then turn to some more fine grained phonotactic data, to which we apply the \(K\) statistic introduced in Section \ref{phylo-sig-quant}, which measures phylogenetic signal in continuous variables. Having ascertained the level of phylogenetic signal in the Pama-Nyungan laminals, we then estimate the phylogenetically-weighted proportion of languages with a laminal contrast in Pama-Nyungan using the ACL and BM methods. To account for phylogenetic uncertainty, we consider results using a set of 100 Pama-Nyungan trees inferred by \textcite{bowern_pama-nyungan_2015} and described in \textcite{macklin-cordes_phylogenetic_2021}.

\hypertarget{phylogenetic-signal-in-the-binary-laminal-contrast}{%
\subsection{Phylogenetic signal in the binary laminal contrast}\label{phylogenetic-signal-in-the-binary-laminal-contrast}}

\begin{figure}
\includegraphics[width=1\linewidth]{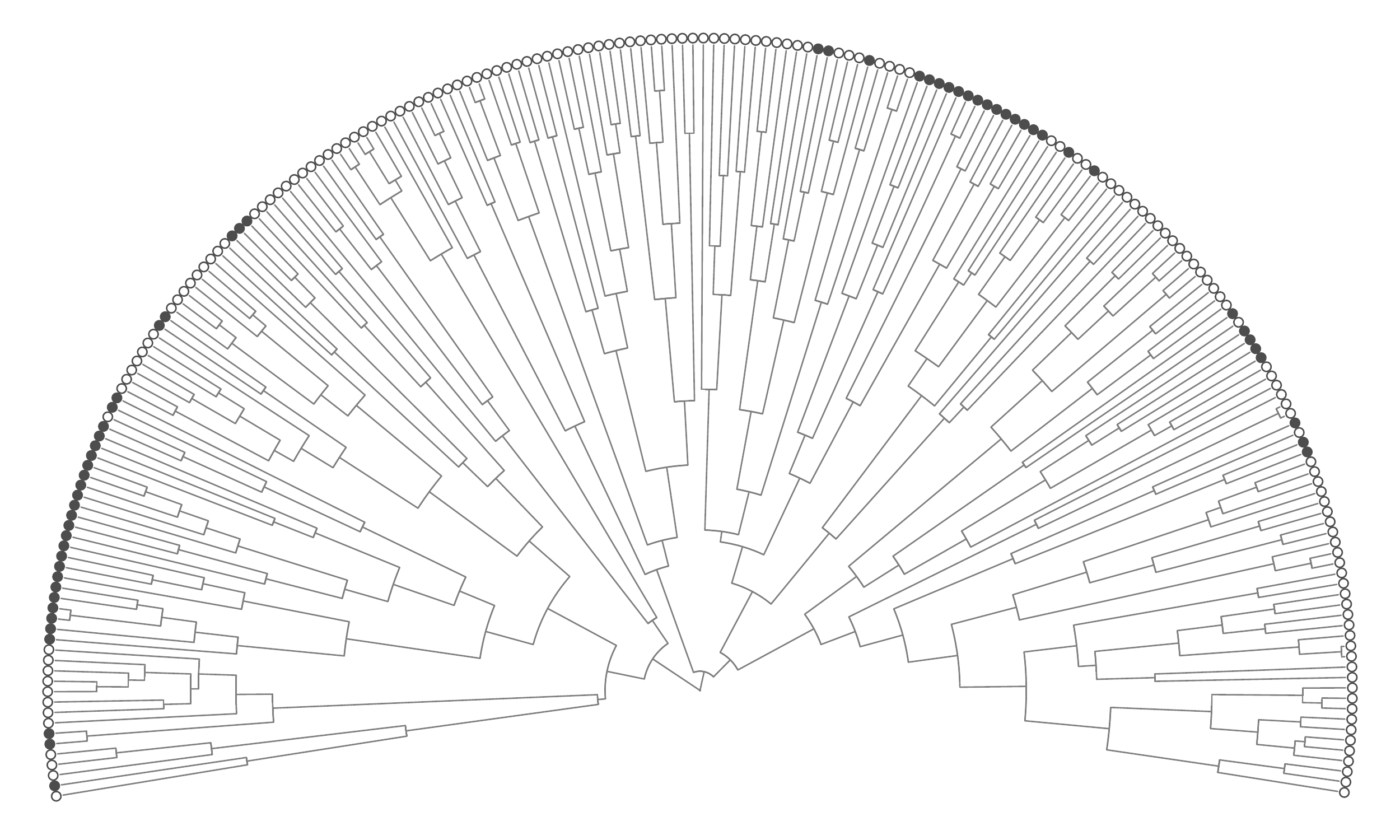} \caption{The distribution of the presence (light) and absence (dark) of a laminal place of articulation contrast across Pama-Nyungan, displayed on a maximum clade credibility (MCC) tree. An MCC tree is a single tree within a tree sample which most adequately represents the highest-probability subgroups in the trees of the sample. This MCC tree is taken from the sample of 100 highly-likely Paman-Nyungan phylogenies used in the current study.}\label{fig:laminal-tree}
\end{figure}

In Figure \ref{fig:maps}b, we saw that the distribution of the laminal contrast in Pama-Nyungan hews closely to major subgroup boundaries, so we will not be surprised if a \(D\) test returns a strong confirmation of phylogenetic signal. Figure \ref{fig:laminal-tree}, which plots the presence and absence of a laminal contrast against the Pama-Nyungan tree, reinforces this expectation. We tested a set of 216 Pama-Nyungan languages \autocite{round_phonemic_2019}, each coded for the binary presence/absence of the phonemic laminal contrast. To account for phylogenetic uncertainty, the statistic is calculated using 100 individual reference phylogenies.

\begin{table}

\caption{\label{tab:d-results}Phylogenetic signal in the binary presence/absence of a phonemic laminal contrast in 216 Pama-Nyungan languages. $D$ statistic using a sample of 100 reference trees, and $p$ values for the hypotheses of randomness (rejected) and phylogenetic signal (not rejected).}
\centering
\begin{tabular}[t]{ccc}
\toprule
\textbf{D statistic} & \textbf{p (randomness)} & \textbf{p (phylogenetic signal)}\\
\midrule
-0.439 (SD 0.019) & 0.000 (SD 0.000) & 0.987 (SD 0.005)\\
\bottomrule
\end{tabular}
\end{table}

The 100 results are summarised in Table \ref{tab:d-results}. The mean \(D\) statistic obtained is low, at \(-0.439\), indicating that the data is phylogenetically clumped to an even greater degree than expected under a Brownian motion model of evolution. Results like this can emerge when the variable under study has changed only rarely, and the changes have mostly been deep within the tree. This is the case in Pama-Nyungan, where variation in the presence/absence of the laminal contrast is mainly \emph{between} major subgroups rather than within them. Returning to the statistical results, the hypothesis of randomness is rejected \((p < 0.001)\) and the hypothesis of phylogenetic signal is not rejected (\(p = 0.987 \pm 0.005\)). The values of the \(D\) statistic have a small standard deviation (\(0.019\)), indicating that a similar result is obtained for all 100 reference trees. In sum, the \(D\) test results confirm, in a quantitative manner and taking into account our uncertainty in the Pama-Nyungan phylogenetic tree, what our inspection of the map in Figure \ref{fig:maps}b could only suggest: that the binary presence/absence of the laminal contrast in Pama-Nyungan has strong phylogenetic signal.

\hypertarget{phylogenetic-signal-in-continuously-valued-phonotactic-variables}{%
\subsection{Phylogenetic signal in continuously-valued phonotactic variables}\label{phylogenetic-signal-in-continuously-valued-phonotactic-variables}}

Languages vary not only in what contrastive segments they have but also in how frequently they use them \autocites{frisch_similarity_2004}{hall_probabilistic_2009}{wedel_high_2013}{macklin-cordes_re-evaluating_2020}. For example, Pitta Pitta \autocite{blake_pitta_1990} and Burduna \autocite{burgman_burduna_2007} are similar in that they both contrast laminal stops, nasals and laterals in word-initial position. However, a closer examination reveals notable differences. In word-initial position before /u/, 29\% of the consonantal laminals in Pitta Pitta are pre-palatal while 71\% are dental, whereas in Burduna the frequencies are reversed, with 68\% pre-palatal and just 32\% dental. Frequency measures such as these can be viewed as continuous variables that can be investigated for phylogenetic signal \autocite{macklin-cordes_phylogenetic_2021}. In this section we examine continuous variables of this kind, which describe the relative predominance of pre-palatals versus dentals in nine phonotactic positions, across 76 languages that possess the contrast. Data is from a phonemicised lexical database of Australian languages, which is under development \autocite{round_ausphon-lexicon_2017}, and which extends and enhances the Chirila database \autocite{bowern_chirila_2016}. Raw data tables and details of the primary language documentation sources are provided in Section S2 of the Supplementary Materials.

Our choice of nine variables is informed by the typological literature on Australian phonology. One long-established characteristic of Australian laminals is that their relative frequencies are sensitive to the quality of neighbouring vowels \autocites{dixon_proto-australian_1970}{dixon_languages_1980}.\footnote{The palatal semi-vowel /j/ patterns more freely. In this section we set it aside and examine the consonantal laminals, i.e., laterals, nasals and obstruents.} Most Australian languages have three contrastive vowel qualities \autocite{round_segment_2022}, with /i/ contexts favouring the laminal pre-palatal, /u/ contexts favouring the dental, and /a/ contexts somewhere in between. Here we examine the relative predominance of pre-palatals in word-initial position before /i,a,u/ and in intervocalic position before /i,a,u/ and after /i,a,u/.\footnote{To minimise error in the values of the variables, we include observations only from those languages in whose lexicons at least 20 consonantal laminals are attested in the relevant phonotactic context (see further, Section S2 the Supplementary Materials).} We apply the randomisation test described in Section \ref{phylo-sig-quant} and then calculate a \(K\) statistic. As in our \(D\) test, we address phylogenetic uncertainty by applying the statistical tests using a sample of 100 reference trees.

\begin{table}

\caption{\label{tab:k-results}Phylogenetic signal in nine continuous variables describing the proportion of laminals which are pre-palatal, in specific phonotactic contexts. $K$ statistic using a sample of 100 reference trees, and $p$ values for the hypothesis of randomness (rejected in all cases).}
\centering
\begin{tabular}[t]{lll}
\toprule
\textbf{Context} & \textbf{K} & \textbf{p (randomness)}\\
\midrule
\#\_a & 0.827 (SD 0.052) & 0.001 (SD 0.000)\\
\#\_i & 1.322 (SD 0.055) & 0.001 (SD 0.000)\\
\#\_u & 0.783 (SD 0.040) & 0.001 (SD 0.000)\\
\addlinespace
a\_V & 0.480 (SD 0.038) & 0.002 (SD 0.009)\\
i\_V & 0.536 (SD 0.031) & 0.002 (SD 0.001)\\
u\_V & 0.615 (SD 0.018) & 0.001 (SD 0.000)\\
\addlinespace
V\_a & 0.337 (SD 0.019) & 0.015 (SD 0.011)\\
V\_i & 0.696 (SD 0.025) & 0.001 (SD 0.000)\\
V\_u & 0.620 (SD 0.019) & 0.003 (SD 0.002)\\
\bottomrule
\end{tabular}
\end{table}

Results are summarised in Table \ref{tab:k-results}. The randomisation test finds phylogenetic signal to be statistically significant (\(p < 0.05\)) in all 9 variables and 100 reference trees except in two cases: these were the a\_V and V\_a contexts, for the same, one tree. Given that both contexts are judged to have significant phylogenetic signal in all other 99 trees in the 100-tree sample, we conclude that phylogenetic signal is present at a stastically significant level in all nine phonotactic variables.

The findings for the \(K\) statistic differ among the variables. For the word-initial variables, \(K\) is high, ranging from \(0.783\) to \(1.322\), whereas for the intervocalic variables it is uniformly lower, ranging from \(0.337\) to \(0.696\). In all cases, the standard deviation is low, indicating that similar results are obtained for all 100 reference trees. To put these \(K\) values in perspective, \textcite{blomberg_testing_2003} examined 121 biological traits of a wide variety of plant and animal organisms, finding mean \(K\) of 0.35 for behavioral traits, 0.54 for physiology and 0.83 for traits related to body size. \textcite{macklin-cordes_phylogenetic_2021} estimated \(K\) for biphones (sequences of two adjacent phonemes) in Pama-Nyungan and found mean \(K\) of 0.52 for biphones of individual segments, and \(K\) of 0.63 when segments are binned into groups by place or manner of articulation. This suggests that our laminal phonotactic variables exhibit a level of phylogenetic signal at least as high as many evolved, biological traits, as well as the Pama-Nyungan biphone variables investigated in \textcite{macklin-cordes_phylogenetic_2021}.

The highest \(K\) value, at \(1.322\), is for laminals in word-initial position before /i/. A \(K\) value well above \(1\) is consistent with a scenario in which a linguistic property varies between deep branches of the tree, but much less so within the subgroups below those branches. This is true of Pama-Nyungan laminals word-initially before /i/. In the western half of the family, this position favours pre-palatals, reflecting a typical effect of the neighbouring vowel, whereas in the eastern half, the initial position in a word is one which favours dentals, irrespective of the following vowel.

A novel and consistent finding was that laminals exhibit stronger phylogenetic signal in word-initial position than intervocalically. There are many reasons why this might be so, and here we consider just one. \emph{Pertinacity} \autocite{dresher_main_2005} refers the perpetuation of linguistic patterns even as the items that instantiate them change. For instance, though a borrowed word may be new, its phonology is often reshaped to conform to the existing patterns in the recipient language \autocite{hyman_role_1970}, which then perpetuates the phonological patterns even as the set of items instantiating them changes. Similarly, if neologisms conform to existing statistical patterns in the lexicon, they too will contribute to pertinacity. Because our phonotactic variables are based on whole lexicons, and not merely a basic vocabulary list, lexical turnover will have been an important contributor to their historical dynamics. If it is the case that word-initial laminals have been subject to more-pertinacious changes than intervocalic laminals, such as more reshaping of borrowed words, or neologism which more closely replicates existing statistical patterns in the lexicon, then this could potentially lead to the difference in phylogenetic signal that we find. Whether there is additional evidence to support this hypothesis remains a question for future research, however the fact that such a hypothesis is able to emerge, illustrates how phylogenetic analysis can supplement the typologist's existing toolkit for generating theoretically interesting hypotheses from the analysis of cross-linguistic data.

\hypertarget{genealogically-sensitive-proportions-of-languages-with-a-laminal-contrast}{%
\subsection{Genealogically-sensitive proportions of languages with a laminal contrast}\label{genealogically-sensitive-proportions-of-languages-with-a-laminal-contrast}}

We turn now to examine the phylogenetically-weighted proportion of Pama-Nyungan languages that have a laminal contrast. We know already, just by counting, that the simple proportion of Pama-Nyungan languages with a laminal contrast is \(157/216 = 0.727\). Our question here is, what is the proportion when genealogy is taken into account? As discussed in Section \ref{weighting}, there are different methods available for calculating this phylogenetic quantity, just as there are different kinds of non-phylogenetic averages. Here we compare the ACL and BM methods introduced earlier. We account for phylogenetic uncertainty by calculating them with respect to a sample of 100 reference trees. Table \ref{tab:weighting-results} reports the results. In this case the answer is broadly similar according to all three methods: the simple proportion is \(0.727\), the ACL-weighted proportion is somewhat higher, at \(0.761\) (SD \(0.009\)) and the BM-weighted proportion marginally lower, at \(0.705\) (SD \(0.003\)). The standard deviations of the phylogenetically weighted proportions are low, indicating that a similar result is obtained for all 100 reference trees. As mentioned in Section \ref{weighting}, an ACL proportion is more sensitive to genealogical structure deep within the tree than the BM method is, thus if we wish to remain conservative about our confidence in deep tree structure, we could conclude that a figure of around 71\% (but perhaps as high as 76\%) provides a good representation of the proportion of Pama-Nyungan languages that possess a laminal contrast. Note that unlike for balanced sampling, we did not need to discard any data, meaning that our results provide a faithful reflection of the evidence provided by all 216 languages and they do so while taking phylogenetic autocorrelation, including our uncertainty about Pama-Nyungan genealogy, into account.

\begin{table}

\caption{\label{tab:weighting-results}Genealogically sensitive proportions of Pama-Nyungan languages with a laminal contrast.}
\centering
\begin{tabular}[t]{lll}
\toprule
\textbf{Simple proportion} & \textbf{ACL weighting} & \textbf{BM weighting}\\
\midrule
0.727 & 0.761 (SD 0.009) & 0.705 (SD 0.003)\\
\bottomrule
\end{tabular}
\end{table}

Our case study has illustrated the application of methods and principles introduced in earlier sections. We have confirmed that the presence/absence of a laminal contrast in Pama-Nyungan has significant phylogenetic signal, notwithstanding a long history in the literature of emphasising its apparent areality. An examination of phylogenetic signal in continuously-valued phonotactic variables prompted us to notice a major east-west split in the treatment of word-initial laminals before /i/ and suggested a potential difference in the pertinacity of laminals and their statistical frequencies in word-initial versus intervocalic positions. Finally, having first confirmed the presence of phylogenetic signal, we then calculated genealogically-weighted proportions of the Pama-Nyungan languages which have the laminal contrast. This was done taking into account phylogenetic uncertainty in the Pama-Nyungan tree, and using two weighing methods which allow us to compare the consequences of investing a more conservative or less conservative degree of confidence in the deep-time branching structure of the trees.

\hypertarget{discussion}{%
\section{Discussion}\label{discussion}}

Phylogenetic autocorrelation has long challenged the analysis of comparative data both in linguistics and in other comparative sciences, such as comparative anthropology and comparative biology. The core problem is that many statistical methods require observations that are independent, yet languages, cultures and species are inherently non-independent owing to the way they develop historically. For several decades, comparative fields explored methodological approaches which were broadly parallel, focussed on balanced sampling. Obvious drawbacks of such approaches are that the vast majority of available comparative data must be ignored, and that even then, complete independence remains elusive. In 1985, Felsenstein showed that by focussing on phylogenetically independent contrasts it is possible even under conditions of phylogenetic autocorrelation to extract truly independent observations for subsequent analysis. We have argued that it is nothing more than historical accident that this breakthrough occurred in biology and not in linguistic typology or anthropology, since it is the solution to a problem that is shared across disciplinary boundaries. One of the motivations behind this article, is that while phylogenetic comparative methods have been gaining currency in linguistics, their essential relationship to balanced sampling in linguistic typology has not been clearly articulated, and we hope to have achieved that here.

In Sections \ref{phylo-sig}--\ref{laminals} we introduced concepts and related methods for reckoning with phylogenetic signal, phylogenetic uncertainty and genealogically-sensitive averages. A leitmotif running through that presentation was that phylogenetic comparative methods do not lock the typologist into any single assumption about a phylogeny. On the contrary, because these methods require a precise statement of one's hypothesised phylogeny, it is possible to compare multiple hypotheses and explicitly examine their impacts on the analysis. In this section we expand on some of our earlier points in relation to two topics of central importance in typology: comparison across families and areality.

\hypertarget{deep-time}{%
\subsection{Comparison across families and deep-time genealogy}\label{deep-time}}

Throughout our paper, we have discussed phylogenetic comparative methods primarily within the scope of a single family. In this single-family, single-tree context we have examined phylogenetic uncertainty, testing for phylogenetic signal and the estimation of genealogically-sensitive averages and proportions. However, in Section \ref{phylo-uncertainty} at the end of our discussion of uncertainty in phylogentic trees, we mentioned the problem of comparing across language families. We noted that logically, if it is believed that multiple families ultimately are related genealogically, then it is not possible to compare them without implicating a grand phylogeny that links them all. Methods which place all families on equal footing merely do this by positing a rake tree. Thus, as radical as it may sound to say that we must hypothesise a deep-time tree which links currently-distinct families together, this is in fact something linguists have been doing for decades, covertly. Consequently, the question is not whether to use a grand, supra-familial tree but instead, which grand tree to use. Until now, linguists have generally declined to engage in positing grand trees that span beyond the reach of the comparative method, for the eminently good reason, that such trees cannot be demonstrated to be correct. However, as we have emphasised, trees do not need to be verifiably correct to be gainfully used with phylogenetic comparative methods. Instead, trees are hypotheses. Even if we do not, or cannot, know what the correct tree is, we surely can distinguish between more or less plausible hypotheses. Once we view the creation of grand trees as a matter of \emph{hypothesis generation}, then there is every reason to begin working with them earnestly. For readers who find themselves still skeptical, consider the issue presented in the form of this question: Is a rake tree truly the best hypothesis that linguists could come up with about deep-time relatedness, entailing that every language family everywhere in the world is exactly equally related to every other? If our answer is anything other than an unequivocal yes, then we are effectively, tacitly entertaining the existence of other, more plausible grand trees.

To summarise so far, in order to apply phylogenetic comparative methods not only within but also across known families, we join the families in a grand tree. If the grand tree is a rake, then we are effectively continuing current practice in supra-familial language sampling. If the grand tree is otherwise, then we are beginning to explore alternative hypotheses for deep-time relatedness. As with the examples discussed earlier in the paper, phylogenetic comparative methods can be applied to multiple, alternative grand trees in order to reflect phylogenetic uncertainty and to investigate its implications.

Given this state of affairs, it strikes us that an important task for linguistic typology in coming years will be to establish an inventory of deep-time genealogical hypotheses, represented as phylogenies, as key ingredients for phylogenetic typological research, much in the way that the field in previous decades developed a variety of sampling techniques. Hypotheses within this inventory might come from many sources, whether from detailed interdisciplinary studies such as \textcite{matsumae2021} or novel linguistic attempts such as \textcite{jager_global-scale_2018}, or more prosaically in the form of random samples of plausible hypotheses that meet certain constraining assumptions. There is ample scope for innovation. In Section S1 of the Supplementary Materials, we provide an extended description of a set of tools \autocite{glottoTrees} designed specifically with linguists in mind, for generating hypotheses about linguistic genealogy either within or across families, by creating and adjusting explicit linguistic phylogenies (see also \autocite{dediu_making_2018} for constructing within-family trees).

\hypertarget{areality}{%
\subsection{Areality}\label{areality}}

In scientific discussions with colleagues, we have encountered the concern that phylogenetic comparative methods cannot work, because they do not take into account the effects of areality \autocites[similarly, in published work see e.g.][]{blench_new_2015}{francois_trees_2014}. We believe that this concern may follow from a partial misapprehension about what phylogenetic comparative methods ought to be able to achieve. By way of comparison, it would be amiss to argue that a good model of gender should not be incorporated into a sociolinguistic analysis, merely because it does not account for geography. One could argue with good justification that we also desire an account of geography, but that is not the same thing as rejecting the successful model of gender. Similarly, we should not dismiss the breakthrough that Felsenstein achieved, dealing with genealogy far more effectively than in previous methods, merely because areality remains as difficult a problem as it always was. Here we briefly discuss why areality remains a hard problem and what can be done about it.

Viewed in mathematical and statistical terms, phylogenies are rather simple geometric objects. One consequence of their simplicity is that PICs can also be defined in a simple and effective manner. In contrast, the relationships implied by thousands of years of areality, including interactions with languages that have left no direct descendants, are significantly more complex. As mentioned in Section \ref{phylo-uncertainty}, comparative biology is also confronted with similarities shaped by areality, including in high-stakes fields such as bacteriology. Thus it is not for lack of motivation or interest that mathematical biologists are yet to produce methodological solutions to areality that match the solutions for phylogeny. The work is well underway, but the mathematics of \emph{historical networks}, which such phenomena imply, is truly challenging \autocite{elworth2019advances}.

In this context, it is imperative for typologists to continue grappling with the problem of areality, though not by rejecting phylogenetic comparative methods, but instead by supplementing them. Recent methodological work that addresses areality in concert with phylogenetic comparative methods includes \textcite{cathcart_areal_2018} on areality in grammatical change, and \textcite{verkerk_detecting_2019} on estimating areality effects in relation to phylogenetic uncertainty. Similarly, it will be important to continue to learn more about the empirical facts of areality and its typological implications, to better understand its expected quantitative impact on the performance of phylogenetic comparative methods. For example, in the domain of lexical phylogenetic inference, \textcite{bowern_does_2011} clarified empirical levels of lexical borrowing among hunter-gatherer and small-scale agriculturalist societies, providing crucial empirical knowledge about areality which could then be compared with the results of robustness studies \autocite{greenhill_does_2009}, to suggest that at known empirical rates of borrowing, quantitative inference of phylogenies from lexical data should not suffer from significant impairment.

In all likelihood, areality will remain a tough challenge for linguistic typology, as it is for comparative biology, for some time to come. The problems that areality presents are different to and more complex than phylogeny. However, the mere fact that areality is hard is no sound reason to reject the advances offered by phylogenetic comparative methods. Instead, as always, the best available methods for handling genealogy must be supplemented with the current best attempts at handling areality.

\hypertarget{conclusions}{%
\section{Conclusions}\label{conclusions}}

Typologists are deeply invested in the methodology of balanced sampling, because traditionally it has been our best response to the fundamental challenge of phylogenetic autocorrelation. However, phylogenetic comparative methods provide a better solution to the same problem. The fact that these methods were invented in biology is an accident of history; they could just as well have been invented in linguistics. While phylogenetic comparative methods do not solve all of the problems of typological analysis, they do solve the core challenge of phylogeny. For this reason, we see little reason not to adopt them, apart from inertia and perhaps a little professional envy (given that a linguist did not, in fact, discover them). To assist typologists who are interested in exploring these methods, here we introduced some fundamental concepts and methodological tools, and provided an illustration of their application in a typological case study. In Section S1 of the Supplementary Materials, we introduce computational tools for converting genealogical hypotheses into trees, and using the trees to calculate genealogically-sensitive averages. See also footnotes in Section \ref{phylo-sig} for references to other, free computational tools for examining phylogenetic signal. Phylogenetic comparative methods will enable typologists for the first time to use all available documentary data when drawing inferences about the diversity of human language, and to begin a far richer discussion on how competing hypotheses about linguistic genealogy---whether in shallow or in deep time---can alter the inferences we draw about the nature of human language from the empirical evidence granted us by today's seven thousand tongues.

\hypertarget{data-availability-statement}{%
\section*{Data availability statement}\label{data-availability-statement}}
\addcontentsline{toc}{section}{Data availability statement}

Data and results files are available on Zenodo at \url{https://doi.org/10.5281/zenodo.5602216}. Documentation and code for performing the analysis is available in Supplementary Materials Section S2. The R packages \emph{glottoTrees} \autocite{glottoTrees} and \emph{phyloWeights} \autocite{phyloWeights} referred to in Supplementary Materials Section S1 are available at \url{https://github.com/erichround/glottoTrees} and \url{https://github.com/erichround/phyloWeights}.

\printbibliography

@book{nichols_linguistic_1992,
	address = {Chicago},
	title = {Linguistic Diversity in Space and Time},
	isbn = {978-0-226-58056-2},
	pagetotal = {388},
	publisher = {University of Chicago Press},
	author = {Nichols, Johanna},
	year = {1992}
}

@article{naroll_two_1961,
	title = {Two Solutions to {Galton's} Problem},
	volume = {28},
	issn = {0031-8248},
	doi = {10.1086/287778},
	pages = {15--39},
	number = {1},
	journal = {Philosophy of Science},
	shortjournal = {Philosophy of Science},
	author = {Naroll, Raoul},
	year = {1961}
}

@article{perkins_statistical_1989,
	title = {Statistical techniques for determining language sample size},
	volume = {13},
	doi = {10.1075/sl.13.2.04per},
	pages = {293--315},
	number = {2},
	journal = {Studies in Language},
	author = {Perkins, Revere D.},
	year = {1989},
	address = {Amsterdam}
}

@article{dryer_large_1989,
	title = {Large linguistic areas and language sampling},
	volume = {13},
	issn = {0378-4177},
	doi = {10.1075/sl.13.2.03dry},
	pages = {257--292},
	number = {2},
	journal = {Studies in Language},
	author = {Dryer, Matthew S.},
	year = {1989}
}

@incollection{bell_language_1978,
	address = {Stanford, California},
	title = {Language samples},
	volume = {1},
	pages = {123--156},
	booktitle = {Universals of Human Language},
	publisher = {Stanford University Press},
	author = {Bell, Alan},
	editor = {Greenberg, Joseph H.},
		year = {1978}
}

@phdthesis{perkins_evolution_1980,
	address = {Buffalo, New York},
	type = {Ph.{D}. thesis},
	title = {The Evolution of Culture and Grammar},
	school = {State University of New York},
	author = {Perkins, Revere D.},
	year = {1980}
}

@incollection{perkins_covariation_1988,
	address = {Amsterdam},
	title = {The covariation of culture and grammar},
	isbn = {978-90-272-2891-8},
	booktitle = {Studies in Syntactic Typology},
	publisher = {John Benjamins Publishing},
	author = {Perkins, Revere D.},
	editor = {Hammond, Michael and Moravcsik, Edith A. and Wirth, Jessica R.},
	year = {1988}
}

@book{bybee_morphology_1985,
	address = {Amsterdam},
	series = {Typological {Studies} in {Language}},
	title = {Morphology: {A} Study of the Relation Between Meaning and Form},
	isbn = {978-90-272-2878-9},
	shorttitle = {Morphology},
	doi = {10.1075/tsl.9},
	number = {9},
	publisher = {John Benjamins Publishing},
	author = {Bybee, Joan L.},
	month = jan,
	year = {1985}
}

@article{voegelin_index_1966,
	title = {Index of languages of the world},
	volume = {8},
	issn = {0003-5483},
	url = {http://www.jstor.org/stable/30029439},
	pages = {1--222},
	number = {6},
	journal = {Anthropological Linguistics},
	author = {Voegelin, C. F. and Voegelin, F. M.},
	year = {1966}
}

@phdthesis{kenny_numerical_1975,
	address = {Bloomington},
	type = {Ph.{D}. thesis},
	title = {A numerical taxonomy of ethnic units using {Murdock}'s 1967 world sample.},
	school = {Indiana University},
	author = {Kenny, James Andrew},
	year = {1975}
}

@book{murdock_ethnographic_1967,
	address = {Pittsburgh},
	title = {Ethnographic Atlas},
	publisher = {University of Pittsburgh Press},
	author = {Murdock, George Peter},
	year = {1967}
}

@article{rijkhoff_method_1993,
	title = {A method of language sampling},
	volume = {17},
	issn = {0378-4177, 1569-9978},
	doi = {10.1075/sl.17.1.07rij},
	number = {1},
	journal = {Studies in Language},
	author = {Rijkhoff, Jan and Bakker, Dik and Hengeveld, Kees and Kahrel, Peter},
	month = jan,
	year = {1993},
	pages = {169--203}
}

@article{rijkhoff_language_1998,
	title = {Language sampling},
	volume = {2},
	number = {3},
	journal = {Linguistic Typology},
	author = {Rijkhoff, Jan and Bakker, Dik},
	year = {1998},
	pages = {263--314}
}

@incollection{bakker_language_2011,
	address = {Oxford},
	title = {Language sampling},
	booktitle = {The {Oxford} Handbook of Linguistic Typology},
	publisher = {Oxford University Press},
	author = {Bakker, Dik},
	editor = {Song, Jae Jung},
	year = {2011},
	pages = {100--127}
}

@article{miestamo_sampling_2016,
	title = {Sampling for variety},
	volume = {20},
	issn = {1430-0532},
	doi = {10.1515/lingty-2016-0006},
	number = {2},
	journal = {Linguistic Typology},
	author = {Miestamo, Matti and Bakker, Dik and Arppe, Antti},
	year = {2016},
	pages = {233--296}
}

@book{dryer_wals_2013,
	address = {Leipzig},
	title = {{WALS} Online},
	url = {http://wals.info/},
	publisher = {Max Planck Institute for Evolutionary Anthropology},
	editor = {Dryer, Matthew S. and Haspelmath, Martin},
	year = {2013}
}

@article{bickel_refined_2009,
	title = {A refined sampling procedure for genealogical control},
	volume = {61},
	doi = {10.1524/stuf.2008.0022},
	pages = {221},
	number = {3},
	journaltitle = {{STUF} - Language Typology and Universals (Sprachtypologie und Universalienforschung)},
	shortjournal = {stuf},
	author = {Bickel, Balthasar},
	year = {2009}
}

@article{roberts_robust_2018,
	title = {Robust, causal, and incremental approaches to investigating linguistic adaptation},
	volume = {9},
	issn = {1664-1078},
	doi = {10.3389/fpsyg.2018.00166},
	
	journal = {Frontiers in Psychology},
	author = {Roberts, Seán G.},
	year = {2018}
}

@article{felsenstein_phylogenies_1985,
	title = {Phylogenies and the Comparative Method},
	volume = {125},
	issn = {0003-0147},
	doi = {10.1086/284325},
	pages = {1--15},
	number = {1},
	journal = {The American Naturalist},
	shortjournal = {The American Naturalist},
	author = {Felsenstein, Joseph},
	year = {1985}
}

@incollection{harvey_comparisons_1982,
	address = {Cambridge},
	title = {Comparisons between taxa and adaptive trends: {P}roblems of methods},
	shorttitle = {Comparisons between taxa and adaptive trends},
	pages = {343--361},
	booktitle = {Current Problems in Sociobiology},
	publisher = {Cambridge University Press},
	author = {Harvey, Paul H. and Mace, Georgina M.},
	editor = {{King's College Sociobiology Group, Cambridge}},
	year = {1982}
}

@article{clutton-brock_primate_1977,
	title = {Primate ecology and social organization},
	volume = {183},
	issn = {1469-7998},
	doi = {10.1111/j.1469-7998.1977.tb04171.x},
	pages = {1--39},
	number = {1},
	journal = {Journal of Zoology},
	author = {Clutton-Brock, T. H. and Harvey, Paul H.},
	year = {1977}
}

@article{baker_evolution_1979,
	title = {The evolution of bird coloration},
	volume = {287},
	rights = {Scanned images copyright © 2017, Royal Society},
	issn = {0080-4622, 2054-0280},
	doi = {10.1098/rstb.1979.0053},
	pages = {63--130},
	number = {1018},
	journal = {Philosophical Transactions of the Royal Society of London B: Biological Sciences},
	shortjournal = {Phil. Trans. R. Soc. Lond. B},
	author = {Baker, R. R. and Parker, G. A.},
	year = {1979}
}

@book{nunn_comparative_2011,
	address = {Chicago},
	title = {The Comparative Approach in Evolutionary Anthropology and Biology},
	publisher = {University of Chicago Press},
	author = {Nunn, Charles L.},
	year = {2011}
}

@article{grafen_phylogenetic_1989,
	title = {The phylogenetic regression},
	volume = {326},
	issn = {0962-8436},
	pages = {119--157},
	number = {1233},
	journal = {Philosophical Transactions of the Royal Society of London. Series B, Biological Sciences},
	shortjournal = {Philos. Trans. R. Soc. Lond., B, Biol. Sci.},
	author = {Grafen, A.},
	year = {1989},
	pmid = {2575770}
}

@article{dunn_evolved_2011,
	title = {Evolved structure of language shows lineage-specific trends in word-order universals},
	volume = {473},
	copyright = {2011 Nature Publishing Group},
	issn = {1476-4687},
	doi = {10.1038/nature09923},
	number = {7345},
	journal = {Nature},
	author = {Dunn, Michael and Greenhill, Simon J. and Levinson, Stephen C. and Gray, Russell D.},
	month = may,
	year = {2011},
	pages = {79--82}
}

@article{maurits_tracing_2014,
	title = {Tracing the roots of syntax with {Bayesian} phylogenetics},
	volume = {111},
	copyright = {©  . Freely available online through the PNAS open access option.},
	issn = {0027-8424, 1091-6490},
	doi = {10.1073/pnas.1319042111},
	number = {37},
	journal = {Proceedings of the National Academy of Sciences},
	author = {Maurits, Luke and Griffiths, Thomas L.},
	month = sep,
	year = {2014},
	pmid = {25192934},
	pages = {13576--13581}
}

@article{verkerk_diachronic_2014,
	title = {Diachronic change in {Indo}-{European} motion event encoding},
	volume = {4},
	issn = {2210-2116, 2210-2124},
	doi = {10.1075/jhl.4.1.02ver},
	number = {1},
	journal = {Journal of Historical Linguistics},
	author = {Verkerk, Annemarie},
	month = jan,
	year = {2014},
	pages = {40--83}
}

@article{bouckaert2012mapping,
	title={Mapping the origins and expansion of the {Indo-European} language family},
	author={Bouckaert, Remco and Lemey, Philippe and Dunn, Michael and Greenhill, Simon J. and Alekseyenko, Alexander V. and Drummond, Alexei J. and Gray, Russell D. and Suchard, Marc A. and Atkinson, Quentin D.},
	journal={Science},
	volume={337},
	number={6097},
	pages={957--960},
	year={2012},
	publisher={American Association for the Advancement of Science}
}

@article{cathcart_areal_2018,
	title = {Areal pressure in grammatical evolution},
	volume = {35},
	number = {1},
	journal = {Diachronica},
	author = {Cathcart, Chundra and Carling, Gerd and Larsson, Filip and Johansson, Niklas and Round, Erich R.},
	year = {2018},
	pages = {1--34}
}

@article{chang2015ancestry,
	title={Ancestry-constrained phylogenetic analysis supports the {Indo-European} steppe hypothesis},
	author={Chang, Will and Hall, David and Cathcart, Chundra and Garrett, Andrew},
	journal={Language},
	pages={194--244},
	year={2015},
	publisher={JSTOR}
}

@incollection{elworth2019advances,
	title={Advances in computational methods for phylogenetic networks in the presence of hybridization},
	author={Elworth, R. A. Leo and Ogilvie, Huw A. and Zhu, Jiafan and Nakhleh, Luay},
	booktitle={Bioinformatics and Phylogenetics},
	editor={Warnow, Tandy},
	pages={317--360},
	year={2019},
	publisher={Springer},
	location = {Cham, Switzerland}
}

@article{birchall_comparison_2015,
	title = {A comparison of verbal person marking across {Tupian} languages},
	volume = {10},
	issn = {1981-8122},
	doi = {10.1590/1981-81222015000200007},
	number = {2},
	journal = {Boletim do Museu Paraense Emílio Goeldi. Ciências Humanas},
	author = {Birchall, Joshua},
	month = aug,
	year = {2015},
	pages = {325--345}
}

@article{zhou_quantifying_2015,
	title = {Quantifying uncertainty in the phylogenetics of {Australian} numeral systems},
	volume = {282},
	issn = {0962-8452, 1471-2954},
	doi = {10.1098/rspb.2015.1278},
	number = {1815},
	journal = {Proceedings of the Royal Society B},
	author = {Zhou, Kevin and Bowern, Claire},
	month = sep,
	year = {2015},
	pmid = {26378214},
	pages = {20151278}
}

@article{calude_typology_2016,
	title = {The typology and diachrony of higher numerals in {Indo}-{European}: {A} phylogenetic comparative study},
	volume = {1},
	issn = {2058-4571},
	shorttitle = {The typology and diachrony of higher numerals in {Indo}-{European}},
	doi = {10.1093/jole/lzw003},
	number = {2},
	journal = {Journal of Language Evolution},
	author = {Calude, Andreea S. and Verkerk, Annemarie},
	month = jul,
	year = {2016},
	pages = {91--108}
}

@article{dunn_dative_2017,
	title = {Dative sickness: {A} phylogenetic analysis of argument structure evolution in {Germanic}},
	volume = {93},
	issn = {1535-0665},
	shorttitle = {Dative sickness},
	doi = {10.1353/lan.2017.0012},
	number = {1},
	journal = {Language},
	author = {Dunn, Michael and Dewey, Tonya Kim and Arnett, Carlee and Eythórsson, Thórhallur and Barðdal, Jóhanna},
	month = mar,
	year = {2017},
	pages = {e1--e22}
}

@article{kolipakam2018bayesian,
	title={A Bayesian phylogenetic study of the Dravidian language family},
	author={Kolipakam, Vishnupriya and Jordan, Fiona M. and Dunn, Michael and Greenhill, Simon J. and Bouckaert, Remco and Gray, Russell D. and Verkerk, Annemarie},
	journal={Royal Society open science},
	volume={5},
	number={3},
	pages={171504},
	year={2018},
	publisher={The Royal Society Publishing}
}

@inproceedings{verkerk_phylogenetic_2017,
	address = {Australian National University, Canberra, Australia},
	title = {Phylogenetic comparative methods for typologists ({Focusing} on families and regions: {A} plea for using phylogenetic comparative methods in linguistic typology)},
	booktitle = {Quantitative {Analysis} in {Typology}: {The} logic of choice among methods (workshop at the 12th {Conference} of the {Association} for {Linguistic} {Typology}},
	author = {Verkerk, Annemarie},
	month = dec,
	year = {2017}
}

@article{purvis_polytomies_1993,
	title = {Polytomies in comparative analyses of continuous characters},
	volume = {42},
	issn = {1063-5157},
	doi = {10.2307/2992489},
	number = {4},
	journal = {Systematic Biology},
	author = {Purvis, Andy and Garland, Theodore},
	year = {1993},
	pages = {569--575}
}

@article{purvis_truth_1994,
	title = {Truth or consequences: {Effects} of phylogenetic accuracy on two comparative methods},
	volume = {167},
	issn = {0022-5193},
	shorttitle = {Truth or {Consequences}},
	doi = {10.1006/jtbi.1994.1071},
	number = {3},
	journal = {Journal of Theoretical Biology},
	author = {Purvis, Andy and Gittleman, John L. and Luh, Hang-Kwang},
	month = apr,
	year = {1994},
	pages = {293--300}
}

@article{symonds_effects_2002,
	title = {The effects of topological inaccuracy in evolutionary trees on the phylogenetic comparative method of independent contrasts},
	volume = {51},
	issn = {1063-5157},
	doi = {10.1080/10635150290069977},
	number = {4},
	journal = {Systematic Biology},
	author = {Symonds, Matthew R. E. and Page, Rod},
	month = jul,
	year = {2002},
	pages = {541--553}
}

@article{freckleton_phylogenetic_2002,
	title = {Phylogenetic analysis and comparative data: {A} test and review of evidence.},
	volume = {160},
	issn = {0003-0147},
	shorttitle = {Phylogenetic {Analysis} and {Comparative} {Data}},
	doi = {10.1086/343873},
	number = {6},
	journal = {The American Naturalist},
	author = {Freckleton, Robert P. and Harvey, Paul H. and Pagel, Mark},
	month = dec,
	year = {2002},
	pages = {712--726}
}

@article{blomberg_testing_2003,
	title = {Testing for phylogenetic signal in comparative data: behavioral traits are more labile},
	volume = {57},
	shorttitle = {Testing for phylogenetic signal in comparative data},
	doi = {doi: 10.1111/j.0014-3820.2003.tb00285.x.},
	number = {4},
	journal = {Evolution},
	author = {Blomberg, Simon P. and Garland, Theodore and Ives, Anthony R.},
	year = {2003},
	pages = {717--745}
}

@article{blomberg_tempo_2002,
	title = {Tempo and mode in evolution: phylogenetic inertia, adaptation and comparative methods},
	volume = {15},
	issn = {1420-9101},
	shorttitle = {Tempo and mode in evolution},
	doi = {10.1046/j.1420-9101.2002.00472.x},
	number = {6},
	journal = {Journal of Evolutionary Biology},
	author = {Blomberg, Simon. P. and Garland, Theodore},
	year = {2002},
	pages = {899--910}
}

@article{revell_phylogenetic_2008,
	title = {Phylogenetic Signal, Evolutionary process, and rate},
	volume = {57},
	issn = {1063-5157},
	doi = {10.1080/10635150802302427},
	number = {4},
	journal = {Systematic Biology},
	author = {Revell, Liam J. and Harmon, Luke J. and Collar, David C. and Oakley, Todd},
	month = aug,
	year = {2008},
	pages = {591--601}
}

@article{irschick_comparison_1997,
	title = {A comparison of evolutionary radiations in mainland and {Caribbean} anolis lizards},
	volume = {78},
	copyright = {© 1997 by the Ecological Society of America},
	issn = {1939-9170},
	doi = {10.1890/0012-9658(1997)078[2191:ACOERI]2.0.CO;2},
	number = {7},
	journal = {Ecology},
	author = {Irschick, Duncan J. and Vitt, Laurie J. and Zani, Peter A. and Losos, Jonathan B.},
	year = {1997},
	pages = {2191--2203}
}

@article{balisi_dietary_2018,
	title = {Dietary specialization is linked to reduced species durations in {North} {American} fossil canids},
	volume = {5},
	copyright = {© 2018 The Authors.. Published by the Royal Society under the terms of the Creative Commons Attribution License http://creativecommons.org/licenses/by/4.0/, which permits unrestricted use, provided the original author and source are credited.},
	issn = {2054-5703},
	doi = {10.1098/rsos.171861},
	number = {4},
	journal = {Royal Society Open Science},
	author = {Balisi, Mairin and Casey, Corinna and Valkenburgh, Blaire van},
	month = apr,
	year = {2018},
	pages = {171861}
}

@article{hutchinson_contemporary_2018,
	title = {Contemporary ecological interactions improve models of past trait evolution},
	volume = {67},
	doi = {10.1093/sysbio/syy012},
	number = {5},
	journal = {Systematic Biology},
	author = {Hutchinson, Matthew C. and Gaiarsa, Marília P. and Stouffer, Daniel B.},
	year = {2018},
	pages = {1--13}
}

@article{felsenstein_phylogenies_1988,
	title = {Phylogenies and quantitative characters},
	volume = {19},
	issn = {0066-4162},
	doi = {10.1146/annurev.es.19.110188.002305},
	journal = {Annual Review of Ecology and Systematics},
	author = {Felsenstein, Joseph},
	year = {1988},
	pages = {445--471}
}

@article{garland_phylogenetic_1993,
	title = {Phylogenetic analysis of covariance by computer simulation},
	volume = {42},
	issn = {1063-5157},
	doi = {10.1093/sysbio/42.3.265},
	number = {3},
	journal = {Systematic Biology},
	author = {Garland, Theodore and Dickerman, Allan W. and Janis, Christine M. and Jones, Jason A.},
	month = sep,
	year = {1993},
	pages = {265--292}
}

@article{hansen_translating_1996,
	title = {Translating between microevolutionary process and macroevolutionary patterns: {The} correlation structure of interspecific data},
	volume = {50},
	copyright = {© 1996 The Society for the Study of Evolution},
	issn = {1558-5646},
	shorttitle = {Translating {Between} {Microevolutionary} {Process} and {Macroevolutionary} {Patterns}},
	doi = {10.1111/j.1558-5646.1996.tb03914.x},
	number = {4},
	journal = {Evolution},
	author = {Hansen, Thomas F. and Martins, Emília P.},
	year = {1996},
	pages = {1404--1417}
}

@article{lavin_morphometrics_2008,
	title = {Morphometrics of the avian small intestine compared with that of nonflying mammals: {A} phylogenetic approach},
	volume = {81},
	issn = {1522-2152},
	shorttitle = {Morphometrics of the {Avian} {Small} {Intestine} {Compared} with {That} of {Nonflying} {Mammals}},
	doi = {10.1086/590395},
	number = {5},
	journal = {Physiological and Biochemical Zoology},
	author = {Lavin, Shana R. and Karasov, William H. and Ives, Anthony R. and Middleton, Kevin M. and Garland, Theodore},
	month = sep,
	year = {2008},
	pages = {526--550}
}

@article{moran_notes_1950,
	title = {Notes on continuous stochastic phenomena},
	volume = {37},
	issn = {0006-3444},
	doi = {10.2307/2332142},
	number = {1/2},
	journal = {Biometrika},
	author = {Moran, P. A. P.},
	year = {1950},
	pages = {17--23}
}

@article{gittleman_adaptation:_1990,
	title = {Adaptation: {Statistics} and a null model for estimating phylogenetic effects},
	volume = {39},
	issn = {1063-5157},
	shorttitle = {Adaptation},
	doi = {10.2307/2992183},
	number = {3},
	journal = {Systematic Biology},
	author = {Gittleman, John L. and Kot, Mark},
	month = sep,
	year = {1990},
	pages = {227--241}
}

@article{abouheif_method_1999,
	title = {A method for testing the assumption of phylogenetic independence in comparative data},
	volume = {1},
	issn = {1522–0613},
	
	number = {8},
	journal = {Evolutionary Ecology Research},
	author = {Abouheif, Ehab},
	year = {1999},
	pages = {895--909}
}

@article{munkemuller_how_2012,
	title = {How to measure and test phylogenetic signal},
	volume = {3},
	copyright = {© 2012 The Authors. Methods in Ecology and Evolution © 2012 British Ecological Society},
	issn = {2041-210X},
	doi = {10.1111/j.2041-210X.2012.00196.x},
	number = {4},
	journal = {Methods in Ecology and Evolution},
	author = {Münkemüller, Tamara and Lavergne, Sébastien and Bzeznik, Bruno and Dray, Stéphane and Jombart, Thibaut and Schiffers, Katja and Thuiller, Wilfried},
	year = {2012},
	pages = {743--756}
}

@article{fritz_selectivity_2010,
	title = {Selectivity in mammalian extinction risk and threat types: {A} new measure of phylogenetic signal strength in binary traits},
	volume = {24},
	issn = {1523-1739},
	shorttitle = {Selectivity in {Mammalian} {Extinction} {Risk} and {Threat} {Types}},
	doi = {10.1111/j.1523-1739.2010.01455.x},
	number = {4},
	journal = {Conservation Biology},
	author = {Fritz, Susanne A. and Purvis, Andy},
	month = feb,
	year = {2010},
	pages = {1042--1051}
}

@article{zheng_new_2009,
	title = {New multivariate tests for phylogenetic signal and trait correlations applied to ecophysiological phenotypes of nine {Manglietia} species},
	volume = {23},
	copyright = {© 2009 The Authors. Journal compilation © 2009 British Ecological Society},
	issn = {1365-2435},
	doi = {10.1111/j.1365-2435.2009.01596.x},
	number = {6},
	journal = {Functional Ecology},
	author = {Zheng, Li and Ives, Anthony R. and Garland, Theodore and Larget, Bret R. and Yu, Yang and Cao, Kunfang},
	year = {2009},
	pages = {1059--1069}
}

@article{adams_generalized_2014,
	title = {A generalized {K} statistic for estimating phylogenetic signal from shape and other high-dimensional multivariate data},
	volume = {63},
	issn = {1063-5157},
	doi = {10.1093/sysbio/syu030},
	number = {5},
	journal = {Systematic Biology},
	author = {Adams, Dean C.},
	month = sep,
	year = {2014},
	pages = {685--697}
}

@article{bouckaert_origin_2018,
	title = {The origin and expansion of {Pama}–{Nyungan} languages across {Australia}},
	volume = {2},
	copyright = {2018 The Author(s)},
	issn = {2397-334X},
	doi = {10.1038/s41559-018-0489-3},
	number = {4},
	journal = {Nature Ecology \& Evolution},
	author = {Bouckaert, Remco R. and Bowern, Claire and Atkinson, Quentin D.},
	month = apr,
	year = {2018},
	pages = {741--749}
}

@article{everett_climate_2015,
	title = {Climate, vocal folds, and tonal languages: {Connecting} the physiological and geographic dots},
	volume = {112},
	issn = {0027-8424, 1091-6490},
	shorttitle = {Climate, vocal folds, and tonal languages},
	doi = {10.1073/pnas.1417413112},
	number = {5},
	journal = {Proceedings of the National Academy of Sciences},
	author = {Everett, Caleb and Blasi, Damián E. and Roberts, Seán G.},
	month = feb,
	year = {2015},
	pmid = {25605876},
	pages = {1322--1327}
}

@article{everett_languages_2017,
	title = {Languages in drier climates use fewer vowels},
	volume = {8},
	issn = {1664-1078},
	doi = {10.3389/fpsyg.2017.01285},
	
	journal = {Frontiers in Psychology},
	author = {Everett, Caleb},
	year = {2017}
}

@article{blasi_grammars_2017,
	title = {Grammars are robustly transmitted even during the emergence of creole languages},
	volume = {1},
	rights = {2017 The Author(s)},
	issn = {2397-3374},
	doi = {10.1038/s41562-017-0192-4},
	pages = {723--729},
	number = {10},
	journal = {Nature Human Behaviour},
	author = {Blasi, Damián E. and Michaelis, Susanne Maria and Haspelmath, Martin},
	year = {2017}
}

@article{pagel_inferring_1999,
	title = {Inferring the historical patterns of biological evolution},
	volume = {401},
	doi = {10.1038/44766},
	number = {6756},
	journal = {Nature},
	author = {Pagel, Mark},
	year = {1999},
	pages = {877--884}
}

@article{macklin-cordes_phylogenetic_2021,
	title = {Phylogenetic signal in phonotactics},
	doi = {10.1075/dia.20004.mac},
	journal = {Diachronica},
	author = {Macklin-Cordes, Jayden L. and Bowern, Claire and Round, Erich R.},
	year = {2021}
}

@article{macklin-cordes_re-evaluating_2020,
	title = {Re-evaluating phoneme frequencies},
	volume = {11},
	journal = {Frontiers in psychology},
	author = {Macklin-Cordes, Jayden L. and Round, Erich R.},
	year = {2020},
	pages = {3181}
}

@book{burgman_burduna_2007,
	address = {South Hedland, Western Australia},
	title = {Burduna dictionary: {English}-{Burduna} wordlist and thematic wordlist},
	isbn = {1-921312-05-X},
	shorttitle = {Burduna dictionary},
	publisher = {Wangka Maya Pilbara Aboriginal Language Centre},
	author = {Burgman, Albert},
	collaborator = {{Wangka Maya Pilbara Aboriginal Language Centre}},
	year = {2007}
}

@article{hyman_role_1970,
	title = {The role of borrowing in the justification of phonological grammars},
	volume = {1},
	issn = {00390-3533},
	url = {https://journals.linguisticsociety.org/elanguage/sal/article/view/927.html},
	pages = {1--48},
	number = {1},
	journaltitle = {Studies in African Linguistics},
	author = {Hyman, Larry M.},
	date = {1970}
}

@incollection{dresher_main_2005,
	location = {Amsterdam},
	title = {Main stress left in {Early Middle English}},
	isbn = {978-90-272-9477-7},
	pages = {76--85},
	booktitle = {Historical Linguistics 2003: {S}elected Papers from the 16th International Conference on Historical Linguistics},
	publisher = {John Benjamins},
	author = {Dresher, B. Elan and Lahiri, Aditi},
	editor = {Fortescue, Michael and Mogensen, Jens Erik and Schøsler, Lene},
	date = {2005-01},
	doi = {10.1075/cilt.257}
}

@article{dixon_proto-australian_1970,
	title = {Proto-{Australian} laminals},
	volume = {9},
	number = {2},
	journal = {Oceanic Linguistics},
	author = {Dixon, R. M. W.},
	year = {1970},
	pages = {79--103}
}

@book{dixon_languages_1980,
	address = {Cambridge},
	title = {The languages of {Australia}},
	publisher = {Cambridge University Press},
	author = {Dixon, R. M. W.},
	year = {1980}
}

@incollection{evans_current_1995,
	address = {Cambridge, MA},
	title = {Current issues in the phonology of {Australian} languages},
	booktitle = {The handbook of phonological theory},
	publisher = {Blackwell},
	author = {Evans, Nicholas},
	editor = {Goldsmith, John A.},
	year = {1995},
	pages = {723--761}
}

@inproceedings{round_ausphon-lexicon_2017,
	location = {Poznań, Poland},
	title = {The {AusPhon}-Lexicon project: 2 million normalized segments across 300 {Australian} languages},
	eventtitle = {47th Poznań Linguistic Meeting},
	booktitle = {47th Poznań Linguistic Meeting},
	author = {Round, Erich R.},
	date = {2017-09}
}

@article{bowern_chirila_2016,
	title = {Chirila: {Contemporary} and historical resources for the {Indigenous} languages of {Australia}},
	volume = {10},
	rights = {Creative Commons Attribution-{NonCommercial} 4.0 International License},
	issn = {1934-5275},
	url = {http://hdl.handle.net/10125/24685},
	shorttitle = {Chirila},
	journaltitle = {Language Documentation and Conservation},
	author = {Bowern, Claire},
	date = {2016-03}
}

@incollection{round_segment_2022,
	address = {Oxford},
	title = {Segment inventories},
	booktitle = {Oxford {Guide} to {Australian} languages},
	publisher = {Oxford University Press},
	author = {Round, Erich R.},
	editor = {Bowern, Claire},
	year = {2022}
}

@article{frisch_similarity_2004,
	title = {Similarity avoidance and the {OCP}},
	volume = {22},
	doi = {10.1023/B:NALA.0000005557.78535.3c},
	number = {1},
	journal = {Natural Language \& Linguistic Theory},
	author = {Frisch, Stefan A. and Pierrehumbert, Janet B. and Broe, Michael B.},
	year = {2004},
	pages = {179--228}
}

@phdthesis{hall_probabilistic_2009,
	type = {Ph.{D}. {Dissertation}},
	title = {A probabilistic model of phonological relationships from contrast to allophony},
	school = {The Ohio State University},
	author = {Hall, Kathleen Currie},
	year = {2009}
}

@article{wedel_high_2013,
	title = {High functional load inhibits phonological contrast loss: {A} corpus study},
	volume = {128},
	number = {2},
	journal = {Cognition},
	author = {Wedel, Andrew and Kaplan, Abby and Jackson, Scott},
	year = {2013},
	pages = {179--186}
}

@incollection{round_phonemic_2019,
	address = {Jena},
	title = {Phonemic inventories of {Australia} [{Database} of 392 languages]},
	booktitle = {{PHOIBLE} 2.0},
	publisher = {Max Planck Institute for the Science of Human History},
	author = {Round, Erich R.},
	editor = {Moran, Steven and McCloy, Daniel},
	year = {2019}
}

@article{altschul_weights_1989,
	title = {Weights for data related by a tree},
	volume = {207},
	number = {4},
	journal = {Journal of Molecular Biology},
	author = {Altschul, Stephen F. and Carroll, Raymond J. and Lipman, David J.},
	year = {1989},
	pages = {647--653}
}

@article{stone_constructing_2007,
	title = {Constructing a meaningful evolutionary average at the phylogenetic center of mass},
	volume = {8},
	number = {1},
	journal = {BMC bioinformatics},
	author = {Stone, Eric A. and Sidow, Arend},
	year = {2007},
	pages = {222}
}

@article{vingron_weighting_1993,
	title = {Weighting in sequence space: {A} comparison of methods in terms of generalized sequences},
	volume = {90},
	number = {19},
	journal = {Proceedings of the National Academy of Sciences},
	author = {Vingron, Martin and Sibbald, Peter R.},
	year = {1993},
	pages = {8777--8781}
}

@article{de_maio_phylogenetic_2020,
	title = {Phylogenetic Novelty Scores: {A} New Approach for Weighting Genetic Sequences},
	doi = {10.1101/2020.12.03.410100},
	journal = {bioRxiv},
	author = {De Maio, Nicola and Alekseyenko, Alexander V. and Coleman-Smith, William J. and Pardi, Fabio and Suchard, Marc A. and Tamuri, Asif U. and Truszkowski, Jakub and Goldman, Nick},
	year = {2020}
}

@unpublished{blake_pitta_1990,
	location = {Canberra},
	title = {{Pitta Pitta} wordlist},
	rights = {Access - Open access for reading and copying in accordance with copyright. Contact {AIATSIS} Library},
	type = {word list},
	author = {Blake, Barry J.},
	date = {1990},
	howpublished = {Australian Institute of Aboriginal and Torres Strait Islander Studies, Australian Indigenous Languages Collection},
	note = {ASEDA 0275}
}

@article{bentz_evolution_2018,
	title = {The evolution of language families is shaped by the environment beyond neutral drift},
	volume = {2},
	rights = {2018 The Author(s), under exclusive licence to Springer Nature Limited},
	issn = {2397-3374},
	doi = {10.1038/s41562-018-0457-6},
	pages = {816--821},
	number = {11},
	journaltitle = {Nature Human Behaviour},
	author = {Bentz, Christian and Dediu, Dan and Verkerk, Annemarie and Jäger, Gerhard},
	date = {2018-11}
}

@inproceedings{bowern_pama-nyungan_2015,
	location = {Leiden University, Leiden, Netherlands},
	title = {{Pama-Nyungan} phylogenetics and beyond [plenary address]},
	doi = {10.5281/zenodo.3032846},
	eventtitle = {Lorentz Center workshop on phylogenetic methods in linguistics},
	booktitle = {Lorentz Center workshop on phylogenetic methods in linguistics},
	author = {Bowern, Claire},
	date = {2015}
}

@article{jager2021phylogenetic,
	title={Phylogenetic typology},
	author={J{\"a}ger, Gerhard and Wahle, Johannes},
	journal={arXiv preprint arXiv:2103.10198},
	year={2021}
}

@article{cathcart2020numeral,
	title={Numeral classifiers and number marking in {Indo-Iranian}: {A} phylogenetic approach},
	author={Cathcart, Chundra and H{\"o}lzl, Andreas and J{\"a}ger, Gerhard and Widmer, Paul and Bickel, Balthasar},
	journal={Language Dynamics and Change},
	volume={1},
	pages={1--53},
	year={2020},
	publisher={Brill}
}

@Manual{Rcore,
	title = {R: A Language and Environment for Statistical Computing},
	author = {{R Core Team}},
	organization = {R Foundation for Statistical Computing},
	address = {Vienna, Austria},
	year = {2021},
	url = {https://www.R-project.org/}
}

@Article{phytools,
	title = {phytools: {An} {R} package for phylogenetic comparative biology (and other things)},
	author = {Liam J. Revell},
	journal = {Methods in Ecology and Evolution},
	year = {2012},
	volume = {3},
	pages = {217-223}
}

@misc{blench_new_2015,
	address = {Nijmegen, Netherlands},
	title = {'{New} mathematical methods' in linguistics constitute the greatest intellectual fraud in the discipline since {Chomsky}},
	author = {Blench, Roger},
	year = {2015},
}

@incollection{francois_trees_2014,
	address = {Abingdon, UK},
	title = {Trees, {Waves}, and {Linkages}: {Models} of {Language} {Diversification}},
	booktitle = {The {Routledge} {Handbook} of {Historical} {Linguistics}},
	publisher = {Routledge},
	author = {François, Alexandre},
	editor = {Bowern, Claire and Evans, Bethwyn},
	year = {2014},
	pages = {161--189},
}

@article{greenhill_does_2009,
	title = {Does horizontal transmission invalidate cultural phylogenies?},
	volume = {276},
	copyright = {© 2009 The Royal Society},
	issn = {0962-8452, 1471-2954},
	doi = {10.1098/rspb.2008.1944},
	language = {en},
	number = {1665},
	journal = {Proceedings of the Royal Society of London B: Biological Sciences},
	author = {Greenhill, Simon J. and Currie, Thomas E. and Gray, Russell D.},
	month = jun,
	year = {2009},
	pmid = {19324763},
	pages = {2299--2306},
}

@article{bowern_does_2011,
	title = {Does lateral transmission obscure inheritance in hunter-gatherer languages?},
	volume = {6},
	issn = {1932-6203},
	doi = {10.1371/journal.pone.0025195},
	number = {9},
	journal = {PLoS ONE},
	author = {Bowern, Claire and Epps, Patience and Gray, Russell and Hill, Jane and Hunley, Keith and McConvell, Patrick and Zentz, Jason},
	month = sep,
	year = {2011},
	pages = {e25195},
}

@article{verkerk_detecting_2019,
	title = {Detecting non-tree-like signal using multiple tree topologies},
	volume = {9},
	doi = {10.1075/jhl.17009.ver},
	number = {1},
	urldate = {2021-01-01},
	journal = {Journal of Historical Linguistics},
	author = {Verkerk, Annemarie},
	month = jul,
	year = {2019},
	pages = {9--69},
}

@article{levinson_universal_2011,
	title = {Universal typological dependencies should be detectable in the history of language families},
	volume = {15},
	doi = {10.1515/lity.2011.034},
	number = {2},
	author = {Levinson, Stephen C. and Greenhill, Simon J. and Gray, Russell D. and Dunn, Michael},
	year = {2011},
	pages = {509--534},
}

@incollection{cysouw_quantitative_2005,
	address = {Berlin},
	title = {Quantitative methods in typology = {Quantitative} {Methoden} in {Der} {Typologie}},
	booktitle = {Quantitative {Linguistik}: {Ein} {Internationales} {Handbuch}},
	publisher = {Walter de Gruyter},
	author = {Cysouw, Michael},
	editor = {Köhler, Reinhard and Altman, Gabriel and Piotrowski, Rajmund G.},
	year = {2005},
	pages = {554--578},
}

@article{jaeger_mixed_2011,
	title = {Mixed effect models for genetic and areal dependencies in linguistic typology},
	volume = {15},
	doi = {10.1515/lity.2011.021},
	number = {2},
	author = {Jaeger, T. Florian and Graff, Peter and Croft, William and Pontillo, Daniel},
	year = {2011},
	pages = {281--319},
}

@article{piantadosi_quantitative_2014,
	title = {Quantitative Standards for Absolute Linguistic Universals},
	volume = {38},
	doi = {10.1111/cogs.12088},
	number = {4},
	journal = {Cognitive Science},
	author = {Piantadosi, Steven T. and Gibson, Edward},
	year = {2014},
	pages = {736--756},
}

@article{maslova_stochastic_2000,
	title = {Stochastic models in typology: {Obstacle} or prerequisite?},
	volume = {4},
	number = {3},
	journal = {Linguistic Typology},
	author = {Maslova, Elena},
	year = {2000},
	pages = {357--364},
}

@article{maslova_dynamic_2000,
	title = {A dynamic approach to the verification of distributional universals},
	volume = {4},
	issn = {1613-415X},
	url = {http://www.degruyter.com/document/doi/10.1515/lity.2000.4.3.307/html},
	doi = {10.1515/lity.2000.4.3.307},
	language = {en},
	number = {3},
	urldate = {2021-09-09},
	author = {Maslova, Elena},
	month = jan,
	year = {2000},
	note = {Publisher: De Gruyter Mouton
Section: Linguistic Typology},
	pages = {307--333},
}

@article{everett_sound_2021,
	title = {The Sound Systems of Languages Adapt, But to What Extent?},
	volume = {2},
	doi = {10.25189/2675-4916.2021.v2.n1.id342},
	number = {1},
	journal = {Cadernos de Linguística},
	author = {Everett, Caleb},
	month = apr,
	year = {2021},
	pages = {01--23},
}

@article{mace_comparative_1994,
	title = {The Comparative Method in Anthropology [and Comments and Reply]},
	volume = {35},
	doi = {10.1086/204317},
	number = {5},
	journal = {Current Anthropology},
	author = {Mace, Ruth and Pagel, Mark and Bowen, John R. and Otterbein, Keith F. and Ridley, Mark and Schweizer, Thomas and Voland, Eckart},
	year = {1994},
	pages = {549--564},
}

@article{holden_phylogenetic_2009,
	title = {Phylogenetic Analysis of the Evolution of Lactose Digestion in Adults},
	volume = {81},
	doi = {10.3378/027.081.0609},
	number = {5/6},
	journal = {Human Biology},
	author = {Holden, Clare J. and Mace, Ruth},
	year = {2009},
	pages = {597--619},
}

@incollection{symonds_primer_2014,
	address = {Berlin},
	title = {A Primer on Phylogenetic Generalised Least Squares},
	booktitle = {Modern Phylogenetic Comparative Methods and Their Application in Evolutionary Biology: {Concepts} and Practice},
	publisher = {Springer},
	author = {Symonds, Matthew R. E. and Blomberg, Simon P.},
	editor = {Garamszegi, László Zsolt},
	year = {2014},
	doi = {10.1007/978-3-662-43550-2_5},
	pages = {105--130},
}

@article{jordan_matrilocal_2009,
	title = {Matrilocal residence is ancestral in {Austronesian} societies},
	volume = {276},
	doi = {10.1098/rspb.2009.0088},
	number = {1664},
	journal = {Proceedings of the Royal Society B: Biological Sciences},
	author = {Jordan, Fiona M. and Gray, Russell D. and Greenhill, Simon J. and Mace, Ruth},
	year = {2009},
	pages = {1957--1964},
}

@article{holden_spread_2003,
	title = {Spread of cattle led to the loss of matrilineal descent in {Africa}: {A} coevolutionary analysis},
	volume = {270},
	doi = {10.1098/rspb.2003.2535},
	number = {1532},
	journal = {Proceedings of the Royal Society of London. Series B: Biological Sciences},
	author = {Holden, Clare J. and Mace, Ruth},
	year = {2003},
	pages = {2425--2433},
}

@article{revell_phylogenetic_2010,
	title = {Phylogenetic signal and linear regression on species data},
	volume = {1},
	doi = {10.1111/j.2041-210X.2010.00044.x},
	number = {4},
	journal = {Methods in Ecology and Evolution},
	author = {Revell, Liam J.},
	year = {2010},
	pages = {319--329},
}

@article{borges_measuring_2019,
	title = {Measuring phylogenetic signal between categorical traits and phylogenies},
	volume = {35},
	doi = {10.1093/bioinformatics/bty800},
	number = {11},
	journal = {Bioinformatics},
	author = {Borges, Rui and Machado, João Paulo and Gomes, Cidália and Rocha, Ana Paula and Antunes, Agostinho},
	year = {2019},
	pages = {1862--1869},
}

@article{jager_global-scale_2018,
	title = {Global-scale phylogenetic linguistic inference from lexical resources},
	volume = {5},
	doi = {10.1038/sdata.2018.189},
	number = {1},
	journal = {Scientific Data},
	author = {Jäger, Gerhard},
	month = oct,
	year = {2018},
	pages = {180189},
}

@Manual{phyloWeights,
    title = {{phyloWeights}: Calculation of Genealogically-sensitive Proportions and Averages},
    author = {Erich R. Round},
    year = {2021},
    note = {R package version 0.3},
    url = {https://github.com/erichround/phyloWeights},
  }

@Manual{glottoTrees,
    title = {{glottoTrees}: Phylogenetic trees in Linguistics.},
    author = {Erich R. Round},
    year = {2021},
    note = {R package version 0.1},
    url = {https://github.com/erichround/glottoTrees},
  }

@article{keeling2008horizontal,
  title={Horizontal gene transfer in eukaryotic evolution},
  author={Keeling, Patrick J and Palmer, Jeffrey D},
  journal={Nature Reviews Genetics},
  volume={9},
  number={8},
  pages={605--618},
  year={2008},
  publisher={Nature Publishing Group}
}

@article{matsumae2021,
	author = {Hiromi Matsumae  and Peter Ranacher  and Patrick E. Savage  and Damián E. Blasi  and Thomas E. Currie  and Kae Koganebuchi  and Nao Nishida  and Takehiro Sato  and Hideyuki Tanabe  and Atsushi Tajima  and Steven Brown  and Mark Stoneking  and Kentaro K. Shimizu  and Hiroki Oota  and Balthasar Bickel },
	title = {Exploring correlations in genetic and cultural variation across language families in northeast Asia},
	journal = {Science Advances},
	volume = {7},
	number = {34},
	pages = {eabd9223},
	year = {2021},
	doi = {10.1126/sciadv.abd9223}
}

@article{rohlf_comment_2006,
	title = {A Comment on Phylogenetic Correction},
	volume = {60},
	doi = {10.1111/j.0014-3820.2006.tb01229.x},
	number = {7},
	journal = {Evolution},
	author = {Rohlf, F. James},
	year = {2006},
	pages = {1509--1515},
}

@article{dediu_making_2018,
	title = {Making genealogical language classifications available for phylogenetic analysis: Newick trees, unified identifiers, and branch length},
	volume = {8},
	doi = {10.1163/22105832-00801001},
	pages = {1--21},
	number = {1},
	journaltitle = {Language Dynamics and Change},
	author = {Dediu, Dan},
	date = {2018-06-22},
}

\newpage
\setlength{\voffset}{0cm}
\setlength{\hoffset}{0cm}


\section*{Supplementary material}

# Supplementary information, Section S1

Computational methods for calculating genealogically-sensitive averages and proportions

# Contents

# 1 Introduction

This document provides a guide to calculating genealogically-sensitive proportions and averages. Because a key part of the analysis is the preparation of a phylogenetic tree, considerable space is given over to how such trees can be prepared. Much of the discussion here is an introduction to the functionality of two R packages, *glottoTrees* (Round 2021a) and *phyloWeights* (Round 2021b), which have been specifically written for these tasks.

Since typologists may have little familiarity with R, we begin with a short introduction to it in Section 2. We then discuss the calculation of genealogically-sensitive proportions and averages in Section 3. Section 4 explains how typologists can prepare phylogenetic trees by adapting resources freely available from glottolog.com (Hammarström et al. 2021). Section 5 provides a worked example used in a typological investigation of sonority sequencing by Yin (2020).

# 2 A short orientation to R

The code presented here runs in R. Here we quickly introduce R objects (such as variables and dataframes) in Section 2.1, R packages in Section 2.2 and how R works with tree objects in Section 2.3. Readers already familiar with the basics of R may wish to skip directly to Section 2.3.

Below, chunks of R code appear against a grey background. The results that R produces from each chunk is shown below it, either as lines of text preceded by `##` or as a plot that R generates, or both. The code within the grey boxes can be copied and pasted into R. Within the code chunks, where a single R command runs longer than one printed line, we have indented the lines after the first; to run these in R, you will need to copy and paste the whole command.

## 2.1 R objects

In R, values can be assigned to objects (such as variables) using the assignment operator, `<-`:

```r
x <- 7.5
y <- 2 / 9
z <- sqrt(2)
a <- "hello"
```

The content of a simple object can be displayed just by entering its name. The `[1]` at the start of the output is telling you this is item number 1 in the object:

```r
z
```

```
## [1] 1.414214
```

A common object in R is a one-dimensional vector, such as a sequence of numbers or characters strings. Here is a vector of numbers. The `c()` here is a function, while the numbers 3, 4, 5 and 6 are its arguments. The function takes these four individual arguments and returns a single vector.

```r
c(3,5,4,6)
```

```
## [1] 3 5 4 6
```

A dataframe is the object in R that most resembles an Excel spreadsheet. It has columns that are named and rows that may or may not be named. Here we see the creation of a dataframe using the function `data.frame()`. In this instance, the first two arguments of the function are a vector of strings, `my_letters`, and a vector of numbers `my_numbers`.[1] These vectors need to be of equal length, as they become the columns of the resulting dataframe. Here we have assigned the result to the object named `my_dataframe`:

```r
my_dataframe <- data.frame(my_letters = c("x", "y", "z"),
                           my_numbers = c(7,1,43),
                           stringsAsFactors = FALSE)
```

The contents of `my_dataframe` are displayed like this:

```r
my_dataframe
```

```
##   my_letters my_numbers
## 1          x          7
## 2          y          1
## 3          z         43
```

In a dataframe, the contents of any one column will all be of the same class, e.g. all character strings, or all numbers, but as in `my_dataframe`, different columns can contain items of different classes.

## 2.2 R packages

R provides a range of basic statistical functions, but it is most powerful when extended by the addition of *packages* which contain additional functions. Here we will use the packages *ape* "Analyses of Phylogenetics

---

[1] The third argument `stringsAsFactors = FALSE` is used to ensure R reads the first argument as character strings, and not as another kind object (which won't be of concern to us here), called a `factor`. If you are using R version 4.0 or later, the argument `stringsAsFactors = FALSE` isn't strictly necessary, since R will assume it by default. In earlier versions of R, it is necessary, since the default assumption was `stringsAsFactors = TRUE`.

and Evolution" (Paradis & Schliep 2018) to work with trees, *dplyr* (Wickham et al. 2021) to manipulate dataframes, *glottoTrees* (Round 2021a) to prepare linguistic phylogenies and *phyloWeights* (Round 2021b) to perform the analysis of genealogically-sensitive averages and proportions.

Packages need to be *installed*, i.e., downloaded and unpacked, just once. Later, they are *loaded* during each work session as needed.

To install the packages we will use, run these commands. You will only ever need to do this once. Since installation involves downloading, you will need an active internet connection.

```
install.packages("dplyr")
install.packages("ape")
install.packages("devtools")
library(devtools)
install_github("erichround/phyloWeights")
install_github("erichround/glottoTrees")
```

To load the installed packages, ready for use, run these commands:

```
library(dplyr)
library(ape)
library(phyloWeights)
library(glottoTrees)
```

## 2.3   Trees in R

Here we discuss how trees are created, manipulated and plotted in R. Terminology we will use includes: *tips* at the ends of trees, which in a linguistic tree would usually be the languages or lects; the *branches* of a tree; the *interior nodes* or just *nodes* of a tree, where branches join together; and the *root* of the tree, its deepest node. R will represent trees as complex objects, in which the tips, nodes and branches all appear, along with labels for the tips and nodes.

One of the simplest methods of constructing a tree in R begins with a description of the tree using a form a bracketing notation known as the *Newick* standard (Felsenstein n.d.). In its simplest form, a tree is represented in Newick format by a set of tip labels grouped by parentheses, separated by commas, and ending with a semicolon. For example, here is a string that represents a tree with four tips, A, B, C and D, which we assign to the object `my_newick`:

```
my_newick <- "(((A,B),C),D);"
```

A Newick-formatted string can then be converted to a tree object by supplying it as the `text` argument of the function `read.tree()`, from the *ape* package:

```
my_tree <- read.tree(text = my_newick)
```

We can plot the tree using the `plot()` function:

```
plot(my_tree)
```

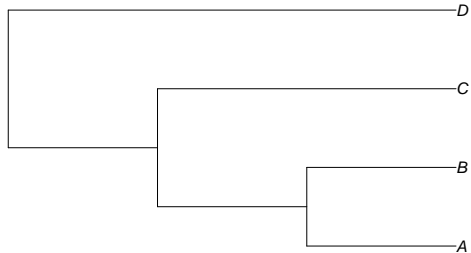

By default, trees in R are plotted horizontally following the convention in biology. The *glottoTrees* package provides a function `plot_glotto()` which plots trees in a more typical, downward-running linguistic format, as below.

```
plot_glotto(my_tree)
```

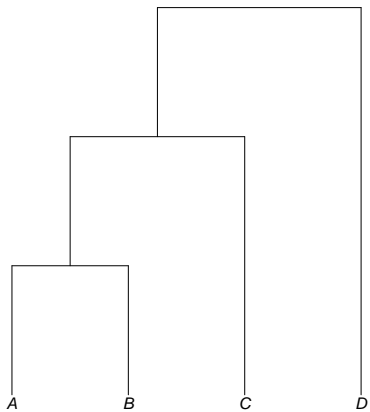

The tree object `my_tree` which we defined above did not include information about branch lengths. In Newick format, branch lengths are written with a preceding colon and appear directly after a language or the closing bracket for a subgroup:

```
my_newick2 <- "(((A:4,B:4):1,C:5):3,D:8);"
```

Converting this to a tree object and plotting it:

```
my_tree2 <- read.tree(text = my_newick2)
plot_glotto(my_tree2)
```

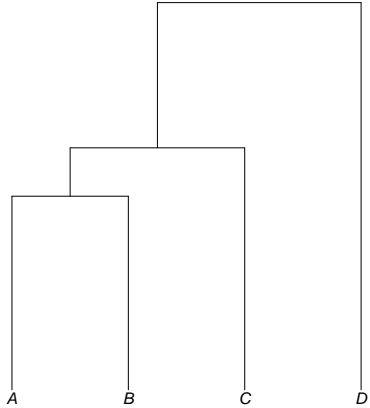

Trees can have labels not only for their tips, but also for their internal nodes. In linguistics, an internal node of a tree may be interpreted taxonomically, as representing a subgroup and labeled accordingly, or genealogically, as a proto-language from which the subgroup descends and labeled accordingly. In Newick format, labels for internal nodes are placed directly after a closing parenthesis. For example, here we add labels that reflect a genealogical interpretation of the nodes:

```
my_newick3 <- "(((A:4,B:4)proto-AB:1,C:5)proto-ABC:3,D:8)proto-ABCD;"
my_tree3 <- read.tree(text = my_newick3)
plot_glotto(my_tree3)
```

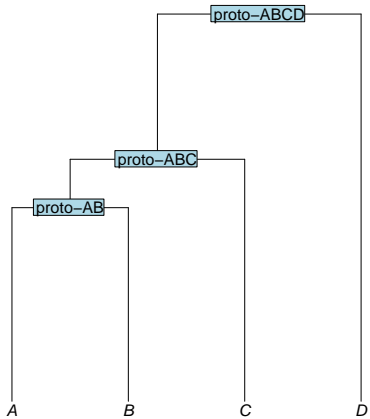

Technically speaking, trees are represented by R as objects with a customised class called `phylo`.

```
class(my_tree3)
```

## [1] "phylo"

A `phylo` object stores information about the tree *topology* (i.e., its branching structure), the branch lengths, and the labels of the tips and nodes. In R we often use the `$` operator to access one object that is contained inside another. For instance, the object `y` contained within the larger object `x` would be referred to as `x$y`.

Here are some examples:

```
my_tree3$edge.length
```

```
## [1] 3 1 4 4 5 8
```

```
my_tree3$tip.label
```

```
## [1] "A" "B" "C" "D"
```

```
my_tree3$node.label
```

```
## [1] "proto-ABCD" "proto-ABC"  "proto-AB"
```

Another object class in R, related to the `phylo` class, is the `multiPhylo` class. Objects of the `multiPhylo` class are used to store multiple `phylo` trees in a single, larger object.

```
newick_a <- "(((A:4,B:4):1,C:5):3,D:8);"
newick_b <- "((A:2,B:2):1,(C:1,D:1,E:1):2);"
tree_a <- read.tree(text = newick_a)
tree_b <- read.tree(text = newick_b)
my_multiPhylo <- c(tree_a, tree_b)
class(my_multiPhylo)
```

```
## [1] "multiPhylo"
```

For reasons we won't go into here, the `phylo` trees inside a `multiPhylo` object are not accessed using the `$` operator but using double square brackets. For example, here we refer to the second tree inside the object `my_multiPhylo` by writing `my_multiPhylo[[2]]`:

```
plot_glotto(my_multiPhylo[[2]])
```

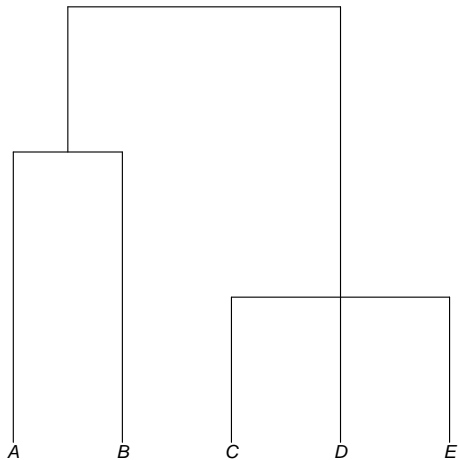

Before concluding this section, a final word is in order about the irrelevance of the left-to-right arrangement of trees. In a tree, there is no meaningful difference between (A,B) and (B,A): in both, A and B are sisters under a shared parent node. Similarly, these are all equivalent: (A,B,C), (A,C,B), (B,A,C), (B,C,A), (C,A,B) and (C,B,A). And likewise, these are all equivalent: (A,(B,C)), (A,(C,B)), ((B,C),A) and ((C,B),A).

# 3   Genealogically-sensitive averages and proportions

We now turn to the calculation of genealogically-sensitive proportions and averages. To run the code in this and subsequent sections, ensure that the packages described in Section 2.2 have been installed and loaded.

Calculating averages and proportions will require two key components:

1. A `phylo` or `multiPhylo` object containing one or more trees.
2. A dataframe, which (i) contains the typological data to be averaged and (ii) relates that data to the tips of the trees.

The `phylo` or `multiPhylo` object can be manually defined, as described in Section 2.3, can be read from a file,[2] or it can be constructed by using and adjusting materials freely available from glottolog.com, as described below in Section 4.

The dataframe can be manually defined or be read from a file. Perhaps the easiest method is to read from a file that you have created and saved in "CSV" (comma separated value) format. CSV files can be created in commercial spreadsheet software like Excel, and then read by R using the `read.csv()` function like this:

```
my_dataframe <- read.csv("my_csv_file.csv")
```

The dataframe must contain one column named `tip` (note that in R, names of columns and other objects are case sensitive) plus at least one column containing numerical data. The contents of the `tip` column must be the same as the tip labels of the tree(s) in the `phylo` or `multiPhylo` object. The contents of the numerical columns will depend on whether a proportion or an average is desired. To calculate a proportion, fill a numerical column with 1 if the language possesses the property and 0 if it does not. To calculate an average, fill a numerical column with the values of the variable for each language.

As an example, in order to apply these analyses to the four languages in Figure 2a,b,c of the main paper, here are the trees that would be needed, which are placed inside a single `multiPhylo` object:

```
newick_Fig2a <- "(((A:1,B:1,C:1):1,D:2):0.3);"
newick_Fig2b <- "(((A:0.2,B:0.2,C:0.2):1.8,D:2):0.3);"
newick_Fig2c <- "(((A:1.8,B:1.8,C:1.8):0.2,D:2):0.3);"
tree_Fig2a <- read.tree(text = newick_Fig2a)
tree_Fig2b <- read.tree(text = newick_Fig2b)
tree_Fig2c <- read.tree(text = newick_Fig2c)
multiPhylo_Fig2 <- c(tree_Fig2a, tree_Fig2b, tree_Fig2c)
```

The dataframe required is shown below. In addition to the `tip` column, it contains two numerical columns, `is_SOV` and `n_consonants`. As good housekeeping, we recommend using column names of the form `is_X` or `has_X` for columns that contain data for proportions, and names of the form `n_X` for columns that contain counts to be averaged.

```
data_Fig2 <- data.frame(tip = c("A", "B", "C", "D"),
                        is_SOV = c(1, 1, 1, 0),
                        n_consonants = c(18, 20, 22, 40),
                        stringsAsFactors = FALSE)
```

Genealogically-sensitive averages and proportions are obtained using the *phyloWeights* function `phylo_average()`, specifying its arguments `phy` and `data` as in the example below. Here we have assigned the output of this function to a new object `results_Fig2`. We recommend always assigning the output of `phylo_average()` to an object. We will see below how to extract from it the various parts of the results.

```
results_Fig2 <- phylo_average(phy = multiPhylo_Fig2, data = data_Fig2)
```

The function `phylo_average()` may take up to several minutes to run if the tree is large, or many trees are provided. It will return error messages if the inputs provided to it are not what is required.

---

[2] See online documentation of the *ape* package for reading trees from various common file formats.

The results object will contain several parts, which can be accessed using the `$` operator. In `$phy` will be the tree(s) that were supplied and in `$data` will be the dataframe that was supplied, e.g.:

```
results_Fig2$data
```

```
##    tip is_SOV n_consonants
## 1   A      1           18
## 2   B      1           20
## 3   C      1           22
## 4   D      0           40
```

In `$ACL_weights` is a dataframe containing one column for each tree provided, in which appear the phylogenetic weights obtained using the ACL method. The dataframe also contains all of the non-numeric columns of the input `data` dataframe. In `$BM_weights` is a similar dataframe, with the phylogenetic weights obtained using the BM method:

```
results_Fig2$ACL_weights
```

```
##   tip tree1     tree2     tree3
## 1   A   0.2 0.1724138 0.2380952
## 2   B   0.2 0.1724138 0.2380952
## 3   C   0.2 0.1724138 0.2380952
## 4   D   0.4 0.4827586 0.2857143
```

```
results_Fig2$BM_weights
```

```
##   tip     tree1     tree2     tree3
## 1   A 0.2317460 0.1856200 0.2489451
## 2   B 0.2317460 0.1856200 0.2489451
## 3   C 0.2317460 0.1856200 0.2489451
## 4   D 0.3047619 0.4431401 0.2531646
```

In `$ACL_averages` is a dataframe with one row per tree and one column for each numerical column in the `data` dataframe. These are filled with the genealogically-sensitive averages or proportions obtained using the ACL method. In `$BM_averages` appear the genealogically-sensitive averages or proportions obtained using the BM method:

```
results_Fig2$ACL_averages
```

```
##    tree    is_SOV n_consonants
## 1 tree1 0.6000000     28.00000
## 2 tree2 0.5172414     29.65517
## 3 tree3 0.7142857     25.71429
```

```
results_Fig2$BM_averages
```

```
##    tree    is_SOV n_consonants
## 1 tree1 0.6952381     26.09524
## 2 tree2 0.5568599     28.86280
## 3 tree3 0.7468354     25.06329
```

It is possible to save any of these dataframes to a file using the `write.csv()` function, for example:

```
write.csv(results_Fig2$ACL_averages, file = "my_ACL_averages.csv")
```

# 4 Using and adapting trees from glottolog.com

Glottolog.com (Hammarström et al. 2021) contains many useful resources for quantitative typology and the package *glottoTrees* (Round 2021a) has been written to help linguists make the most of these resources,

including by modifying them as they desire. This section covers the glottolog data itself and the functionality of *glottoTrees*. We introduce glottolog's genealogical data in Section 4.1, discussing how to locate metadata about languages and families of interest, and how to view glottolog's linguistic family trees. Since genealogically-sensitive averages and proportions require languages to be represented within a single tree, we then discuss how glottolog's individual trees can be combined in Section 4.2. Since typological studies will often examine language varieties at a level of granularity that differs from glottolog's own, in Section 4.3 we discuss how to add and remove languages from trees. In Section 4.4 we discuss how to add branch lengths to trees, since branch lengths are necessary for the calculation of genealogically-sensitive averages and proportions. Section 4.5 discusses how to export trees for use with other software.

## 4.1 Glottolog's genealogical data

Glottolog provides metadata about the world's language varieties, their division into language families and the hierarchical subgrouping of languages inside those families. Naturally, there are many points of contention in linguistics about what the world's stock of languages and dialects actually is, how it groups into families, and how the families themselves are subgrouped. Glottolog provides one set of answers, and structures them in a way which provides typologists with a basis for carrying out changes to suit their own hypotheses. In later sections we will see how this can be done. In this section we describe glottolog's own global linguistic metadata.

At time of writing, the current version of glottolog is v4.4. The *glottoTrees* package contains a copy of the v4.4 metadata covering language names, language identification codes, family names, geographical groupings, and family trees. The original metadata files that contain this information are currently available at https://glottolog.org/meta/downloads, where a file named `tree_glottolog_newick.txt`[3] contains glottolog's trees, and `languages_and_dialects_geo.csv` provides geographical metadata.

Language metadata can be accessed using the *glottoTrees* function `get_glottolog_languages()`. This function returns a dataframe of close to twenty-six thousand rows. To view it in full, we suggest saving it to a CSV file and opening it in spreadsheet software such as Excel:

```
language_metadata <- get_glottolog_languages()
write.csv(language_metadata, "language_metadata.csv")
```

Here are the first ten rows:

```
language_metadata <- get_glottolog_languages()
head(language_metadata, n = 10)
```

```
##    glottocode isocodes        name name_in_tree position tree          tree_name
## 1    3adt1234           3Ad-Tekles   3Ad-Tekles      tip  391        Afro-Asiatic
## 2    aala1237               Aalawa       Aalawa      tip   94        Austronesian
## 3    aant1238            Aantantara   Aantantara      tip   90 NuclearTransNewGuinea
## 4    aari1238    <NA>         <NA>   Aari-Gayil     node   22          SouthOmotic
## 5    aari1239     aiw         Aari         Aari      tip   22          SouthOmotic
## 6    aari1240     aay       Aariya       Aariya     <NA>   NA              <NA>
## 7    aasa1238     aas        Aasax        Aasax      tip  391        Afro-Asiatic
## 8    aasd1234             Aasdring     Aasdring      tip  269        Indo-European
## 9    aata1238            Aatasaara    Aatasaara      tip   90 NuclearTransNewGuinea
## 10   abaa1238               Rngaba       Rngaba      tip  345        Sino-Tibetan
```

Listed here are glottolog's languages, dialects, subgroups and families. These entities are identified by a name, an ISO-639-3 code if available (format: three letters) and a glottolog-specific *glottocode* (format: four letters followed by four digits[4]). Also described is the entity's relationship to a glottolog tree: the representation of

---

[3]Although this file is named `tree_glottolog_newick.txt`, it is not in true Newick format due to its use of square brackets in node and tip labels (Felsenstein n.d.). In *glottoTrees*, the square brackets in glottolog's file are converted to angled brackets (i.e., greater-than and less-than symbols), to bring them into conformity with the Newick standard.

[4]There are two exceptional glottocodes with numbers in the initial four characters: b10b1234 and 3adt1234.

its name in the tree (which may differ slightly from the name used elsewhere by glottolog[5]), its position (as tip or node), and the tree's number and name.

By default, the metadata functions in *glottoTrees*, such as `get_glottolog_languages()`, will return information about the most recent version of glottolog which the package contains. To access older versions, supply the version number via the `glottolog_version` argument:[6]

```
language_metadata_v4.3 <- get_glottolog_languages(glottolog_version = "4.3")
head(language_metadata_v4.3, n = 10)
```

```
##    glottocode isocodes        name name_in_tree position tree           tree_name
## 1   3adt1234          3Ad-Tekles   3Ad-Tekles      tip  186          Afro-Asiatic
## 2   aala1237             Aalawa       Aalawa        tip  205           Austronesian
## 3   aant1238          Aantantara   Aantantara       tip  145 NuclearTransNewGuinea
## 4   aari1238     <NA>       <NA>   Aari-Gayil      node   85            SouthOmotic
## 5   aari1239      aiw       Aari         Aari        tip   85            SouthOmotic
## 6   aari1240      aay     Aariya       Aariya      <NA>   NA                  <NA>
## 7   aasa1238      aas      Aasax        Aasax        tip  186          Afro-Asiatic
## 8   aasd1234          Aasdring     Aasdring        tip  179          Indo-European
## 9   aata1238          Aatasaara   Aatasaara        tip  145 NuclearTransNewGuinea
## 10  abaa1238            Rngaba       Rngaba         tip  329          Sino-Tibetan
```

Briefer metadata about glottolog's language families can be accessed using the *glottoTrees* function `get_glottolog_families()`. This returns a dataframe of 420 rows, so to view it in full, we also suggest saving it to a CSV file and opening it in spreadsheet software. Here are the first ten rows:

```
family_metadata <- get_glottolog_families()
head(family_metadata, n = 10)
```

```
##    tree             tree_name n_tips n_nodes main_macroarea
## 1     1                   Yam     33      18       Papunesia
## 2     2       Mongolic-Khitan     66      25         Eurasia
## 3     3 Kol{PapuaNewGuinea}     2       1       Papunesia
## 4     4         Namla-Tofanma      2       1       Papunesia
## 5     5            Tanahmerah      1       1       Papunesia
## 6     6           Jarawa-Onge      2       1         Eurasia
## 7     7            Ta-Ne-Omotic    29      15          Africa
## 8     8                Pomoan     10       7   North America
## 9     9           WesternDaly     14       7       Australia
## 10   10             Yangmanic      3       1       Australia
```

Glottolog v4.4 divides the world's languages into 420 families, including 138 isolates, and it provides a tree for each. Together, the 420 trees contain 8,209 internal nodes and 17,008 tips, many of which represent varieties that would typically be considered dialects. Geographically, glottolog assigns each language variety to one of six *macroareas*: Africa, Australia, Eurasia, Papunesia, South America or North America. The *glottoTrees* metadata includes a column `main_macroarea`. This is the one macroarea which contains more of the family's language varieties than any other. We will see how this information can be useful in Section 4.2.

Glottolog's 420 family trees are stored in a `multiPhylo` object named `glottolog_trees_v4.4`. For example, here is glottolog's representation of the Great Andamanese family, which is tree 340 within the object `glottolog_trees_v4.4`. For readability, we plot this tree horizontally:

```
tree_GA <- glottolog_trees_v4.4[[340]]
plot(tree_GA, x.lim = c(-0.3, 14))
```

---

[5]The differences are systematic and are made in order to conform with the permissible Newick format of tree labels: spaces and apostrophes are removed, parentheses are replaced by braces, and commas are replaced by forward slashes.

[6]*glottoTrees* currently contains information from glottolog versions 4.0, 4.1, 4.2, 4.3 and 4.4 (which is current at time of writing). Our intention is to update *glottoTrees* as glottolog updates in the future.

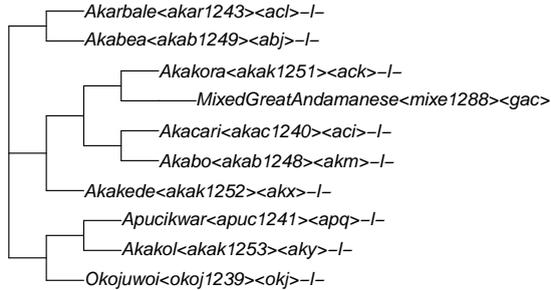

Just above, we obtained the tree for Great Andamanese by referring to its tree number (340) in the `glottolog_trees_v4.4` object. The package *glottoTrees* also provides a function `get_glottolog_trees()` which enables trees to be obtained using the glottolog name for their families, for instance:

```
tree_GA <- get_glottolog_trees("GreatAndamanese")
plot(tree_GA, x.lim = c(-0.3, 14))
```

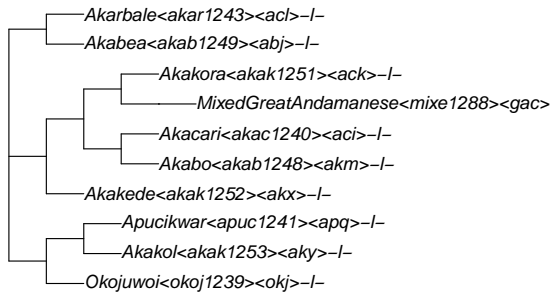

If you know the name of one or more families and would like to now the number of their trees, use `which_tree()`:

```
which_tree("GreatAndamanese")
```

```
## GreatAndamanese
##             340
```

```
which_tree(c("Turkic", "Tupian", "Tuu"))
```

```
## Turkic Tupian    Tuu
##    217     32     76
```

Both `get_glottolog_trees()` and `which_tree()` allow the usage of a `glottolog_version` argument, to refer to older versions of glottolog. For instance, here are the tree numbers of the same families in version 4.1:

```
which_tree("GreatAndamanese", glottolog_version = "4.1")
```

```
## GreatAndamanese
##               6
```

```
which_tree(c("Turkic", "Tupian", "Tuu"), glottolog_version = "4.1")
```

```
## Turkic Tupian    Tuu
##     66    297     80
```

In glottolog's trees, the tip labels are rather long, consisting of a name followed by a glottocode in angled brackets, an ISO code in angled brackets (if one exists) and possibly the string "-l-". Node labels (not shown in the tree above) have the same structure. The *glottoTrees* function `abridge_labels()` will shorten labels to just the glottocode, for example:

```
tree_GA_abr <- abridge_labels(tree_GA)
plot_glotto(tree_GA_abr)
```

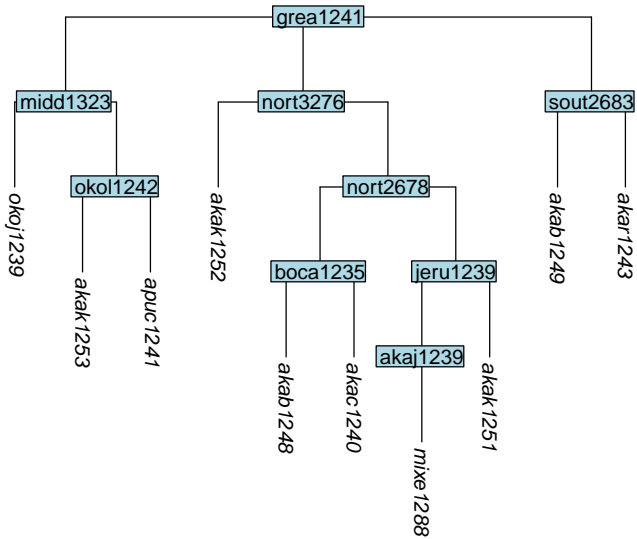

The function `abridge_labels()` will issue a warning if there are tip or node labels in which it is unable to identify a glottocode. We will see an example of this shortly below.

In glottolog's trees, the branches are all of equal length. We will discuss how to assign more realistic branch lengths in Section 4.4.

## 4.2 How to combine trees

As discussed in the main paper, the comparison of languages across language families unavoidably carries a commitment to a genealogical hypothesis, even if that hypotheses is, tacitly, that all families are equally (un)related. Given that making such hypotheses is unavoidable, it will be most beneficial for progress in the field to make them explicit. To enable typologists to explore genealogical hypotheses and to make those hypotheses explicit, *glottoTrees* provides tools for combining multiple glottolog trees into one.

To begin with a small example, here we combine five glottolog families to represent the hypothesised Arnhem group in northern Australia (Green 2003). First we create a `multiPhylo` object containing the five glottolog language families (Gunwinyguan, Mangarrayi-Maran, Maningrida, and the isolates Kungarakany and Gaagudju):

```
arnhem_family_names <-
  c("Gunwinyguan", "Mangarrayi-Maran", "Maningrida", "Kungarakany", "Gaagudju")
multiPhylo_arnhem <- get_glottolog_trees(arnhem_family_names)
```

The *glottoTrees* function `assemble_rake()` enables the trees in a `multiPhylo` object to be assembled into a single tree with a rake structure at its root. Here we apply `assemble_rake()` to our `multiPhylo` object and

assign the resulting, single tree to the object `tree_arnhem`.

```
tree_arnhem <- assemble_rake(multiPhylo_arnhem)
```

For plotting, it will be convenient to shorten the labels in the tree to just the glottocodes, using `abridge_labels()`. However, the root node in our newly created tree `tree_arnhem` has no label. Accordingly, the function `abridge_labels()` will issue a warning, that it encountered one node without a glottocode. This is not an error message, and `abridge_labels()` still shortens all labels to glottocodes where it can; it is just flagging the fact that it was not able to do so in all cases.

```
tree_arnhem_abr <- abridge_labels(tree_arnhem)
```

```
## Warning in abridge_labels(tree_arnhem): Labels without glottocodes were detected and
## left unchanged for: 0 tip(s); 1 node(s):
```

Plotting the resulting tree enables us to inspect our newly created Arnhem tree. Note how all five families are joined to the root in a rake-like structure, without any additional subgrouping.

```
plot_glotto(tree_arnhem_abr, nodelabels = FALSE)
```



It is possible to give a combined tree more structure, by using `assemble_rake()` iteratively. For instance, suppose we wished to hypothesise that Gunwinyguan, Mangarrayi-Maran and Maningrida form their own subgroup. First, we create a `multiPhylo` object containing those three trees, and combine it into a single rake tree, which we call `tree_A`:

```
multiPhylo_A <- get_glottolog_trees(c("Gunwinyguan", "Mangarrayi-Maran", "Maningrida"))
tree_A <- assemble_rake(multiPhylo_A)
```

Then we create the final tree by combining `tree_A` with the two isolate family trees:

```
multiPhylo_arnhem2 <- c(tree_A, get_glottolog_trees(c("Kungarakany", "Gaagudju")))
tree_arnhem2 <- assemble_rake(multiPhylo_arnhem2)
tree_arnhem2_abr <- abridge_labels(tree_arnhem2)
```

```
## Warning in abridge_labels(tree_arnhem2): Labels without glottocodes were detected and
## left unchanged for: 0 tip(s); 2 node(s): ,
```

```
plot_glotto(tree_arnhem2_abr, nodelabels = FALSE)
```

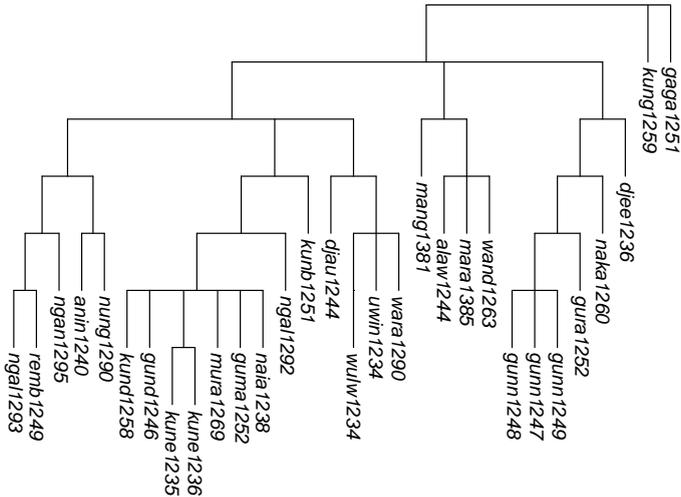

Typological studies often examine languages from very many families. To group all 420 families into a single 'supertree', *glottoTrees* provides the function `assemble_supertree()`. By default, the function returns a supertree that divides first into glottolog's six macroareas, with an internal node for each, and directly below these macroarea nodes appear all of the glottolog families, grouped by their `main_macroarea` mentioned in Section 4.1 above. This tree is enormous, so we do not plot it here. It is obtained like this:

```
my_supertree <- assemble_supertree()
```

The highest-level, macroarea groupings can also be controlled through the function's argument `macro_groups`. For instance, to group all of the world's families directly into a 420-pronged rake structure, set `macro_groups = NULL`:

```
my_supertree <- assemble_supertree(macro_groups = NULL)
```

It is also possible to group macroareas together, for example, to combine North and South America into a single group. Grouping of macroareas is achieved by setting the `macro_groups` argument to a *list* whose items are the desired groups of macroareas. Each group will then appear as one of the highest-level nodes of the tree, and all of its families below it. For instance, to keep all of glottolog's macroareas separate, but to combine North and South America into a single group, the following code would be used. First we define a list, which we've called `my_list`, within which any groupings containing more than one macroarea are represented as a vector, using the `c()` function:

```
my_list <- list("Africa", "Australia", "Eurasia", "Papunesia",
                c("South America", "North America"))
```

We then use that list as the `macro_groups` argument of `assemble_supertree()`:

```
my_supertree <- assemble_supertree(macro_groups = my_list)
```

Taking a second example, to create a supertree containing only the families whose `main_macroarea` is either Africa or Eurasia, and to place Africa and Eurasia under separate, highest-level nodes, we would use:

```
my_list <- list("Africa", "Eurasia")
my_supertree <- assemble_supertree(macro_groups = my_list)
```

## 4.3 How to modify trees

There are several reasons why typologists may wish to use a tree that departs from the glottolog trees. Most commonly, a typological study will cover a set of languages that differs from the set of tips in any single glottolog tree, either through the exclusion of some of the lects that glottolog represents as tips or through the distinction of additional lects. A third case that can arise is when glottolog places one or more dialects at the tree's tips and more a general, language node above them. The typologist may have data that applies to the language (an internal node) rather than the dialects (the tips), yet the calculation of genealogically-sensitive averages and proportions requires one's typological variables to be related to the tips of trees, not to internal nodes. In these cases and many others, the typologist may wish to alter the glottolog tree to suit the purposes of the research. The *glottoTrees* package supplies a set of functions to aid in performing each of these tree manipulations. In the section we introduce them and illustrate their use.

In the following examples, we will make use glottolog's representation of the Great Andamanese family, whose labels we shorten to just the glottocodes using `abridge_labels()`:

```
tree_GA <- get_glottolog_trees("GreatAndamanese")
tree_GA_abr <- abridge_labels(tree_GA)
plot_glotto(tree_GA_abr)
```

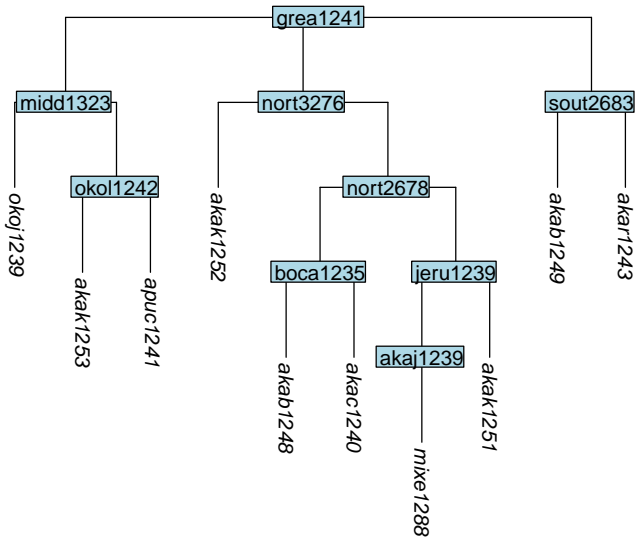

### 4.3.1 How to remove tips

Firstly, we illustrate the removal of tips from a tree. There are two functions in *glottoTrees* for doing this. The function `remove_tip()` works by specifying which tips are to be removed, while the function `keep_tip()` works by specifying which tips are to be retained. First we will remove three of the original ten tips in the Great Andamanese tree. We do this by setting the `label` argument of `remove_tip()` to a vector containing the labels of the tips to be removed. Within the vector, the labels can appear in any order.

```
tree_GAa <- remove_tip(tree_GA_abr, label = c("akab1249", "akak1251", "apuc1241"))
plot_glotto(tree_GAa)
```

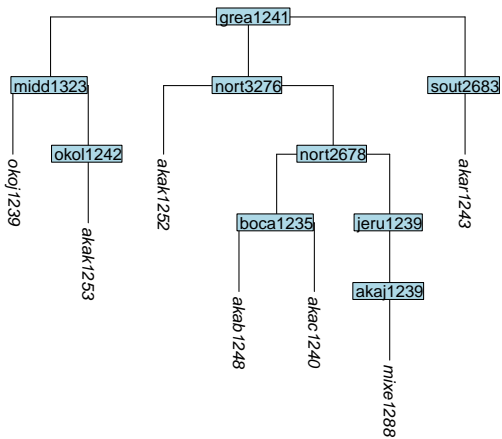

In this next example, we remove the tips `akab1249` and `akar1243`. These tips are the only tips that sit below the internal node `sout2683`. This is significant, because it triggers a convention in tree manipulation, that if all tips below a node are removed, then the node is removed also. We see that here:

```
tree_GAb <- remove_tip(tree_GA_abr, label = c("akab1249", "akar1243"))
plot_glotto(tree_GAb)
```

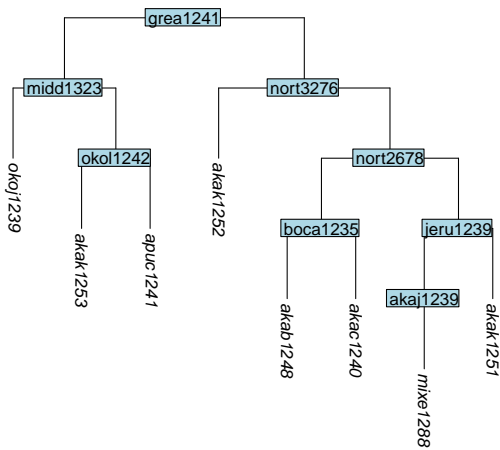

We now illustrate the usage of the *glottoTrees* function `keep_tip()`. Here we use it to retain six of the original ten tips in the Great Andamanese. We do this by setting the `label` argument to a vector containing the labels of the six desired tips.

```
tree_GAc <- keep_tip(tree_GA_abr, label = c("akar1243", "akak1251", "akac1240",
                                            "akak1252", "apuc1241", "okoj1239"))
```

```
plot_glotto(tree_GAc)
```

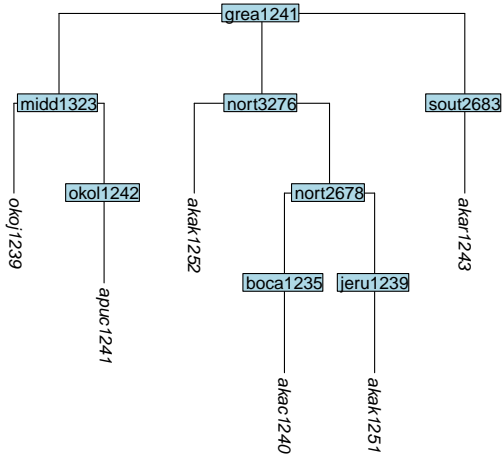

As before, if our use of `keep_tip()` results in a node having all of the tips below it removed, then the node will also be removed automatically. This is illustrated here, where the node `boca1235` is removed automatically because neither of the tips below it are kept:

```
tree_GAd <- keep_tip(tree_GA_abr, label = c("akar1243", "akak1251", "akak1252",
                                            "apuc1241", "okoj1239"))
plot_glotto(tree_GAd)
```

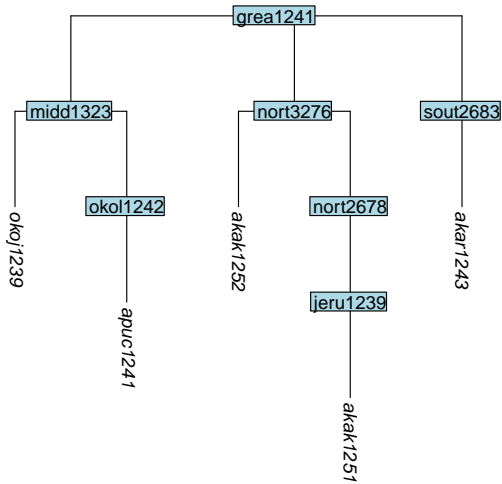

### 4.3.2 How to remove tips and convert nodes to tips

As mentioned earlier, many of the tips in glottolog's trees correspond to dialects, with languages represented as nodes above the dialectal tips. One usage case we foresee is that a typologist will wish to study a set of language varieties, some of which correspond to glottolog's tips and some of which correspond to nodes. The *glottoTrees* function `keep_as_tips()` takes an argument `label` which can contain both tip labels and node labels. Any tips will be kept, and any nodes will be converted into tips, with all of the structure below them being removed. Be mindful when using `keep_as_tips()` that it is not possible to both convert a node into a tip and also retain the structure below it, such as tips that it dominates. Here we keep the same tips as in the tree above, while also converting the node `boca1235` into a tip:[7]

```
tree_GAe <- keep_as_tip(tree_GA_abr, label = c("akar1243", "akak1251", "akak1252",
                                               "apuc1241", "okoj1239", "boca1235"))

plot_glotto(tree_GAe)
```

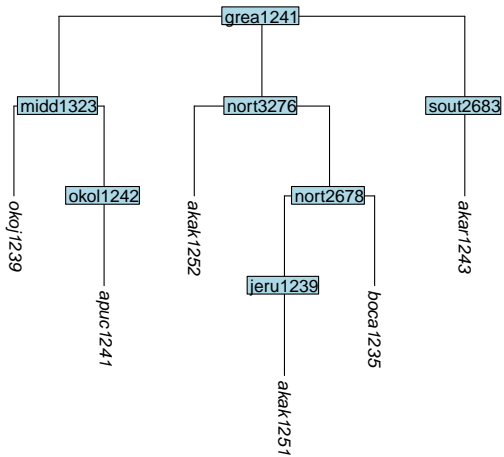

One workflow that we envision for `keep_as_tip()` is that the typologist has prepared a CSV file, one of whose columns is named `tip` and contains the glottocodes of all the language varieties in the study. This CSV file can be loaded in R and assigned to a dataframe, and then its `tip` column can be passed to `keep_as_tip()` as the value of the `labels` argument, like this:

```
my_dataframe <- read.csv("my_data_file.csv", stringsAsFactors = FALSE)
my_new_tree <- keep_as_tip(my_old_tree, label = my_dataframe$tip)
```

To just convert one or more nodes into tips, use `convert_to_tip()`, as we do here to convert the nodes `okol1242` and `sout2683` to tips:

```
tree_GAf <- convert_to_tip(tree_GA_abr, label = c("okol1242", "sout2683"))
plot_glotto(tree_GAf)
```

---

[7]Converting nodes into tips may cause them to move to the left or right in the tree plot. The movement is meaningless, since a subgroup (A, B) is exactly the same as subgroup (B, A). Since the movement is meaningless, it's also harmless.

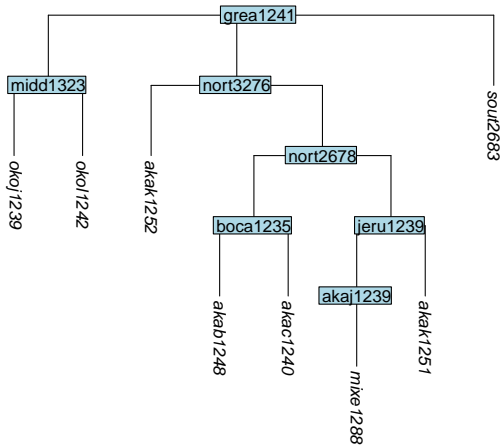

### 4.3.3 How to remove internal nodes

Sometimes, the removal of tips will cause one or more of the remaining tips to sit below a node which dominates only it. This reflects that fact that `remove_tip()`, `keep_tip()` and `keep_as_tip()` all preserve the original *depth* of any tips that remain in the tree (the reader may like to confirm this by reviewing the plots above). Depending on the researcher's needs, this outcome may or may not be desirable. If it is undesirable, then non-branching, internal nodes can be removed using the *glottoTrees* function `collapse_node()`. For instance, here we remove two of the non-branching nodes from the tree `tree_GAc` above, by naming them in the `label` argument of `collapse_node()`. In the resulting tree, these nodes have been removed, thus reducing the depth of the tips below them:

```
tree_GAg <- collapse_node(tree_GAc, label = c("boca1235", "okol1242"))
plot_glotto(tree_GAg)
```

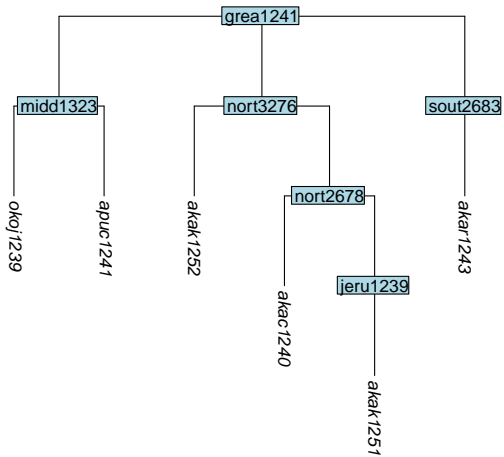

When deciding whether to collapse nodes, in can be handy to know which nodes in a tree that have only one child below them. The function `nonbranching_nodes()` will return a vector of all such nodes, for example:

```
nonbranching_nodes(tree_GAc)
```

```
## [1] "okol1242" "boca1235" "jeru1239" "sout2683"
```

```
nonbranching_nodes(tree_GAg)
```

```
## [1] "jeru1239" "sout2683"
```

The function `collapse_node()` can also be used to alter a subgrouping hypothesis, and specifically, to remove a layer of subgrouping, converting a nested structure ((A,B),C) into a flat structure (A,B,C). For instance, here we remove the `okol1242` node of the original glottolog Great Andamanese tree, converting its two daughter languages into sisters of `okoj1239`:

```
tree_GAh <- collapse_node(tree_GA_abr, label= "okol1242")
plot_glotto(tree_GAh)
```

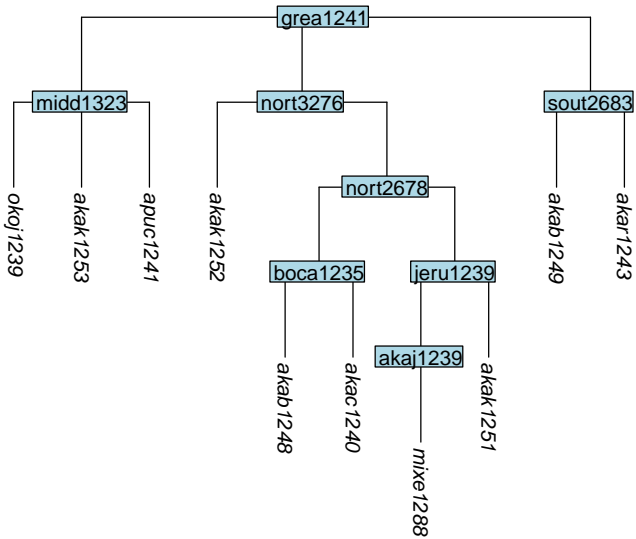

### 4.3.4 How to add tips

The function `add_tip()` allows tips to be added to a tree. The `label` argument specifies the name of the new tip, while `parent_label` specifies the label of the node below which the new tip should appear. Here we add a tip `xxxx1234` below the node `sout2683`:

```
tree_GAi <- add_tip(tree_GA_abr, label = "xxxx1234", parent_label = "sout2683")
plot_glotto(tree_GAi)
```

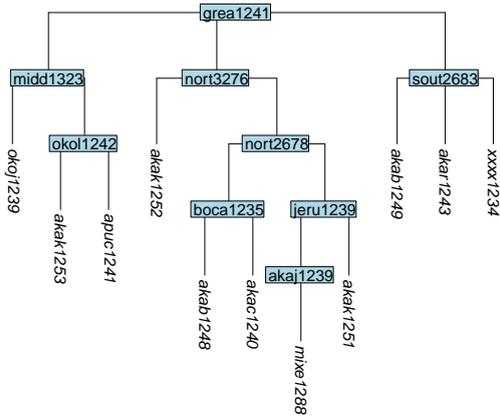

### 4.3.5 How to clone tips

Next we illustrate the cloning of tips. Cloning tips may be useful when glottolog provides only one glottocode, and thus only one tree tip, corresponding to multiple lects in the typologist's study. To clone a tip, use the function `clone_tip()` and in the `label` argument, provide a vector of the tips to be cloned. Here we clone tips `akar1243` and `akak1252`:

```
tree_GAj <- clone_tip(tree_GA_abr, label = c("akar1243", "akak1252"))
plot_glotto(tree_GAj)
```

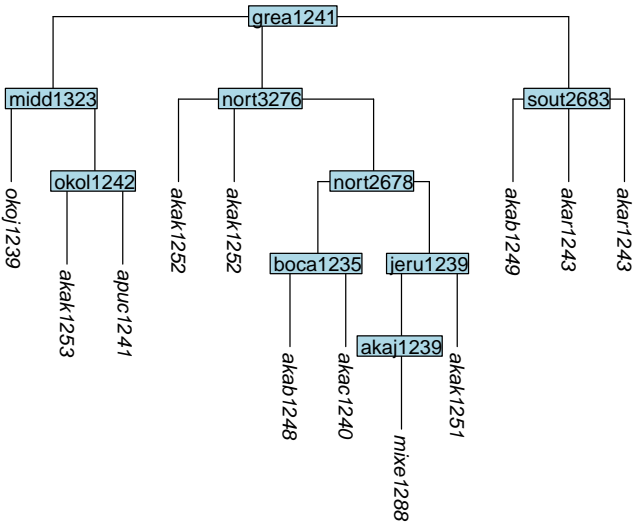

By default, clones are added to the tree as sisters directly beneath the parent node of the original tip. An alternative is to create a new subgroup for each set of sister clones, using the `subgroup` argument and setting it to `subgroup = TRUE`. Each newly created subgroup node is given a label that matches the cloned tips it

21

dominates:

```
tree_GAk <- clone_tip(tree_GA_abr, label = c("akar1243", "akak1252"), subgroup = TRUE)
plot_glotto(tree_GAk)
```

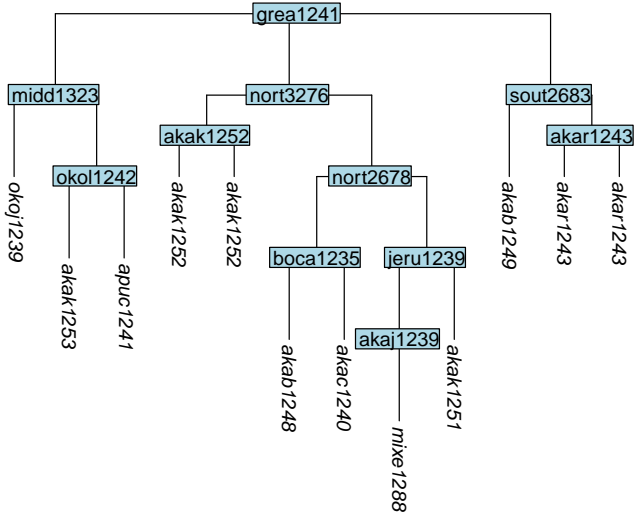

It is also possible to make more than one clone using the `n` argument. Here we create three new clones of `akab1248` and place them in a subgroup:

```
tree_GAl <- clone_tip(tree_GA_abr, label = "akab1248", n = 3, subgroup = TRUE)
plot_glotto(tree_GAl)
```

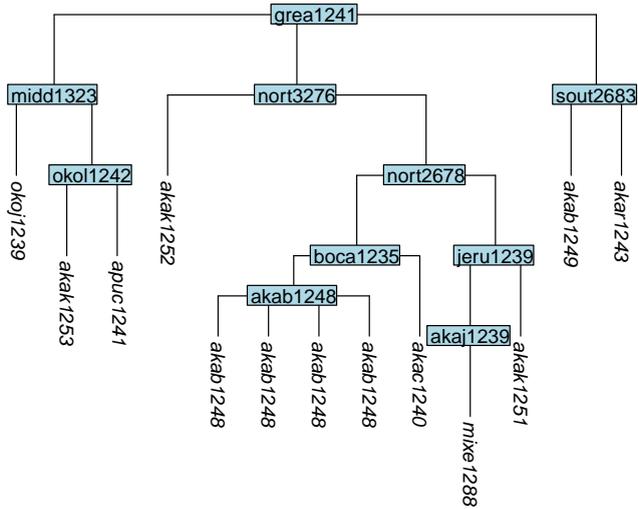

One of the consequences of cloning tips is that the in the resulting tree, not all tips will have distinct names. The function `apply_duplicate_suffixes()` will add a suffix to any tips with duplicate labels, to make them unique.[8] The suffix will consist of a hyphen followed by a number. Here we add suffixes to the tree `tree_GAj`:

```
tree_GAm <- apply_duplicate_suffixes(tree_GAj)
plot_glotto(tree_GAm)
```

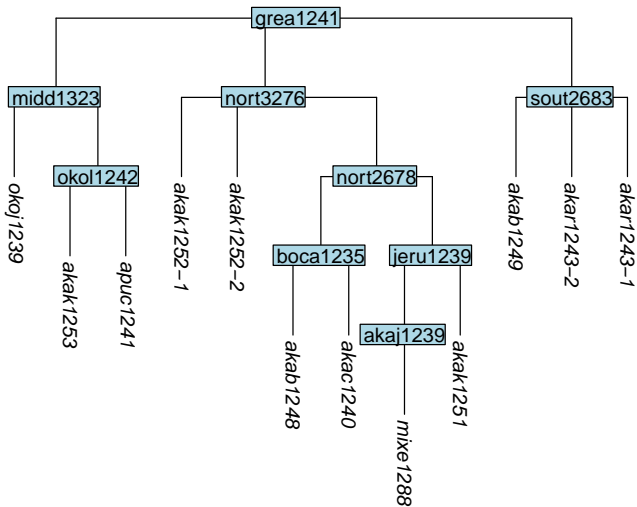

---

[8] `apply_duplicate_suffixes()` will also add suffixes to any nodes with duplicate labels.

### 4.3.6 How to move a tip

Using the function `move_tip()`, a tip can be moved to a new position, beneath a new parent node (one of the nodes already in the tree) which is specified with the `parent_label` argument:

```
tree_GAn <- move_tip(tree_GA_abr, label = "apuc1241", parent_label = "jeru1239")
plot_glotto(tree_GAn)
```

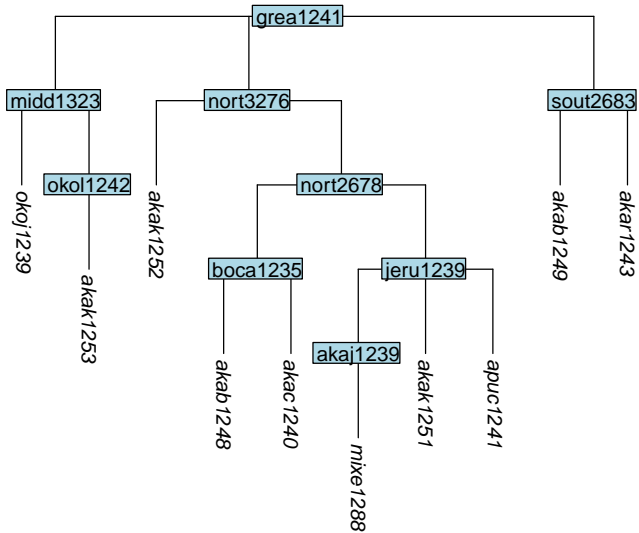

### 4.3.7 How to move a node and its descendants

In a similar fashion, the function `move_node()` is used to move an internal node, along with all of the structure below it, to a position beneath a new parent node:

```
tree_GAo <- move_node(tree_GA_abr, label = "jeru1239", parent_label = "okol1242")
plot_glotto(tree_GAo)
```

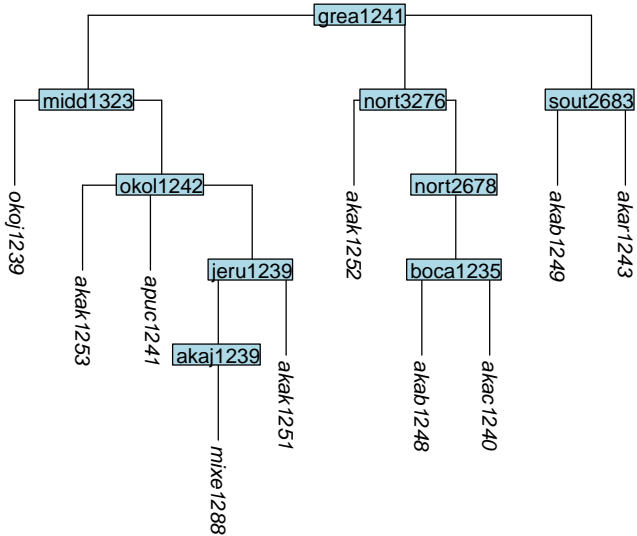

### 4.3.8 Summary: a general-purpose toolkit for curating trees' topology

The functions `remove_tip()`, `keep_tip()`, `keep_as_tip()`, `convert_to_tip()`, `collapse_node()`, `add_tip()`, `clone_tip()`, `move_tip()` and `move_node()` provide a general-purpose toolkit for modifying a single glottolog tree, or a combined tree, or supertree, to make its set of tips, and the subgrouping of those tips, conform to the set of lects that a typologist is analysing in a typological study.

## 4.4 How to add branch lengths

Branch lengths in a tree convey information, and all of the phylogenetic methods discussed in the main paper are sensitive to the information represented by the branch lengths. (To be specific, the methods discussed in the main paper are sensitive to the *relative* lengths of the branches, so multiplying all of the branch lengths in a tree by some constant amount would not affect the results.)

Glottolog's trees contain informative subgrouping structure, but the branch lengths are all equal. Even without knowing what the true branch lengths are for a linguistic tree, we do know that a situation in which all are equal is highly unlikely. A good approximation to the most-likely[9] distribution of branch lengths in a phylogenetic tree, under a variety of assumptions, is exponential (Venditti, Meade & Pagel 2010), i.e., very long branches are rare, and very short ones are frequent. This notion is implemented in the *glottoTrees* package by the function `rescale_branches_exp()`, which sets the deepest branches to length 1/2, then next layer to length 1/4, then the next to 1/8 and so on. This will produce a more plausible set of branch lengths, even in the absence of firm knowledge of exact lengths, and on these grounds we advocate its use if additional information about branch lengths is not available.

Here is an example of the result of applying exponential branch lengths to glottolog's Great Andamanese tree:

---

[9]'Most likely' doesn't mean that we expect to see trees with exactly these branch lengths. Compare this to flipping a coin two million times: although it is unlikely that the outcome will be exactly one million heads and one million tails, it remains true that one million heads and one million tails is the most likely outcome, in the strict sense that it is more likely than any other outcome. The branch lengths discussed here are 'most likely' is a similar sense.

```
tree_GAp <- rescale_branches_exp(tree_GA_abr)
plot_glotto(tree_GAp)
```

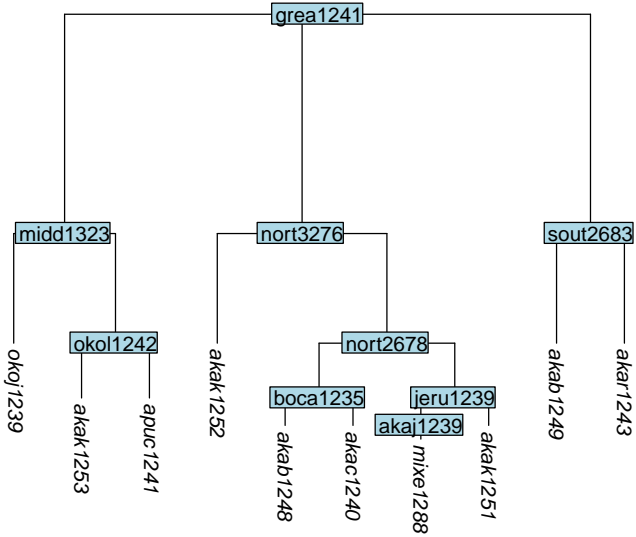

Here is an example of the result of applying them to glottolog's Eskimo-Aleut tree, of 30 tips:

```
tree_EA <- get_glottolog_trees("Eskimo-Aleut")
tree_EA_abr <- abridge_labels(tree_EA)
tree_EAa <- rescale_branches_exp(tree_EA_abr)
plot_glotto(tree_EAa, nodelabels = FALSE)
```

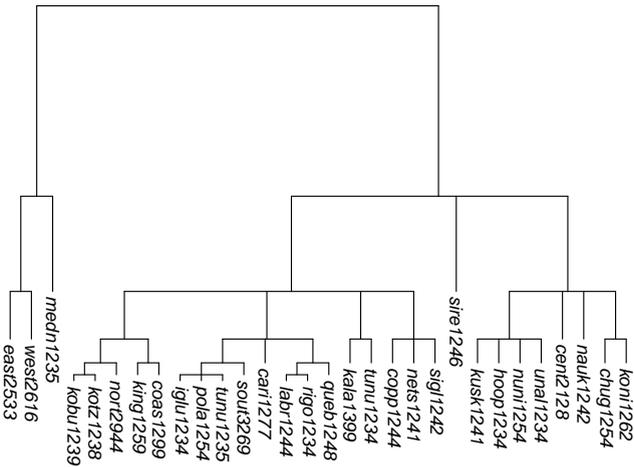

An additional option is to stretch the terminal branches so that all tips are equidistant from the root, creating what is known as an *ultrametric* tree. This is done using the function `ultrametricize()`.

```
tree_EAb <- ultrametricize(tree_EAa)
plot_glotto(tree_EAb, nodelabels = FALSE)
```

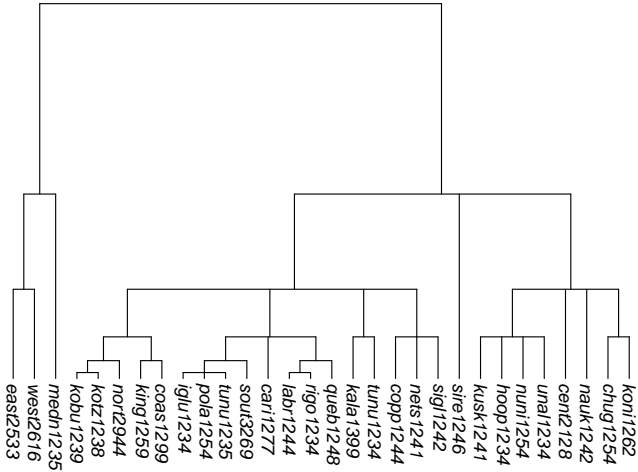

An additional function, `rescale_deepest_branches()`, can be used to adjust just the deepest layer of branches. This may be useful where multiple family trees have been joined together, and there is a desire to manipulate the implied closeness or distance between the first-order branches. For example, here we take the hypothesised Arnhem group from Section 4.2. First we assign exponential branch lengths with `set_branch_lengths_exp()`, which sets the deepest branch length to 1/2. Then we triple the distance of the deepest level of relationships by changing the first branch length to 1.5 using `rescale_deepest_branches()`, before ultrametiricising the tree:

```
tree_arnhem_a <- rescale_branches_exp(tree_arnhem_abr)
tree_arnhem_b <- rescale_deepest_branches(tree_arnhem_a, 1.5)
tree_arnhem_c <- ultrametricize(tree_arnhem_b)
plot_glotto(tree_arnhem_c, nodelabels = FALSE)
```

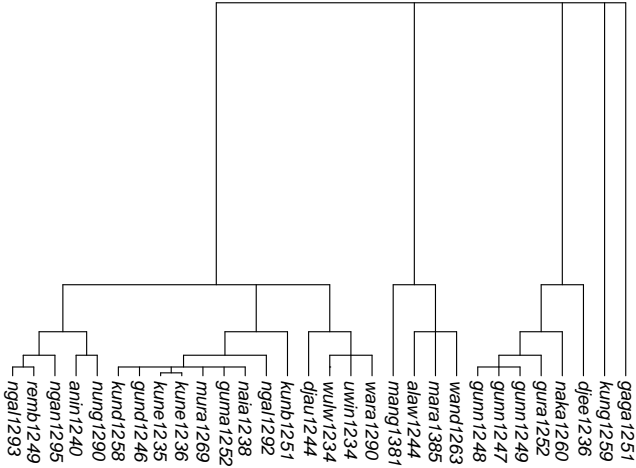

## 4.5 Exporting trees for use with other software

In R, trees can be saved to file in Newick format using the function `write.tree()` in the *ape* package. Files like this can be opened by other software such as FigTree[10], which can be used to interactively generate tree plots that may be useful for publication and dissemination. For instance, here we write the tree `tree_arnhem_c` to a file whose filename ends in the standard file extension, `.tree`:

```
write.tree(tree_arnhem_c, "my_arnhem_tree.tree")
```

Often it will be desirable to reproduce a tree with labels that are more reader-friendly than glottocodes. *glottoTrees* provides the function `relabel_with_names()`, which will replace full glottolog labels, or labels consisting of just a glottocode, with glottolog's corresponding language, dialect, subgroup or family name. Here we relabel the Arnhem tree by the languages' names. As was the case with `abridge_labels()`, warnings are given by `relabel_with_names()` if a tree contains any nodes that cannot be relabeled in this way; these are not errors, just alerts.

```
tree_arnhem_c_namelabels <- relabel_with_names(tree_arnhem_c)
```

```
## Warning in relabel_with_names(tree_arnhem_c): Labels without glottocodes were detected
## and left unchanged for: 0 tip(s); 1 node(s):
```

```
plot_glotto(tree_arnhem_c_namelabels, nodelabels = FALSE)
```

---

[10]https://github.com/rambaut/figtree/releases

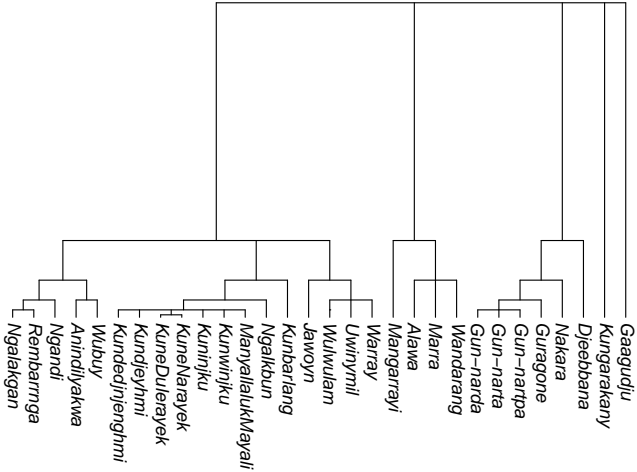

# 5 Putting it together: A worked example

In this section we provide a real worked example of the use of the methods described above.

Yin (2020) examined violations of the sonority sequencing principle in 496 languages, and calculated the genealogically-sensitive proportions of languages in which various violations occurred. The language sample consisted of 496 languages in the CLICS2 database (Anderson et al. 2018) and the AusPhon-Lexicon database (Round 2017). The language sample was not balanced in the traditional sense, and phylogenetic methods were used to help produce a principled interpretation of the data.

Yin's raw data consisted of a table of languages' names and glottocodes and indications of whether or not the languages had consonant clusters in word-initial onsets or word-final codas that contained sonority reversals, coded as 1 for yes and 0 for no. This dataset is provided with the *phyloWeights* package as a dataframe named `yin_2020_data` whose columns are `name`, `tip`, `has_onset_violation` and `has_coda_violation`. The first ten rows are shown here:

```
head(yin_2020_data, n = 10)
```

```
##                    name      tip has_onset_violation has_coda_violation
## 1                Abkhaz abkh1244                   1                  1
## 2                  Abui abui1241                   0                  0
## 3               Achagua acha1250                   0                  1
## 4                 Adang adan1251                   0                  1
## 5          Adnyamathanha adny1235                   0                  0
## 6                Adyghe adyg1241                   1                  1
## 7        Hokkaidoainu ainu1240                   0                  0
## 8                 Alawa alaw1244                   1                  0
## 9      Standardalbanian alba1267                   1                  1
## 10                Aleut aleu1260                   1                  1
```

## 5.1 Preparing a tree

The tree for Yin's study was constructed from a glottolog supertree, using glottolog version 4.2. Yin's supertree made use of glottolog's macroareas. Since the language sample covered relatively few families in

the Americas, a single group was used for South America and North America. Additionally, the only African language available in the sample was Arabic, so Africa and Eurasia were grouped together:

```
yin_macro <- list(c("South America", "North America"), c("Africa", "Eurasia"),
                  "Papunesia", "Australia")
supertree <- assemble_supertree(macro_groups = yin_macro, glottolog_version = "4.2")
supertree_abr <- abridge_labels(supertree)
```

```
## Warning in abridge_labels(supertree): Labels without glottocodes were detected and left
## unchanged for: 0 tip(s); 5 node(s): World, SouthAmerica-NorthAmerica, Africa-Eurasia,
## Papunesia, Australia
```

Five tips were cloned, in cases where Yin had data for two varieties corresponding to just one tip in the glottolog supertree:

```
supertree_a <- clone_tip(supertree_abr, subgroup = TRUE,
                         label = c("ayab1239", "basu1242", "biri1256",
                                   "ikar1243", "peri1265"))
supertree_b <- apply_duplicate_suffixes(supertree_a)
```

Eight tips were added, in case where for sister lects (A,B), glottolog placed A as a node above B. In such cases, in new tip A was placed below the existing glottolog node A:

```
supertree_c <- supertree_b
nodes_to_add_as_tips <- c("alor1249", "gami1243", "guri1247", "mand1415",
                          "sins1241", "wang1291", "warl1254", "yand1253")
# Loop through these nodes, and use add_tip() to add the new tip:
for (node_i in nodes_to_add_as_tips) {
  supertree_c <- add_tip(supertree_c, label = node_i, parent_label = node_i)
}
```

From this supertree, the 496 languages in Yin's dataset were kept. The internal node `mada1298` was collapsed, as were all non-branching internal nodes:

```
supertree_d <- keep_as_tip(supertree_c, label = yin_2020_data$tip)
supertree_e <- collapse_node(supertree_d, label = "mada1298")
supertree_f <- collapse_node(supertree_e, label = nonbranching_nodes(supertree_e))
```

Finally, branch lengths were assigned. Branches were first assigned exponential lengths. Then, in order to diminish the importance of the macro groups, the branches above them were shortened to a length of 1/40. The effect of this decision is that the implied distance between families in different macro groups is only marginally greater than between families within a single macro group.

```
supertree_g <- rescale_branches_exp(supertree_f)
yin_2020_tree <- rescale_deepest_branches(supertree_g, 1/40)
```

The resulting tree appears as in Figure 1, which is plotted with the following code:

```
full_names <- yin_2020_data$name[match(yin_2020_tree$tip.label, yin_2020_data$tip)]
name_tree <- yin_2020_tree
name_tree$tip.label <- full_names
plot(ladderize(name_tree, right = FALSE), type = "fan",
     cex = 0.3, label.offset = 0.002, edge.width = 0.5)
```

## 5.2   Preparing the dataframe of typological data

In order to calculate phylogenetic weights and genealogically-sensitive proportions, in addition to the tree (or a set of trees) we require a dataframe with (a) one column `tip`, whose contents match the tip labels in the trees, and (b) other columns containing numerical data to be averaged. The dataframe `yin_2020_data`

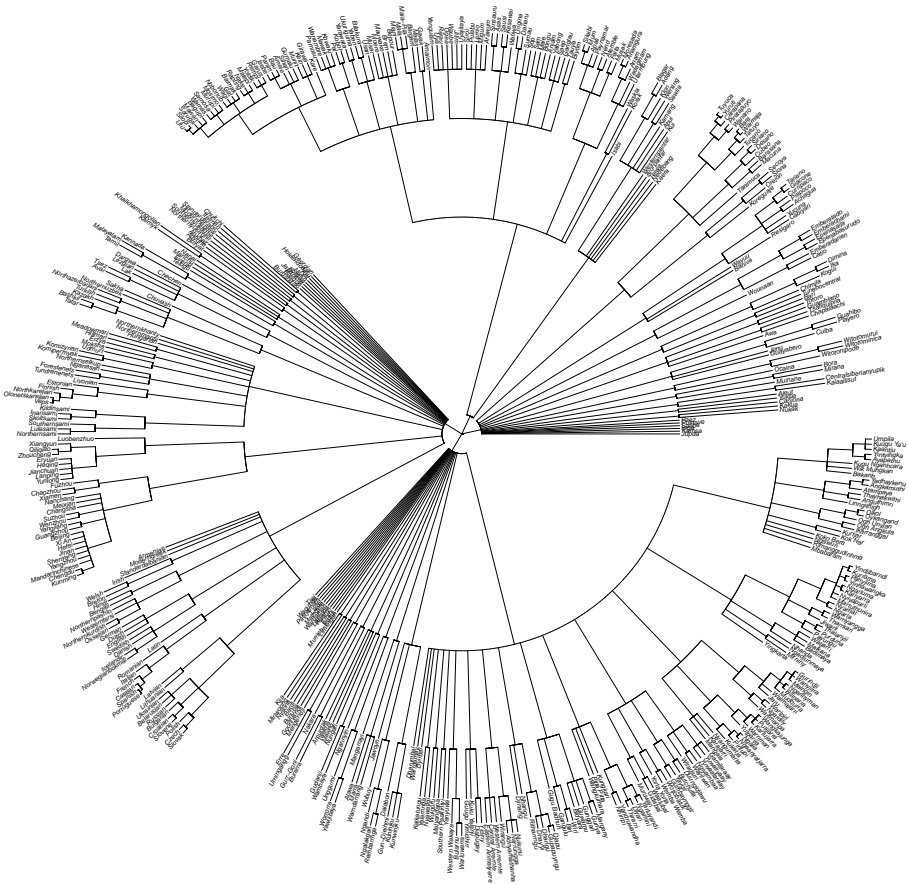

Figure 1: Supertree of 496 languages used in Yin (2020).

has a column `tip` and two columns of numerical data `has_onset_violation` and `has_coda_violation`, and thus it meets the requirements we need. It also contains a column, `names`, of non-numeric data. Columns of non-numeric data (other than `tip`) are ignored by `phylo_average()`, so we do not need to remove them.

## 5.3 Calculating genealogically-sensitive proportions

The results are calculated using `phylo_average()`, setting its `phy` argument to the tree we have constructed, `yin_2020_tree`, and its `data` argument to the dataframe we have prepared, `yin_2020_data`. A warning is issued altering us that the dataframe contains a non-numeric column that gets ignored:

```
yin_2020_results <- phylo_average(phy = yin_2020_tree, data = yin_2020_data)
```

```
## Warning in phylo_average(phy = yin_2020_tree, data = yin_2020_data): `data` contains
## non-numeric columns other than `tip`, which have been ignored: name.
```

Results are in the format described in Section 3. The first ten rows of phylogenetic weights according to the ACL and BM methods are:

```
head(yin_2020_results$ACL_weights, n = 10)
```

```
##                 name       tip        tree1
## 1             Abkhaz  abkh1244 0.0069858330
## 2               Abui  abui1241 0.0012637162
## 3            Achagua  acha1250 0.0002456679
## 4              Adang  adan1251 0.0002197767
## 5       Adnyamathanha adny1235 0.0000637717
## 6             Adyghe  adyg1241 0.0069858330
## 7       Hokkaidoainu  ainu1240 0.0174645825
## 8              Alawa  alaw1244 0.0019652742
## 9    Standardalbanian alba1267 0.0023699724
## 10             Aleut  aleu1260 0.0085266857
```

```
head(yin_2020_results$BM_weights, n = 10)
```

```
##                 name       tip        tree1
## 1             Abkhaz  abkh1244 0.0049475265
## 2               Abui  abui1241 0.0021424363
## 3            Achagua  acha1250 0.0009109899
## 4              Adang  adan1251 0.0009391784
## 5       Adnyamathanha adny1235 0.0008929209
## 6             Adyghe  adyg1241 0.0049475265
## 7       Hokkaidoainu  ainu1240 0.0064304759
## 8              Alawa  alaw1244 0.0027423594
## 9    Standardalbanian alba1267 0.0040995307
## 10             Aleut  aleu1260 0.0055527679
```

The genealogically-sensitive proportions according to the ACL and BM methods are the following. Recall from the main paper that ACL proportions are more sensitive to languages on outlier branches, and the BM proportions are less so:

```
yin_2020_results$ACL_averages
```

```
##    tree has_onset_violation has_coda_violation
## 1 tree1           0.3711102           0.409806
```

```
yin_2020_results$BM_averages
```

```
##    tree has_onset_violation has_coda_violation
## 1 tree1           0.4014756          0.3847729
```

As a point of comparison, the raw proportions, which are equal to the means of the columns `has_onset_violation` and `has_coda_violation`, are these:

```
mean(yin_2020_data$has_onset_violation)
```

```
## [1] 0.3649194
```

```
mean(yin_2020_data$has_coda_violation)
```

```
## [1] 0.3145161
```

The estimated proportion of languages with onset violations is similar using both the AM and BM methods. The raw proportion is also similar, with all three results in the range of 36–40%. Turning to languages with coda violations, the estimated proportions are also similar using both the AM (41%) and BM (38%) methods, but the raw proportion (31%) is notably lower, illustrating how our understanding of the commonness of typological phenomena can shift once we take genealogy into account.

As we seek to analyse the empirical diversity of attested languages, genealogy must be part of picture. Since the genealogies of human languages are still incompletely known, it is imperative to make our phylogenetic assumptions as explicit and as testable as possible. In this document, we hope have shown that doing so is not only worthwhile, but also feasible and attainable.

# 6 Using these methods in typological research

We hope that typologists find the arguments in our main paper compelling at the conceptual level and the methods in this supplementary document convenient at the practical level.

If you are interested in trying out the methods described here, please check for updates and additional information at their web pages, https://github.com/erichround/glottoTrees and https://github.com/erichround/phyloWeights.

Through the use of *glottoTrees* and *phyloWeights*, we hope that in the future, linguistic trees and the code used to produce them can be published together with typological studies. This will enable subsequent researchers to replicate the study's findings, and just as importantly, to modify its assumptions by modifying the trees, and thereby to test further hypotheses inspired by the initial research.

If you find these tools useful in your own research, please cite the packages as Round (2021a) and Round (2021b).

# References

Anderson, Cormac et al. 2018. CLICS2: an improved database of cross-linguistic colexifications assembling lexical data with the help of cross-linguistic data formats. *Linguistic Typology* 22(2). 277–306.

Felsenstein, Joseph. N.d. *The Newick tree format.* http://evolution.genetics.washington.edu/phylip/newicktree.html.

Green, Rebecca. 2003. Proto Maningrida with proto Arnhem: evidence from verbal inflectional suffixes. In Nicholas Evans (ed.), *The non-Pama-Nyungan languages of northern Australia : comparative studies of the continent's most linguistically complex region*, 369–421. Pacific Linguistics.

Hammarström, Harald et al. 2021. *Glottolog 4.4.* Leipzig. https://doi.org/10.5281/zenodo.4761960. https://glottolog.org/.

Paradis, Emmanuel & Klaus Schliep. 2018. Ape 5.0: an environment for modern phylogenetics and evolutionary analyses in R. *Bioinformatics* 35. 526–528.

Round, Erich R. 2017. The AusPhon-Lexicon project: 2 million normalized segments across 300 Australian languages. 47th Poznań Linguistic Meeting.

Round, Erich R. 2021a. *glottoTrees: Phylogenetic trees in Linguistics.* R package version 0.1. https://github.com/erichround/glottoTrees.

Round, Erich R. 2021b. *phyloWeights: Calculation of Genealogically-sensitive Proportions and Averages*. R package version 0.3. https://github.com/erichround/phyloWeights.

Venditti, Chris, Andrew Meade & Mark Pagel. 2010. Phylogenies reveal new interpretation of speciation and the red queen. *Nature* 463(7279). 349–352.

Wickham, Hadley et al. 2021. *dplyr: A Grammar of Data Manipulation*. R package version 1.0.5. https://CRAN.R-project.org/package=dplyr.

Yin, Ruihua. 2020. Violations of the sonority sequencing principle: How, and how often? 2020 Conference of the Australian Linguistic Society. https://als.asn.au/Conference/Program.

Supplementary Information, Section S2: Data, sources and code

## Contents

Here we document raw data, its sources and the code used to calculate results in the main paper. Data files are archived at https://doi.org/10.5281/zenodo.5602216. Code is written in R (R Core Team 2021).

# 1 Pama-Nyungan laminal contrast typological data

## 1.1 Discrete variable: presence/absence of contrast

In Section 6.1 we analyse the presence/absence of a binary phonemic laminal contrast in 216 Pama-Nyungan languages. Data is from Round (2019). Languages with the contrast are in (1) and without are in (2). Eight-character glottocodes are cited with the language names.

|  | **Languages** |
|---|---|
| (1) | Adnyamathanha adny1235, Aghu Tharnggala aghu1254, Alngith alng1239, Alyawarr alya1239, Anguthimri angu1242, Antekerrepenhe ande1247, Arabana arab1267, Ayapathu ayab1239, Badimaya badi1246, Bakanh paka1251, Barna biri1256, Bidyara bidy1243, Biri biri1256, Bularnu bula1255, Buwandik bung1264, Darkinyung hawk1239, Dhangu dhan1270, Dharawal thur1254, Dharawala bidy1243, Dharuk sydn1236, Dhay'yi dhal1246, Dhudhuroa dhud1236, Dhurga dhur1239, Diyari dier1241, Djambarrpuyngu djam1256, Djapu djap1238, Eastern Arrernte east2379, Eora sydn1236, Flinders Island flin1247, Gamilaraay gami1243, Gangulu gang1268, Garlali kala1380, Gudjal gudj1237, Gugu Badhun gugu1253, Gumatj guma1253, Gundungurra gund1248, Gunggari kung1258, Gunya guny1241, Gupapuyngu gupa1247, Guugu Yimidhirr gugu1255, Guwa guwa1242, Guwamu guwa1243, Ikarranggal ikar1243, Jiwarli djiw1241, Kaantju kanj1260, Kala Kawaw Ya kala1378, Kala Lagaw Ya kala1377, Kalkatungu kalk1246, Kariyarra kari1304, Kaurna kaur1267, Kaytetye kayt1238, Keramin kera1256, Kok Nar kokn1236, Koko Bera gugu1254, Kugu Nganhcara wikn1246, Kukatj guga1239, Kungardutji wong1246, Kungkari kuun1236, Kurnu kula1275, Kurrama kurr1243, Kurtjar gurd1238, Kuugu Ya'u kuuk1238, Linngithigh leni1238, Malkana malgl1242, Malthanmungu barr1247, Malyangapa maly1234, Margany marg1253, Martuthunira mart1255, Mayi-Kulan mayk1239, Mayi-Kutuna maya1280, Mayi-Thakurti ngaw1240, Mayi-Yapi mayk1239, Mbabaram mbab1239, Mbiywom mbiy1238, Mirniny mirn1243, Mithaka mith1236, Mpalitjanh mpal1238, Muruwari muru1266, Narrungga naru1238, Ndra'ngith tyan1235, Ngadjunmaya ngad1258, Ngamini ngam1284, Ngarigu ngar1297, Ngarluma ngar1287, Ngawun ngaw1240, Ngayawang uppe1415, Ngiyambaa wang1291, Ngkoth ngko1236, Ngunawal ngun1277, Nhanda nhan1238, Nhangu yann1237, Nhirrpi nhir1234, Nukunu nugu1241, Nyiyaparli nija1241, Ogh Unyjan kawa1290, Olkol ulku1238, Oykangand oyka1239, Pallanganmiddang pall1243, Panyjima pany1241, Parnkalla bang1339, Payungu bayu1240, Piangil wadi1260, Pirriya pirr1240, Pitta Pitta pitt1247, Punthamara punt1240, Purduna burd1238, Ritharrngu rita1239, Southern Paakintyi wilj1239, Takalak taga1279, Thaayorre thay1249, Thalanyji dhal1245, Tharrkari dhar1247, Thaynakwithi tyan1235, Umpila umpi1239, Umpithamu umbi1243, Uradhi urad1238, Wadikali wadi1261, Walangama wala1263, Wangkangurru wang1290, Wangkayutyuru wang1289, Wangkumara wong1246, Warluwarra warl1256, Warriyangga wari1262, Warrnambool warr1257, Wathawurrung wath1238, Wathi Wathi wadi1260, Watjarri waja1257, Wayilwan wayi1238, Wemba Wemba wemb1241, Western Anmatyerre west2442, Western Arrernte west2441, Western Wakaya waga1260, Wik Mungkan wikm1247, Wiradjuri wira1262, Wirangu wira1265, Woiwurrung woiw1238, Wulguru wulg1239, Yabula-Yabula yabu1234, Yadhaykenu yadh1237, Yalarnnga yala1262, Yambina biri1256, Yanda yand1251, Yandruwandha yand1253, Yanyuwa yany1243, Yaraldi narr1259, Yardliyawarra yarl1236, Yarluyandi yarl1238, Yawarrawarrka yawa1258, Yindjibarndi yind1247, Yingkarta ying1247, Yinhawangka yinh1234, Yiningay bidy1243, Yinwum yinw1236, Yir Yoront yiry1245, Yitha Yitha lowe1403, Yorta Yorta yort1237, Yuwaalaraay yuwa1242 |
| (2) | Awabakal awab1243, Bilinarra bili1250, Bundjalung yugu1249, Butchulla baty1234, Dharumbal dhar1248, Djabugay dyaa1242, Djabwurrung djab1234, Djinang djin1253, Djinba djin1252, Duungidjawu duun1241, Dyirbal dyir1250, Gidabal gida1240, Gumbaynggir kumb1268, Gureng-gureng gure1255, Gurindji guri1247, Guwar guwa1244, Jaru jaru1254, Kaniyang kani1276, Karajarri kara1476, Kartujarra kart1247, Katthang wori1245, Kolakngat cola1237, Kukatja kuka1246, Kuku Yalanji kuku1273, Madhi-Madhi madh1244, Malngin maln1239, Mangala mang1383, Manjiljarra many1256, Minjungbal minj1242, Mudburra mudb1240, Ngaanyatjarra ngaa1240, Nganyaywana ngan1296, Ngardily west2437, Ngarinyman ngar1235, Ngarla ngar1296, Nyamal nyam1271, Nyangumarta nyan1301, Nyawaygi nyaw1247, Pintupi pint1250, Pitjantjatjara pitj1243, Thanggati dyan1250, Turubul yaga1256, Waalubal band1358, Wajuk nyun1247, Waka Waka waka1274, Walmajarri walm1241, Wangkajunga wang1288, Wangkatja pini1245, Wardandi ward1248, Warlmanpa warl1255, Warlpiri warl1254, Warnman wanm1242, Warrgamay warr1255, Warumungu waru1265, Warungu waru1264, Yagara yaga1262, Yaygir yayg1236, Yidiny yidi1250, Yulparija yulp1239 |

2

## 1.2 Continuous variable: predominance of pre-palatal in nine phonotactic positions

In Section 6.2 we analyse the relative predominance of pre-palatals versus dentals in nine phonotactic contexts in 76 Pama-Nyungan languages. The proportions of pre-palatals in the nine contexts is shown below. NA indicates that the lexicon contained fewer than 20 consonantal laminals in the relevant phonotactic context, in which case the langauge was excluded for that variable.

| Variety | Glottocode | V_a | V_i | V_u | a_V | i_V | u_V | #_a | #_i | #_u |
|---|---|---|---|---|---|---|---|---|---|---|
| Adnyamathanha | adny1235 | 0.2601 | 0.5909 | 0.5217 | 0.2831 | 0.4815 | 0.3538 | 0.0000 | 0.0000 | 0.2400 |
| Anguthimri | angu1242 | NA | 0.2292 | NA | 0.1795 | NA | NA | 0.2727 | NA | NA |
| Badimaya | badi1246 | 0.7303 | NA | NA | 0.7761 | 0.7692 | 0.5333 | 0.2791 | 0.8966 | 0.4884 |
| Bakanh | paka1251 | 0.4333 | 0.8298 | NA | 0.6667 | NA | 0.6857 | 0.0000 | NA | 0.0000 |
| Bidyara | bidy1243 | 0.4030 | 0.4815 | 0.4400 | 0.4681 | 0.3913 | 0.4082 | 0.0189 | 0.0714 | 0.0682 |
| Biri | biri1256 | 0.5000 | NA | NA | 0.3714 | 0.6000 | 0.6071 | 0.0800 | NA | NA |
| Bularnu | bula1255 | 0.3333 | 0.9467 | 0.2308 | 0.5568 | 0.8036 | 0.2632 | 0.1250 | 0.9000 | 0.2000 |
| Dhangu | dhan1270 | NA | NA | NA | 0.4000 | NA | NA | 0.3261 | 0.7436 | 0.2429 |
| Dhay'yi | dhal1246 | 0.4677 | 0.6000 | 0.7115 | 0.4189 | 0.7708 | 0.6552 | 0.2883 | 0.6471 | 0.3978 |
| Diyari | dier1241 | 0.2703 | 0.5000 | NA | 0.3235 | NA | 0.5000 | 0.1481 | 0.1923 | 0.2381 |
| Djapu | djap1238 | 0.3922 | 0.7000 | 0.3784 | 0.2326 | 0.8615 | 0.4655 | 0.4518 | 0.4237 | 0.2621 |
| Gamilaraay | gami1243 | 0.4118 | 0.3125 | 0.0645 | 0.0704 | 0.8824 | 0.0476 | 0.0000 | 0.0930 | 0.0000 |
| Gangulu | gang1268 | 0.2963 | NA | NA | 0.3182 | NA | NA | 0.1364 | NA | NA |
| Gugu Badhun | gugu1253 | NA | NA | 0.2000 | 0.3636 | NA | NA | 0.2162 | 0.7500 | 0.2174 |
| Gunya | guny1241 | 0.3077 | 0.2647 | NA | 0.2000 | 0.5417 | 0.2812 | 0.0000 | 0.0857 | 0.0270 |
| Gupapuyngu | gupa1247 | 0.4930 | 0.9355 | 0.4615 | 0.4902 | 0.9730 | 0.4048 | 0.3970 | 0.7422 | 0.4294 |
| Guugu Yimidhirr | gugu1255 | NA | NA | NA | 0.1000 | NA | NA | NA | NA | NA |
| Guwamu | guwa1243 | 0.3404 | NA | NA | 0.3421 | NA | 0.2308 | 0.0244 | 0.1724 | 0.0385 |
| Jiwarli | djiw1241 | 0.6562 | 0.8108 | 0.6557 | 0.7105 | 0.8125 | 0.6061 | 0.4891 | 0.8750 | 0.4375 |
| Kalkatungu | kalk1246 | 0.2811 | 0.4876 | 0.5217 | 0.3684 | 0.4464 | 0.3750 | 0.2986 | 0.6163 | 0.4048 |
| Kariyarra | kari1304 | 0.3016 | 0.9600 | NA | 0.6341 | 0.4054 | 0.3913 | 0.3571 | 0.9130 | NA |
| Kok Nar | kokn1236 | 0.1739 | NA | NA | NA | NA | NA | 0.0000 | NA | NA |
| Koko Bera | gugu1254 | 0.3571 | NA | NA | 0.3636 | 0.7391 | NA | 0.0182 | NA | NA |
| Kugu Nganhcara | wikn1246 | NA | NA | NA | 0.2857 | NA | NA | 0.0377 | NA | 0.0000 |
| Kukatj | guga1239 | 0.5641 | 0.7500 | NA | 0.5455 | 0.6538 | 0.4000 | 0.0308 | 0.5714 | 0.1600 |
| Kurrama | kurr1243 | 0.4627 | 0.6667 | 0.4595 | 0.4444 | 0.7188 | 0.3939 | 0.5072 | 0.9455 | 0.5000 |
| Kurtjar | gurd1238 | 0.2692 | 0.4688 | 0.5000 | 0.4107 | NA | 0.4000 | 0.3025 | NA | 0.3636 |
| Kuugu Ya'u | kuuk1238 | 0.4444 | 0.6429 | 0.5873 | 0.4672 | 0.7576 | 0.5072 | 0.1774 | 0.6800 | 0.3529 |
| Linngithigh | leni1238 | 0.3500 | 0.0667 | NA | 0.2143 | 0.1389 | NA | 0.5106 | NA | NA |
| Malkana | malg1242 | 0.2667 | 0.9538 | NA | 0.5818 | 0.8276 | 0.2414 | 0.3333 | NA | NA |
| Malyangapa | maly1234 | 0.4074 | NA | NA | 0.4348 | NA | NA | 0.0000 | NA | NA |
| Margany | marg1253 | 0.4643 | NA | NA | NA | NA | NA | 0.0323 | 0.0417 | NA |
| Martuthunira | mart1255 | 0.4248 | 0.8298 | 0.5326 | 0.6042 | 0.5972 | 0.2973 | 0.5182 | 0.9437 | 0.3494 |
| Mirniny | mirn1243 | 0.4259 | NA | NA | 0.4091 | NA | NA | 0.0870 | NA | 0.3478 |
| Muruwari | muru1266 | 0.3409 | 0.6538 | NA | 0.3770 | 0.8750 | 0.1087 | 0.0145 | 0.0566 | 0.0282 |
| Ngadjunmaya | ngad1258 | 0.9133 | 1.0000 | 1.0000 | 0.9292 | 0.9737 | 0.9091 | 0.9412 | 1.0000 | 0.9857 |
| Ngamini | ngam1284 | 0.7538 | 0.9167 | NA | 0.8281 | 0.8824 | NA | NA | NA | NA |
| Ngarluma | ngar1287 | 0.4545 | 0.9412 | 0.5286 | 0.5597 | 0.8158 | 0.4730 | 0.3740 | 0.9857 | 0.4000 |
| Ngawun | ngaw1240 | 0.5833 | NA | NA | 0.5909 | NA | 0.4500 | 0.3000 | NA | NA |
| Nhanda | nhan1238 | 0.7347 | 0.7778 | NA | 0.8286 | 0.9143 | 0.5200 | NA | NA | NA |
| Nhangu | yann1237 | 0.2540 | 1.0000 | NA | 0.4324 | 0.7750 | 0.3438 | 0.2115 | 0.8088 | 0.3370 |
| Nhirrpi | nhir1234 | 0.3182 | NA | NA | NA | NA | NA | NA | NA | NA |
| Nukunu | nugu1241 | 0.5435 | NA | NA | 0.4872 | 0.7500 | NA | 0.0435 | NA | NA |
| Ogh Unyjan | kawa1290 | 0.6364 | NA | NA | 0.5000 | NA | NA | NA | NA | NA |
| Olkol | ulku1238 | 0.3768 | 0.7879 | 0.5185 | 0.5507 | 0.4375 | NA | NA | NA | NA |
| Oykangand | oyka1239 | 0.4510 | 0.7692 | NA | 0.4571 | 0.5000 | 0.8333 | NA | NA | NA |
| Panyjima | pany1241 | 0.4110 | 1.0000 | 0.6452 | 0.5781 | 0.7500 | 0.4737 | 0.5745 | 0.9091 | 0.5862 |
| Payungu | bayu1240 | 0.5360 | 0.8246 | 0.5510 | 0.6211 | 0.6545 | 0.5500 | 0.4352 | 0.8364 | 0.4412 |
| Pitta Pitta | pitt1247 | 0.5200 | 0.6538 | 0.6364 | 0.5091 | 0.7568 | 0.5625 | 0.2653 | 0.3617 | NA |
| Purduna | burd1238 | 0.6897 | 0.7978 | 0.6410 | 0.7206 | 0.8088 | 0.6556 | 0.5138 | 0.8298 | 0.6809 |
| Ritharrngu | rita1239 | 0.5833 | 0.9524 | NA | 0.5476 | 1.0000 | NA | 0.4946 | 0.8025 | 0.3585 |
| Southern Paakintyi | wilj1239 | 0.7180 | 0.4756 | 0.4310 | 0.6082 | 0.6811 | 0.5909 | 0.0076 | 0.0100 | 0.0345 |
| Thaayorre | thay1249 | 0.1739 | 0.4140 | 0.1250 | 0.3981 | 0.0494 | 0.0986 | 0.0059 | 0.0390 | 0.0076 |
| Thalanyji | dhal1245 | 0.5287 | 0.8810 | 0.6486 | 0.6750 | 0.6389 | 0.6000 | 0.5556 | 0.8750 | 0.5208 |
| Tharrkari | dhar1247 | 0.6842 | NA | 0.7619 | 0.7500 | NA | 0.6286 | 0.4314 | 0.8065 | 0.2955 |

| | | | | | | | | | | |
|---|---|---|---|---|---|---|---|---|---|---|
| Umpila | umpi1239 | 0.3981 | 0.6887 | 0.7302 | 0.5115 | 0.6750 | 0.6094 | 0.0833 | 0.3056 | 0.3871 |
| Wangkumara | wong1246 | 0.6824 | 0.5581 | 0.4000 | 0.6308 | 0.7736 | 0.2667 | 0.0159 | 0.2167 | 0.1818 |
| Warluwarra | warl1256 | 0.4156 | 0.8889 | NA | 0.4074 | 0.7045 | 0.2857 | 0.2273 | 0.7727 | 0.3158 |
| Warriyangga | wari1262 | 0.6346 | 0.6970 | 0.6071 | 0.6275 | 0.7000 | 0.6061 | 0.4286 | 0.8182 | 0.4242 |
| Watjarri | waja1257 | 0.5238 | 0.9368 | 0.7671 | 0.6500 | 0.8100 | 0.6545 | 0.5169 | 0.9114 | 0.6000 |
| Wayilwan | wayi1238 | 0.4615 | 0.5625 | NA | 0.1667 | 0.9400 | 0.1739 | 0.0000 | 0.0000 | 0.0488 |
| Wemba Wemba | wemb1241 | 0.5500 | 0.6765 | 0.8889 | 0.5875 | 0.7000 | 0.5745 | 0.8876 | 1.0000 | 0.9459 |
| Western Arrernte | west2441 | NA | NA | NA | 0.4762 | 1.0000 | NA | NA | NA | NA |
| Western Wakaya | waga1260 | 0.6140 | 0.7927 | 0.5818 | 0.6316 | 0.7822 | 0.5800 | 0.2000 | 0.9423 | 0.3600 |
| Wik Mungkan | wikm1247 | 0.5722 | 0.2286 | NA | 0.2937 | 0.7818 | 0.6207 | 0.0878 | 0.4744 | 0.1061 |
| Wirangu | wira1265 | 0.4889 | 0.8889 | NA | 0.6667 | 0.6970 | NA | 0.2222 | 0.9677 | 0.7667 |
| Yadhaykenu | yadh1237 | 0.2500 | 0.3784 | 0.4186 | 0.2941 | 0.2432 | 0.4182 | NA | NA | NA |
| Yalarnnga | yala1262 | 0.4821 | 0.6909 | 0.5769 | 0.5970 | 0.6842 | 0.4375 | 0.1639 | 0.3824 | 0.2759 |
| Yandruwandha | yand1253 | 0.4921 | 0.5946 | NA | 0.4681 | 0.6207 | 0.5429 | 0.1064 | 0.1818 | 0.2712 |
| Yanyuwa | yany1243 | 0.5864 | 0.9155 | 0.6081 | 0.6386 | 0.9557 | 0.6063 | 0.8030 | 1.0000 | 0.5467 |
| Yarluyandi | yarl1238 | 0.4062 | 0.7407 | NA | 0.5161 | 0.4615 | NA | 0.0000 | 0.1923 | NA |
| Yindjibarndi | yind1247 | 0.5833 | 0.8065 | 0.6667 | 0.6508 | 0.7273 | 0.6000 | 0.5000 | 0.9726 | 0.3380 |
| Yinhawangka | yinh1234 | 0.3264 | 0.8375 | 0.6392 | 0.4724 | 0.5546 | 0.4000 | 0.5441 | 0.9783 | 0.4878 |
| Yir Yoront | yiry1245 | 0.1818 | NA | NA | 0.3200 | 0.1739 | NA | 0.0374 | 0.1132 | 0.1489 |
| Yorta Yorta | yort1237 | 0.6875 | NA | NA | 0.5333 | 0.8788 | 0.6364 | 0.1304 | NA | NA |
| Yuwaalaraay | yuwa1242 | 0.3780 | 0.5070 | 0.2500 | 0.1348 | 0.8866 | 0.0877 | 0.0000 | 0.0261 | 0.0000 |

## 1.3 Wordlist sources

The 76 lexical wordlists used in Section 6.2 study are from the Ausphonlex database, under development by Round (2017), also described in Macklin-Cordes, Bowern & Round (2021). All underlying wordlist data is available, either publicly in the CHIRILA database (Bowern 2016) or elsewhere in published or archived form. A list of original and CHIRILA sources for the wordlists is presented below.

**Adnyamathanha** CHIRILA/v2/McEnteeMcKenzie

John McEntee & Pearl McKenzie. 1992. *Adna-mat-na English dictionary.* Adelaide: the authors. 125 pp.

**Anguthimri** CHIRILA/v1/ASEDA0240

Terry Crowley. 1989. Mbakwithi vocabulary. Australian Institute of Aboriginal and Torres Strait Islander Studies, Australian Indigenous Languages Collection. ASEDA 0240. Canberra

**Badimaya**

Doug Marmion. 1995. Badimaya dictionary. Australian Institute of Aboriginal and Torres Strait Islander Studies, Australian Indigenous Languages Collection. ASEDA 0615. Canberra

**Bakanh**

Philip J. Hamilton. 1997a. *Pakanh Alphabetical Search Index.* Oykangand and Olkola Dictionary. http://www.oocities.org/athens/delphi/2970/pakalpha.htm

**Bidyara**

Gavan Breen. 1973. *Bidyara and Gungabula grammar and vocabulary.* Vol. 8 (Linguistic Communications). Melbourne: Monash University. 227 pp.

**Biri** CHIRILA/v1/Terrell

Angela Terrill. 1999. Biri lexicons. Australian Institute of Aboriginal and Torres Strait Islander Studies, Australian Indigenous Languages Collection. ASEDA 0700. Canberra. http://aiatsis.gov.au/sites/default/files/catalogue_resources/0700_access.zip

**Bularnu**

Gavan Breen. 1988. Bularnu grammar and vocabulary machine-readable files. Australian Institute of Aboriginal and Torres Strait Islander Studies, Australian Indigenous Languages Collection. ASEDA 0007. Canberra. http://aiatsis.gov.au/sites/default/files/catalogue_resources/0007_access.zip

**Dhangu**

R. David Zorc. 2004. Yolngu Matha dictionary. Australian Institute of Aboriginal and Torres Strait Islander Studies, Australian Indigenous Languages Collection. ASEDA 0778. Canberra. http://aiatsis.gov.au/sites/default/files/catalogue_resources/0778_Access.zip

**Dhay'yi**

Djarrayang Wunungmurra. 1993. Dhalwangu dictionary. Australian Institute of Aboriginal and Torres Strait Islander Studies, Australian Indigenous Languages Collection. ASEDA 0502. Canberra

**Diyari**

Peter K. Austin. 1981. *A grammar of Diyari, South Australia* (Cambridge Studies in Linguistics 32). Cambridge; New York: Cambridge University Press. 269 pp.

**Djapu** CHIRILA/v1/mor83

Frances Morphy. 1983. Djapu, a Yolngu dialect. In R. M. W. Dixon & Barry Blake (eds.), *Handbook of Australian languages*, vol. 3, 5 vols., 1–188. Amsterdam: John Benjamins

**Gamilaraay** CHIRILA/v1/ash03

Anna Ash, John Giacon & Amanda Lissarrague. 2003. *Gamilaraay, Yuwaalaraay & Yuwaalayaay dictionary*. Alice Springs, NT, Australia: IAD Press. 344 pp.

**Gangulu** CHIRILA/v1/Terrell

Angela Terrill. 1999. Biri lexicons. Australian Institute of Aboriginal and Torres Strait Islander Studies, Australian Indigenous Languages Collection. ASEDA 0700. Canberra. http://aiatsis.gov.au/sites/default/files/catalogue_resources/0700_access.zip

**Gugu Badhun** CHIRILA/v1/sut73

Peter J. Sutton. 1973. Gugu-Badhun and its neighbours. In *Gugu-Badhun and its neighbours: a linguistic salvage study*, 24–67. Sydney: Macquarie University

**Gunya** CHIRILA/v1/dixbla81

Gavan Breen. 1981a. Margany and Gunya. In R. M. W. Dixon & Barry Blake (eds.), *Handbook of Australian languages*, vol. 2, 275–394. Amsterdam: John Benjamins

**Gupapuyngu** CHIRILA/v1/BL

Beulah Lowe & Beulah Lowe. 1976. Temporary Gupapuyngu dictionary. Milingimbi, NT, Australia

**Guugu Yimidhirr**

John B. Haviland. 1979. Guugu Yimidhirr. In R. M. W. Dixon & Barry Blake (eds.), *Handbook of Australian languages*, vol. 1, 5 vols., 26–180. Amsterdam: John Benjamins

**Guwamu** CHIRILA/v1/Austin 1980

Peter K. Austin. 1980. Guwamu vocabulary and English-Guwamu finder list. Cambridge, MA

**Jiwarli** CHIRILA/v2/ASEDA0435

Peter K. Austin. N.d.(a). A dictionary of Jiwarli. Australian Institute of Aboriginal and Torres Strait Islander Studies, Australian Indigenous Languages Collection. ASEDA 0435. Canberra

**Kalkatungu** CHIRILA/v2/ASEDA0205

Barry J. Blake. 1990a. Kalkatungu vocabulary. Australian Institute of Aboriginal and Torres Strait Islander Studies, Australian Indigenous Languages Collection. ASEDA 0205. Canberra

**Kariyarra**

Sue Smythe & Manny Lockyer. N.d. Kariyarra wordlist. Australian Institute of Aboriginal and Torres Strait Islander Studies, Australian Indigenous Languages Collection. ASEDA 0582. Canberra. http://aiatsis.gov.au/sites/default/files/catalogue_resources/0582_access.zip

**Kok Nar**

Bruce A. Sommer. N.d.(a). Koko Narr. Fryer Library Bruce Sommer Collection. UQFL476_b10f03_64, UQFL476_b10f03_65. Brisbane

**Koko Bera**

Paul D. Black & Kokoberrin Tribal Aboriginal Corporation. 2007. The Kokoberrin and their languages. Australian Institute of Aboriginal and Torres Strait Islander Studies, Australian Indigenous Languages Collection

**Kugu Nganhcara** CHIRILA/v1/ASEDA0021

Ian Smith & Steve Johnson. 1989. Kugu Nganchara. Australian Institute of Aboriginal and Torres Strait Islander Studies, Australian Indigenous Languages Collection. ASEDA 0021. Canberra

**Kukatj**

Gavan Breen. 1991. Kukatj grammar machine-readable files. Australian Institute of Aboriginal and Torres Strait Islander Studies, Australian Indigenous Languages Collection. ASEDA 0022. Canberra. http://aiatsis.gov.au/sites/default/files/catalogue_resources/0022_access.zip

**Kurrama**

Alan C. Dench. N.d. Kurrama. Australian Institute of Aboriginal and Torres Strait Islander Studies, Australian Indigenous Languages Collection. ASEDA 0481. Canberra. http://aiatsis.gov.au/sites/default/files/catalogue_resources/0481_access.zip

**Kurtjar** CHIRILA/v1/ASEDA0026

Paul D. Black & Rolly Gilbert. 1988. Kurtjar dictionary. Australian Institute of Aboriginal and Torres Strait Islander Studies, Australian Indigenous Languages Collection. ASEDA 0026. Canberra

**Kuugu Yau**

David A. Thompson. 1988. "sand beach" language: an outline of Kuuku Ya'u and Umpila. Australian Institute of Aboriginal and Torres Strait Islander Studies, Australian Indigenous Languages Collection. ASEDA 0027. Canberra

**Linngithigh**

Kenneth Hale. 1999. A Linngithigh vocabulary. Australian Institute of Aboriginal and Torres Strait Islander Studies, Australian Indigenous Languages Collection. ASEDA 0687. Canberra

**Malkana**

Andrew Gargett. 2011. *A salvage grammar of Malgana, the language of Shark Bay, Western Australia* (Pacific Linguistics 624). Canberra: Pacific Linguistics. 102 pp. https://doi.org/10.15144/PL-624

**Malyangapa**

Luise A. Hercus. 1989. Maljangapa-Wadigali vocabulary. Australian Institute of Aboriginal and Torres Strait Islander Studies, Australian Indigenous Languages Collection. ASEDA 0246. Canberra. http://aiatsis.gov.au/sites/default/files/catalogue_resources/0246_access.zip

**Margany** CHIRILA/v1/bre81

Gavan Breen. 1981a. Margany and Gunya. In R. M. W. Dixon & Barry Blake (eds.), *Handbook of Australian languages*, vol. 2, 275–394. Amsterdam: John Benjamins

**Martuthunira**

Alan C. Dench. 1995. *Martuthunira, a language of the Pilbara region of Western Australia* (Pacific Linguistics Series C 125). Canberra: Pacific Linguistics. 406 pp. https://doi.org/10.15144/PL-C125

**Mbakwithi** CHIRILA/v1/ASEDA0240

Terry Crowley. 1989. Mbakwithi vocabulary. Australian Institute of Aboriginal and Torres Strait Islander Studies, Australian Indigenous Languages Collection. ASEDA 0240. Canberra

**Mirniny**

Geoffrey N. O'Grady & Edward M. Curr. 1988. Mirniny wordlist. Australian Institute of Aboriginal and Torres Strait Islander Studies, Australian Indigenous Languages Collection. ASEDA 0070. Canberra. http://aiatsis.gov.au/sites/default/files/catalogue_resources/0070_access.zip

**Muruwari** CHIRILA/v1/ASEDA0252

Lynette Frances Oates. 1992. *Muruwari (Moo-roo-warri) dictionary: Words of an Aboriginal language of north-western New South Wales*. Albury, NSW, Australia: Graeme van Brummelen, produced with the assistance of the Australian Institute of Aboriginal & Torres Strait Islander Studies. 97 pp.

**Ngadjunmaya**

Wangka Maya Pilbara Aboriginal Language Centre. 2008. *Ngajumaya dictionary*. South Hedland, WA, Australia: Wangka Maya Pilbara Aboriginal Language Centre. 16 pp.

**Ngamini** CHIRILA/v1/brendn

Gavan Breen. 1967. Ngamini material. Australian Institute of Aboriginal and Torres Strait Islander Studies, Australian Indigenous Languages Collection. MS 117. Canberra

**Ngarluma**

Kenneth Hale. 1989. Ngarluma wordlist. Australian Institute of Aboriginal and Torres Strait Islander Studies, Australian Indigenous Languages Collection. ASEDA 0037. Canberra. http://aiatsis.gov.au/sites/default/files/catalogue_resources/0037_access.zip

**Ngawun** CHIRILA/v2/BreenMayi

Gavan Breen. 1981b. *The Mayi languages of the Queensland Gulf Country* (A.I.A.S. New Series 29). Canberra: Australian Institute of Aboriginal Studies. 238 pp.

**Nhanda** CHIRILA/v1/ble01

Juliette Blevins. 2001. *Nhanda: An Aboriginal language of Western Australia* (Oceanic linguistics special publication 30). Honolulu: University of Hawai'i Press. 170 pp.

**Nhirrpi** CHIRILA/v1/bow-nhi

Claire Bowern. 1999. Nhirrpi vocabulary, based on fieldnotes of S. A. Wurm

**Nukunu** CHIRILA/v2/her92

Luise A. Hercus. 1992a. *A Nukunu dictionary*. Canberra: Department of Linguistics, Australian National University. 51 pp.

**Ogh Unyjan**

Bruce A. Sommer. N.d.(b). Ogh Unydjan. Fryer Library Bruce Sommer Collection. UQFL476_b09f03_s05. Brisbane

**Olkol**

Philip J. Hamilton. 1997b. *Uw Olkola and Uw Oykangand Alphabetical Search Index*. Oykangand and Olkola Multimedia Dictionary. http://www.oocities.org/athens/delphi/2970/olkola.htm

**Oykangand**

Philip J. Hamilton. 1997b. *Uw Olkola and Uw Oykangand Alphabetical Search Index*. Oykangand and Olkola Multimedia Dictionary. http://www.oocities.org/athens/delphi/2970/olkola.htm

**Panyjima**

Alan C. Dench. 1991. Panyjima. Australian Institute of Aboriginal and Torres Strait Islander Studies, Australian Indigenous Languages Collection. ASEDA 0375. Canberra. http://aiatsis.gov.au/sites/default/files/catalogue_resources/0375_access.zip

**Payungu** CHIRILA/v1/ASEDA0394

Peter K. Austin. N.d.(d). Payungu - English dictionary. Australian Institute of Aboriginal and Torres Strait Islander Studies, Australian Indigenous Languages Collection. ASEDA 0394. Canberra

**PittaPitta** CHIRILA/v1/bla0275

Barry J. Blake. 1990b. Pitta Pitta wordlist. Australian Institute of Aboriginal and Torres Strait Islander Studies, Australian Indigenous Languages Collection. ASEDA 0275. Canberra

**Purduna**

Albert Burgman. 2007. *Burduna dictionary: English-Burduna wordlist and thematic wordlist.* In collab. with Wangka Maya Pilbara Aboriginal Language Centre. South Hedland, WA, Australia: Wangka Maya Pilbara Aboriginal Language Centre. 86 pp.

**Ritharrngu** CHIRILA/v1/Heath

Jeffrey Heath. 1976. Ritharngu. In R. M. W. Dixon (ed.), *Grammatical categories in Australian languages* (Linguistic series 22), 285–287. Canberra: Australian Institute of Aboriginal Studies

**Southern Paakintyi**

Luise A. Hercus. N.d.(a). Paakantyi dictionary. Australian Institute of Aboriginal and Torres Strait Islander Studies, Australian Indigenous Languages Collection. ASEDA 0525. Canberra. http://aiatsis.gov.au/sites/default/files/catalogue_resources/0525_access.zip

**Thaayorre**

Tom Foote & Allen Hall. 1993. *Kuuk Thaayorre dictionary: Thaayorre/English ; september, 1966-92.* Brisbane: Jolien Press. 239 pp.

**Thalanyji** CHIRILA/v2/ASEDA0437

Peter K. Austin. N.d.(b). A dictionary of Thalanyji. Australian Institute of Aboriginal and Torres Strait Islander Studies, Australian Indigenous Languages Collection. ASEDA 0437. Canberra

**Tharrkari**

Peter K. Austin. 1992. *A dictionary of Tharrgari, Western Australia.* Bundoora, Victoria, Australia: La Trobe University. 60 pp.

**Umpila**

Geoffrey N. O'Grady. 1988. Umpila wordlist. Australian Institute of Aboriginal and Torres Strait Islander Studies, Australian Indigenous Languages Collection. ASEDA 0094. Canberra. http://aiatsis.gov.au/sites/default/files/catalogue_resources/0094_access.zip

**Wangkumara** CHIRILA/v1/robnd

Carol Robertson. 1985. *Wangkumara grammar and dictionary.* Sydney: Department of Technical & Further Education, Aboriginal Education Unit. 90 pp.

**Warluwarra**

Gavan Breen. 1990. Warluwara grammar and wordlist. Australian Institute of Aboriginal and Torres Strait Islander Studies, Australian Indigenous Languages Collection. ASEDA 0253. Canberra

**Warriyangga**

Peter K. Austin. N.d.(c). A dictionary of Warriyangka. Australian Institute of Aboriginal and Torres Strait Islander Studies, Australian Indigenous Languages Collection. ASEDA 0439. Canberra

**Watjarri**

Doreen Mackman (ed.). 2012. *Wajarri dictionary: The language of the Murchison Region of Western Australia.* In collab. with Irra Wangga Language Centre & Yamaji Language Aboriginal Corporation. Geraldton, WA, Australia: Irra Wangga Language Centre. 249 pp. http://www.bundiyarra.com.au/wajarriApp/ (23 July, 2018)

**Wayilwan** CHIRILA/v2/Wail-lex

John Giacon. N.d. Wailwan wordlist, from recordings by Janet Matthews. Canberra

**Wemba Wemba**

Luise A. Hercus. 1992b. *Wembawemba dictionary*. Canberra: L.A. Hercus. 116 pp.

**Western Arrernte**

Gavan Breen. 2000. *Introductory dictionary of Western Arrernte*. In collab. with John Pfitzner. Alice Springs, NT, Australia: IAD Press. 120 pp.

**Wik Mungkan**

Christine Kilham et al. 2011. *Wik Mungkan-English Interactive Dictionary*. AuSIL Interactive Dictionary Series A-6. In collab. with Charles E. Grimes & Maarten Lecompte. http://ausil.org/Dictionary/Wik-Mungkan/lexicon/mainintro.htm (26 July, 2018)

**Wirangu**

Luise A. Hercus. 1999. *A grammar of the Wirangu language from the West Coast of South Australia* (Pacific Linguistics Series C 150). Canberra: Pacific Linguistics. 239 pp. https://doi.org/10.15144/PL-C150

**Yadhaykenu**

Terry Crowley. 1983. Uradhi. In R. M. W. Dixon & Barry J. Blake (eds.), *Handbook of Australian languages*, vol. 3, 5 vols., 307–428. Amsterdam: John Benjamins

**Yalarnnga** CHIRILA/v1/ASEDA0204

Gavan Breen & Barry J Blake. N.d. Yalarnnga vocab. Australian Institute of Aboriginal and Torres Strait Islander Studies, Australian Indigenous Languages Collection. ASEDA 0204. Canberra

**Yandruwandha**

CHIRILA source: CHIRILA/v1/breyandr

Gavan Breen. 2004. *Innamincka talk: A grammar of the Innamincka dialect of Yandruwandha with notes on other dialects* (Pacific Linguistics 558). Canberra: Pacific Linguistics. 245 pp. https://doi.org/10.15144/PL-558

**Yanyuwa**

John Bradley. N.d. Yanyuwa dictionary. Australian Institute of Aboriginal and Torres Strait Islander Studies, Australian Indigenous Languages Collection. ASEDA 0382. Canberra

**Yarluyandi** CHIRILA/v1/ASEDA0251

Luise A. Hercus. N.d.(b). Yarluyandi vocabulary. Australian Institute of Aboriginal and Torres Strait Islander Studies, Australian Indigenous Languages Collection. ASEDA 0251. Canberra. http://aiatsis.gov.au/sites/default/files/catalogue_resources/0251_access.zip

**Yindjibarndi**

Bruce Anderson, E. Richards & Summer Institute of Linguistics. N.d. Yindjibarndi dictionary. Australian Institute of Aboriginal and Torres Strait Islander Studies, Australian Indigenous Languages Collection. ASEDA 0297. Canberra. http://aiatsis.gov.au/sites/default/files/catalogue_resources/0297_access.zip

**Yinhawangka**

Wangka Maya Pilbara Aboriginal Language Centre. 2008. *Yinhawangka dictionary: English-Yinhawangka wordlist and topical wordlists 2008: draft 1*. South Hedland, WA, Australia: Wangka Maya Pilbara Aboriginal Language Centre. 92 pp.

**Yir Yoront**

Barry Alpher. 1991. *Yir-Yoront lexicon: Sketch and dictionary of an Australian language*. Vol. 6 (Trends in Linguistics: Documentation). Berlin, New York: Mouton de Gruyter

**YortaYorta** CHIRILA/v1/bowmor99

Heather Bowe & Stephen Morey. 1999. *The Yorta Yorta (Bangerang) language of the Murray Goulburn including Yabula Yabula* (Pacific Linguistics Series C 154). Canberra: Pacific Linguistics. 286 pp.

**Yuwaalaraay**

CHIRILA source: CHIRILA/v1/ash03

Anna Ash, John Giacon & Amanda Lissarrague. 2003. *Gamilaraay, Yuwaalaraay & Yuwaalayaay dictionary.* Alice Springs, NT, Australia: IAD Press. 344 pp.

## 2 Pama-Nyungan phylogenetic data

### 2.1 Reference phylogeny

Quantifying phylogenetic signal requires a reference phylogeny, and to account for phylogenetic uncertainty, a set of trees can be used (see Section 4 of the main paper). In Section 6, our reference phylogeny is a set of 100 Pama-Nyungan phylogenies with 285-tips from Bowern (2015). The tree file is available at https://doi.org/10.5281/zenodo.5602216. For more detail on the inference of the tree, see Macklin-Cordes, Bowern & Round (2021). The following table matches the language varieties analysed in Section 6 to the tip labels in the tree.

| Variety | Glottocode | Tip label |
|---|---|---|
| Adnyamathanha | adny1235 | Adnyamathanha |
| Aghu Tharnggala | aghu1254 | AghuTharrnggala |
| Alngith | alng1239 | Alngith |
| Alyawarr | alya1239 | Alyawarr |
| Anguthimri | angu1242 | Mbakwithi |
| Antekerrepenhe | ande1247 | Antekerrepenhe |
| Arabana | arab1267 | Arabana |
| Awabakal | awab1243 | Awabakal |
| Ayapathu | ayab1239 | Ayapathu |
| Badimaya | badi1246 | Badimaya |
| Bakanh | paka1251 | Pakanh |
| Barna | biri1256 | Barna |
| Bidyara | bidy1243 | BidyaraGungabula |
| Bilinarra | bili1250 | Bilinarra |
| Biri | biri1256 | Biri |
| Bularnu | bula1255 | Bularnu |
| Bundjalung | yugu1249 | Yugambeh |
| Butchulla | baty1234 | Batyala |
| Buwandik | bung1264 | Bunganditj |
| Darkinyung | hawk1239 | Darkinyung |
| Dhangu | dhan1270 | Dhangu |
| Dharawal | thur1254 | Dharawal |
| Dharawala | bidy1243 | Dharawala |
| Dharuk | sydn1236 | Dharuk |
| Dharumbal | dhar1248 | Dharumbal |
| Dhay'yi | dhal1246 | Dhayyi |
| Dhudhuroa | dhud1236 | Dhudhuroa |
| Dhurga | dhur1239 | Dhurga |
| Diyari | dier1241 | Diyari |
| Djabugay | dyaa1242 | Djabugay |
| Djabwurung | djab1234 | Tjapwurrung |
| Djambarrpuyngu | djam1256 | Djambarrpuyngu |
| Djapu | djap1238 | Djapu |
| Djinang | djin1253 | Djinang |
| Djinba | djin1252 | Djinba |
| Duungidjawu | duun1241 | Duungidjawu |
| Dyirbal | dyir1250 | Dyirbal |
| Eastern Arrernte | east2379 | EasternArrernte |
| Eora | sydn1236 | Iyora |
| Flinders Island | flin1247 | FlindersIsland |
| Gamilaraay | gami1243 | Gamilaraay |
| Gangulu | gang1268 | Gangulu |
| Garlali | kala1380 | Garlali |
| Gidabal | gida1240 | Githabul |
| Gudjal | gudj1237 | Gudjal |

| | | |
|---|---|---|
| Gugu Badhun | gugu1253 | GuguBadhun |
| Gumatj | guma1253 | Gumatj |
| Gumbaynggir | kumb1268 | Gumbaynggir |
| Gundungurra | gund1248 | Gundungurra |
| Gunggari | kung1258 | Gunggari |
| Gunya | guny1241 | Gunya |
| Gupapuyngu | gupa1247 | Gupapuyngu |
| Gureng-gureng | gure1255 | GoorengGooreng |
| Gurindji | guri1247 | Gurindji |
| Guugu Yimidhirr | gugu1255 | GuuguYimidhirr |
| Guwa | guwa1242 | Guwa |
| Guwamu | guwa1243 | Guwamu |
| Guwar | guwa1244 | Guwar |
| Ikarranggal | ikar1243 | Ikarranggal |
| Jaru | jaru1254 | Jaru |
| Jiwarli | djiw1241 | Jiwarli |
| Kaantju | kanj1260 | Kaanju |
| Kala Kawaw Ya | kala1378 | KKY |
| Kala Lagaw Ya | kala1377 | KLY |
| Kalkatungu | kalk1246 | Kalkatungu |
| Kaniyang | kani1276 | Kaniyang |
| Karajarri | kara1476 | Karajarri |
| Kariyarra | kari1304 | Kariyarra |
| Kartujarra | kart1247 | Kartujarra |
| Katthang | wori1245 | Katthang |
| Kaurna | kaur1267 | Kaurna |
| Kaytetye | kayt1238 | Kaytetye |
| Keramin | kera1256 | Keramin |
| Kok Nar | kokn1236 | KokNar |
| Koko Bera | gugu1254 | KokoBera |
| Kolakngat | cola1237 | Colac |
| Kugu Nganhcara | wikn1246 | KuguNganhcara |
| Kukatj | guga1239 | Kukatj |
| Kukatja | kuka1246 | Kukatja |
| Kuku Yalanji | kuku1273 | KukuYalanji |
| Kungardutji | wong1246 | Kungadutyi |
| Kungkari | kuun1236 | Kungkari |
| Kurnu | kula1275 | Kurnu |
| Kurrama | kurr1243 | Kurrama |
| Kurtjar | gurd1238 | Kurtjar |
| Kuugu Ya'u | kuuk1238 | KuukuYau |
| Linngithigh | leni1238 | Linngithigh |
| Madhi-Madhi | madh1244 | MathiMathi |
| Malkana | malg1242 | Malgana |
| Malngin | maln1239 | Malngin |
| Malthanmungu | barr1247 | BarrowPoint |
| Malyangapa | maly1234 | Malyangapa |
| Mangala | mang1383 | MangalaMcK |
| Manjiljarra | many1256 | Manjiljarra |
| Margany | marg1253 | Margany |
| Martuthunira | mart1255 | Martuthunira |
| Mayi-Kulan | mayk1239 | MayiKulan |
| Mayi-Kutuna | maya1280 | MayiKutuna |
| Mayi-Thakurti | ngaw1240 | MayiThakurti |
| Mayi-Yapi | mayk1239 | MayiYapi |
| Mbabaram | mbab1239 | Mbabaram |

13

| Mbiywom | mbiy1238 | Mbiywom |
|---|---|---|
| Minjungbal | minj1242 | Minjungbal |
| Mirniny | mirn1243 | Mirniny |
| Mithaka | mith1236 | Mithaka |
| Mpalitjanh | mpal1238 | Mpalityan |
| Mudburra | mudb1240 | Mudburra |
| Muruwari | muru1266 | Muruwari |
| Narrungga | naru1238 | Narrungga |
| Ndra'ngith | tyan1235 | Ntrangith |
| Ngaanyatjarra | ngaa1240 | Ngaanyatjarra |
| Ngadjunmaya | ngad1258 | Ngadjumaya |
| Ngamini | ngam1284 | Ngamini |
| Nganyaywana | ngan1296 | Nganyaywana |
| Ngardily | west2437 | Ngardily |
| Ngarigu | ngar1297 | Ngarigu |
| Ngarinyman | ngar1235 | Ngarinyman |
| Ngarla | ngar1296 | Ngarla |
| Ngarluma | ngar1287 | Ngarluma |
| Ngawun | ngaw1240 | Ngawun |
| Ngayawang | uppe1415 | Ngaiawang |
| Ngiyambaa | wang1291 | Ngiyambaa |
| Ngkoth | ngko1236 | Nggoth |
| Ngunawal | ngun1277 | Ngunawal |
| Nhanda | nhan1238 | Nhanta |
| Nhangu | yann1237 | Yannhangu |
| Nhirrpi | nhir1234 | Nhirrpi |
| Nukunu | nugu1241 | Nukunu |
| Nyamal | nyam1271 | Nyamal |
| Nyangumarta | nyan1301 | Nyangumarta |
| Nyawaygi | nyaw1247 | Nyawaygi |
| Nyiyaparli | nija1241 | Nyiyaparli |
| Ogh Unyjan | kawa1290 | Kunjen |
| Olkol | ulku1238 | Olkola |
| Oykangand | oyka1239 | UwOykangand |
| Pallanganmiddang | pall1243 | Pallanganmiddang |
| Panyjima | pany1241 | Panyjima |
| Parnkalla | bang1339 | Parnkala |
| Payungu | bayu1240 | Payungu |
| Piangil | wadi1260 | Piangil |
| Pintupi | pint1250 | PintupiLuritja |
| Pirriya | pirr1240 | Pirriya |
| Pitjantjatjara | pitj1243 | Pitjantjatjara |
| Pitta Pitta | pitt1247 | PittaPitta |
| Punthamara | punt1240 | Punthamara |
| Purduna | burd1238 | Purduna |
| Ritharrngu | rita1239 | Ritharrngu |
| Southern Paakintyi | wilj1239 | Paakantyi |
| Takalak | taga1279 | Tagalag |
| Thaayorre | thay1249 | KuukThaayorre |
| Thalanyji | dhal1245 | Thalanyji |
| Thanggati | dyan1250 | Thanggatti |
| Tharrkari | dhar1247 | Tharrgari |
| Thaynakwithi | tyan1235 | Thaynakwith |
| Turubul | yaga1256 | Durubul |
| Umpila | umpi1239 | Umpila |
| Umpithamu | umbi1243 | Umpithamu |

| Uradhi | urad1238 | Uradhi |
|---|---|---|
| Waalubal | band1358 | Bandjalang |
| Wadikali | wadi1261 | Wadikali |
| Wajuk | nyun1247 | Watjuk |
| Waka Waka | waka1274 | WakaWaka |
| Walangama | wala1263 | Walangama |
| Walmajarri | walm1241 | WalmajarriHR |
| Wangkajunga | wang1288 | Wangkajunga |
| Wangkangurru | wang1290 | Wangkangurru |
| Wangkatja | pini1245 | Wangkatja |
| Wangkayutyuru | wang1289 | Wangkayutyuru |
| Wangkumara | wong1246 | Wangkumara |
| Wardandi | ward1248 | Wardandi |
| Warlmanpa | warl1255 | Warlmanpa |
| Warlpiri | warl1254 | Warlpiri |
| Warluwarra | warl1256 | Warluwarra |
| Warnman | wanm1242 | Warnman |
| Warrgamay | warr1255 | Wargamay |
| Warriyangga | wari1262 | Warriyangga |
| Warrnambool | warr1257 | Warrnambool |
| Warumungu | waru1265 | Warumungu |
| Warungu | waru1264 | Warungu |
| Wathawurrung | wath1238 | Wathawurrung |
| Wathi Wathi | wadi1260 | Wathiwathi |
| Watjarri | waja1257 | Wajarri |
| Wayilwan | wayi1238 | Wailwan |
| Wemba Wemba | wemb1241 | WembaWemba |
| Western Anmatyerre | west2442 | CentralAnmatyerr |
| Western Arrernte | west2441 | WesternArrarnta |
| Western Wakaya | waga1260 | Wakaya |
| Wik Mungkan | wikm1247 | WikMungkan |
| Wiradjuri | wira1262 | Wiradjuri |
| Wirangu | wira1265 | Wirangu |
| Woiwurrung | woiw1238 | Woiwurrung |
| Wulguru | wulg1239 | Wulguru |
| Yabula-Yabula | yabu1234 | YabulaYabula |
| Yadhaykenu | yadh1237 | Yadhaykenu |
| Yagara | yaga1262 | Yagara |
| Yalarnnga | yala1262 | Yalarnnga |
| Yambina | biri1256 | Yambina |
| Yanda | yand1251 | Yanda |
| Yandruwandha | yand1253 | Yandruwandha |
| Yanyuwa | yany1243 | Yanyuwa |
| Yaraldi | narr1259 | Ngarrindjeri |
| Yardliyawarra | yarl1236 | Yardliyawarra |
| Yarluyandi | yarl1238 | Yarluyandi |
| Yawarrawarrka | yawa1258 | Yawarrawarrka |
| Yaygir | yayg1236 | Yaygirr |
| Yidiny | yidi1250 | Yidiny |
| Yindjibarndi | yind1247 | Yindjibarndi |
| Yingkarta | ying1247 | Yingkarta |
| Yinhawangka | yinh1234 | Yinhawangka |
| Yiningay | bidy1243 | Yiningay |
| Yinwum | yinw1236 | Yinwum |
| Yir Yoront | yiry1245 | YirYoront |
| Yitha Yitha | lowe1403 | YithaYitha |

| Yorta Yorta | yort1237 | YortaYorta |
| Yulparija | yulp1239 | Yulparija |
| Yuwaalaraay | yuwa1242 | Yuwaalaraay |

# 3 Data files

Data files are archived at https://doi.org/10.5281/zenodo.5602216, in the following directory structure:

- `data`
  - `laminals_binary.csv` a file in CSV format of languages and the presence/absence of the laminal contrast.
  - `laminals_continuous.csv` a file in CSV format of languages and continuous phonotactic variable data.
  - `PN_100_tree_sample.trees` a file in Newick format containing the Pama-Nyungan tree sample.
- `results`
  - `D_results.csv` a file in CSV format of D test results (cf Section 6.1).
  - `K_results.csv` a file in CSV format of K test results (cf Section 6.2).
  - `PamaNyungan_ACL_proportions.csv` a file in CSV format of genealogically-weighted proportions (cf Section 6.3).
  - `PamaNyungan_BM_proportions.csv` a file in CSV format of genealogically-weighted proportions (cf Section 6.3).

## 4    Code

The following code in R was used to calculate results. For further information about the package PhyloWeights, see section S1.

```r
# Load packages
# Any missing packages can be installed with the following command:
# install.packages("packagename")
library(caper)
library(ape)
library(dplyr)
library(stringr)
library(phangorn)
library(phytools)
library(zen4R)

# Uncomment below to install phyloWeights, if it is not already installed:
# devtools::install_github("erichround/phyloWeights")
library(phyloWeights)

# Download and unzip data from Zenodo
download_zenodo("10.5281/zenodo.5602216")
unzip("Supplementary_data.zip")
setwd("Supplementary_data")

# Note that the Zenodo repository already contains original results files.
# Uncomment below to empty the results folder (ready for new output) if desired.
# file.remove(list.files("results/", full.names = TRUE))

# Read the tree sample
PN_sample <- read.tree("data/PN_100_tree_sample.trees")


########
# D TEST
########

# Read the data
laminals_binary <-
  read.csv("data/laminals_binary.csv",
           stringsAsFactors = FALSE)

# Create data dataframe
df <- laminals_binary %>%
  dplyr::select(tip_label, laminal_contrast)

# Plot data
mcc_tree <- phangorn::maxCladeCred(PN_sample) %>% keep.tip(df$tip_label)
tip_order <- match(mcc_tree$tip.label, df$tip_label)
tip_val <- df$laminal_contrast[tip_order]
plot(mcc_tree, type = "fan", open.angle = 160, rotate.tree = -9,
     edge.color = "grey50", show.tip.label = FALSE)
tiplabels(pch = 19 - tip_val * 18, col = "grey30", cex = 0.9, offset = 100)

# Set seed for reproducibility of stochastic processes
set.seed(1)
```

18

```r
# Results dataframe with one row per tree
lapply(1:100,
       function(i){
         # Create caper comparative data object
         cd <- comparative.data(phy = PN_sample[[i]],
                                data = df,
                                names.col = tip_label)

         # Perform D test
         d <- phylo.d(cd,
                      binvar = laminal_contrast,
                      names.col = tip_label)

         cat(i, "")

         # Output results as one-row dataframe
         data.frame(phy = i,
                    D = d$DEstimate,
                    pval_random = d$Pval1,
                    pval_brownian = d$Pval0)
       }
     ) %>%
  bind_rows() %>%
  write.csv(
    "results/D_results.csv", row.names = FALSE)


########
# K TEST
########

# Read the data
laminals_continuous <-
  read.csv("data/laminals_continuous.csv",
           stringsAsFactors = FALSE)

# Set seed for reproducibility of stochastic processes
set.seed(1)

# Create a list of data dataframes, each with
# the number of observations (n) and proportion
# pre-palatal (p) for one variable, plus
# the tip labels
n_vars <- (ncol(laminals_continuous) - 3) / 2
kdat_list <-
  lapply(1:n_vars,
         function(i) {
           # Select the needed columns
           df <- laminals_continuous %>%
             dplyr::select(tip_label, i*2 + 2, i*2 + 3)
           colnames(df)[2:3] <- c("n", "p")
           # Keep only languages with at least
           # 20 observations
```

19

```r
          df <- filter(df, n >= 20)
        })
variable_names <-
  colnames(laminals_continuous)[(1:n_vars)*2 + 3] %>%
  str_sub(3) %>%
  str_replace("\\.", "#")

# Results dataframe with one row per
# tree/variable combination
lapply(
  1:100,
  function(i) {
    # Results dataframe for one tree,
    # with one row per variable
    lapply(
      1:n_vars,
      function(j) {
        if (j==1) { cat(i, "") }
        dat <- kdat_list[[j]]$p
        names(dat) <- tips <-
          kdat_list[[j]]$tip_label

        # Perform K test, including test for
        # significant phylogenetic signal
        k <- phylosig(
          tree = PN_sample[[i]] %>% keep.tip(tips),
          x = dat, test = TRUE)

        # Output results as one-row dataframe
        data.frame(
          phy = i,
          variable = variable_names[j],
          K = k$K,
          pval = k$P,
          n_lang = length(dat),
          mean_n = mean(kdat_list[[j]]$n))
      }) %>%
      bind_rows()
  }) %>%
  bind_rows() %>%
  mutate(
    context = str_extract(variable, "[CV#]_|_[CV]"),
    neighbour_V = str_extract(variable, "[aiu]")
    ) %>%
  write.csv("results/K_results.csv",
            row.names = FALSE)


#####################################
# Genealogically-weighted proportions
#####################################

# Read the typological data
```

```
dat <-
  read.csv("data/laminals_binary.csv", stringsAsFactors = FALSE) %>%
  mutate(has_contrast = 1 * laminal_contrast) %>%
  select(tip = tip_label, has_contrast)

# Remove unused languages from trees
PN_sample_pruned <- PN_sample
for (i in 1:length(PN_sample_pruned)) {
  PN_sample_pruned[[i]] <- PN_sample_pruned[[i]] %>% keep.tip(dat$tip)
}

# Get genealogically-weighted proportions
phy_ave <- phylo_average(phy = PN_sample_pruned, data = dat)

# Write to file
write.csv(
  phy_ave$ACL_averages,
  "results/PamaNyungan_ACL_proportions.csv",
  row.names = FALSE)
write.csv(
  phy_ave$BM_averages,
  "results/PamaNyungan_BM_proportions.csv",
  row.names = FALSE)
```

# References

Bowern, Claire. 2015. Pama-Nyungan phylogenetics and beyond [plenary address]. In *Lorentz center workshop on phylogenetic methods in linguistics*. Leiden University, Leiden, Netherlands. https://doi.org/10.5281/zenodo.3032846.

Bowern, Claire. 2016. Chirila: contemporary and historical resources for the indigenous languages of australia. *Language Documentation and Conservation* 10. http://hdl.handle.net/10125/24685.

Macklin-Cordes, Jayden L, Claire Bowern & Erich R. Round. 2021. Phylogenetic signal in phonotactics. *Diachronica*. https://doi.org/10.1075/dia.20004.mac.

R Core Team. 2021. *R: A Language and Environment for Statistical Computing*. R Foundation for Statistical Computing. Vienna, Austria. https://www.R-project.org/.

Round, Erich R. 2017. The AusPhon-Lexicon project: 2 million normalized segments across 300 Australian languages. 47th Poznań Linguistic Meeting.

Round, Erich R. 2019. Phonemic inventories of Australia [Database of 392 languages]. In Steven Moran & Daniel McCloy (eds.), *PHOIBLE 2.0*. Jena: Max Planck Institute for the Science of Human History.



\end{document}